\documentclass[11pt,letterpaper]{article}
\usepackage[margin=1in]{geometry}
\usepackage[T1]{fontenc}
\usepackage{lmodern,microtype}
\usepackage{amsmath,amssymb,amsthm,mathtools}
\usepackage{enumitem,booktabs,array}
\usepackage{graphicx,xcolor,float,caption}
\usepackage{algorithm,algpseudocode,needspace}
\usepackage[numbers,sort&compress]{natbib}
\usepackage{xurl}
\usepackage[colorlinks=true,linkcolor=blue!45!black,citecolor=blue!45!black,
  urlcolor=blue!45!black]{hyperref}
\hypersetup{pdftitle={Improved Downward-Closed Delegation Bounds},
  pdfsubject={Delegated choice with downward-closed feasibility constraints}}
\setlist{itemsep=2pt,topsep=4pt}
\newtheorem{theorem}{Theorem}[section]
\newtheorem{lemma}[theorem]{Lemma}
\newtheorem{proposition}[theorem]{Proposition}
\newtheorem{corollary}[theorem]{Corollary}

\theoremstyle{definition}
\newtheorem{definition}[theorem]{Definition}
\theoremstyle{remark}

\numberwithin{equation}{section}

\newcommand{\E}{\mathbb E}
\newcommand{\Prb}{\mathbb P}

\newcommand{\PP}{\mathbb P}

\newcommand{\F}{\mathcal F}
\newcommand{\ind}{\mathbf 1}
\newcommand{\Var}{\operatorname{Var}}
\newcommand{\Cov}{\operatorname{Cov}}
\newcommand{\od}{\operatorname{odds}}
\newcommand{\pos}[1]{(#1)_+}

\title{\vspace{-1.0em}Improved Downward-Closed Delegation Bounds}
\author{%
Hamed Abdi\thanks{University of Maryland, College Park. Email: \texttt{hamedabdi37@gmail.com}.}
\and Kiarash Banihashem\thanks{University of Maryland, College Park. Email: \texttt{kiarash@umd.edu}.}
\and MohammadTaghi Hajiaghayi\thanks{University of Maryland, College Park. Email: \texttt{hajiaghayi@gmail.com}.}
\and Danny Mittal\thanks{University of Maryland, College Park. Email: \texttt{dannymittal@gmail.com}.}}

\begin{document}
\maketitle

\begin{abstract}
We study delegated choice under a downward-closed feasibility constraint, where
the principal and agent values of each item may be arbitrarily correlated but
items are independent. We give a deterministic, agent-oblivious menu that
guarantees the principal more than $29/70\approx0.41429$ of the expected offline
optimum under every best response, including ties resolved against the principal.
The menu depends only on the feasible family and the principal's value
distributions, and verifies a proposal using only its items and their principal
values. The previous best guarantee for downward-closed constraints was $0.1098$.

Matroid constraints admit a guarantee of one half, which is tight already in rank
one, but downward-closed constraints do not: a five-item knapsack instance bounds every
deterministic menu by $500/1011$ of the optimum, and no rule with randomized
acceptance, even one that knows the agent's utilities, can guarantee more than
$30/61$. An explicit computer-assisted family of instances lowers the bound to
$0.48105$ for all such rules.

The guarantee is proved by contradiction. If every menu in a fixed family earns at
most a $29/70$ fraction of the optimum, then a normalized instance satisfies a
finite system of affine inequalities among the survival functions, stop-loss
transforms and overshoot means of its clipped optima. An exact rational
certificate shows that the system is infeasible. We also give a short analytic
proof that a single threshold on the reported total value guarantees $3/10$.
\end{abstract}

\section{Introduction}\label{sec:intro}

In the problem of \emph{delegated choice}, a decision maker, known as the \emph{principal}, must select
one of several items, but does not observe their values. The values are drawn from
known distributions and are observed by an agent, whose own value for each item may
differ from the principal's. The principal cannot pay the agent and cannot search
for the items itself. Instead, before the values are realized, it commits to a rule
that specifies which proposals it will accept; a proposal names an item and reports
its value to the principal, which the principal can verify. The agent then proposes
the acceptable item that it values most, or nothing, and the principal compares its
resulting expected value with the expected value of the best item, which it could
obtain by choosing with full information. This model goes back to
Holmstr\"om~\cite{holmstrom1984} and has been studied extensively in
economics~\cite{melumadshibano1991,alonsomatouschek2008,armstrongvickers2010}.
Kleinberg and Kleinberg~\cite{kleinbergkleinberg2018} brought an approximation
viewpoint to the problem and showed, through a connection with prophet
inequalities, that a suitable rule guarantees the principal half of this benchmark.

We study the combinatorial version, in which the agent proposes a set of items
subject to a downward-closed feasibility constraint $\F$. Item $i$ has a principal
value $X_i\ge0$ and an agent value $Y_i>0$. The pairs $(X_i,Y_i)$ are independent
across items, while $X_i$ and $Y_i$ may be arbitrarily correlated. The principal
commits to an acceptance rule that observes the identities and principal values of
the proposed items. The agent observes all values, proposes a feasible set, and
breaks ties against the principal. The benchmark is the expected offline optimum
$V=\E\max_{S\in\F}X(S)$.

For matroid constraints the optimal guarantee is known. Bechtel and
Dughmi~\cite{bechtel2021} and Banihashem, Hajiaghayi, Krysta and
Shin~\cite{banihashem2025} related delegation under combinatorial constraints to
prophet inequalities against an adversary that sees all values, which yields a
guarantee of $1/4$ for matroids. A deterministic menu that depends only on the
matroid and the principal's value distributions guarantees $V/2$~\cite{abhm2026matroid},
and no deterministic menu does better already in rank one~\cite[Proposition~5.3]{bechtel2021}.

Downward-closed families are much richer. They include knapsack constraints,
matchings, independent sets of graphs, and intersections of several matroids. For
the prophet inequality the difference is drastic: no online algorithm achieves a
constant fraction of the optimum for general downward-closed
families~\cite{rubinstein2016}. Banihashem et al.~\cite{banihashem2025} showed that
delegation behaves differently, by giving an agent-oblivious mechanism that
guarantees $0.1098\,V$ for every downward-closed family. The best upper bound has
been the bound of $1/2$ that holds already in rank one~\cite{bechtel2021}.

\subsection{Our results}

Our main result is a constant guarantee by a deterministic menu that uses only the
feasible family and the laws of the principal values.

\begin{theorem}[Restatement of Theorem~\ref{thm:main}]\label{thm:intro-main}
For every downward-closed family and every instance, there is a deterministic
menu, depending only on $\F$ and the laws of the principal values, under which the
principal's expected payoff exceeds $\frac{29}{70}V\approx0.41429\,V$ for every best
response of the agent.
\end{theorem}

The menu checks only the submitted items and their principal values, and it never
needs the agent's utilities, their distributions, or the correlation between the
two values of an item. The proof shows that one of a fixed family of menus works,
and selects it by computing expected scores (Algorithm~\ref{alg:select}). For finite
rational supports this computation is a finite exact enumeration; we make no
claim about its running time.

The constant $1/2$ that is achievable for matroids is out of reach for
downward-closed families.

\begin{theorem}\label{thm:intro-hard}
There are downward-closed instances on which
\begin{enumerate}[label=(\alph*)]
\item every deterministic menu earns at most $\frac{500}{1011}V$, on a knapsack
 instance with five items;
\item every rule with randomized acceptance earns at most $\frac{5011}{10110}V$ on
 an instance of the same form, and no universal guarantee above $30/61$ is
 possible for such rules, even when the rule knows the agent utilities and ties
 are broken in the principal's favor;
\item every rule with randomized acceptance earns less than $0.48105\,V$, on a
 computer-assisted family of instances.
\end{enumerate}
\end{theorem}

Parts~(a) and~(b) are proved in Section~\ref{sec:hardness-small} by short direct
arguments, and part~(c) in Appendix~\ref{app:hardness}. Together,
\[
 \tfrac{29}{70}<\alpha_D\le\alpha_R<0.48105,
\]
where $\alpha_D$ and $\alpha_R$ are the best universal guarantees of deterministic
and randomized agent-oblivious rules. In particular delegation under downward-closed
constraints is strictly harder than under matroid constraints, even for rules that
may randomize and know the agent's utilities.

We also give, in Section~\ref{sec:warmup}, a self-contained analytic proof that a
single threshold on the reported total value guarantees $3V/10$.

\subsection{Techniques}

\paragraph{Scores.}
Because every agent value is positive, an agent never submits an accepted report
that has an accepted strict superset. For a deterministic menu, the payoff under
any best response is therefore at least the \emph{score}: the least principal value
of an inclusion-maximal accepted report (Lemma~\ref{lem:score}). The score depends
only on the principal values, which removes the agent from the analysis. All our
guarantees are lower bounds on expected scores.

\paragraph{The warm-up.}
For a threshold $z$ on the reported total, every accepted report has value at
least $z$, and an item $i$ whose removal makes the threshold unreachable lies in
every accepted report. Its excess over $z$ is credited in full, and the event that
it is forced is independent of its own value. If no threshold earns more than one,
this gives the tail bound $\Prb(W\ge x)\le1/x$ for $x>1$. An Efron--Stein bound shows
that the optimum with values clipped at $3/2$ has variance at most $3/2$ times its
mean, and a one-dimensional calculation then shows that the mean of the optimum is
less than $10/3$.

\paragraph{The main proof.}
The guarantee of $29/70$ follows the same pattern with many more menus. Suppose
every menu in the family $\mathcal M$ of Section~\ref{sec:menus} earns at most
$\frac{29}{70}V$. After a normalization that adds a safe item of value one
(Lemma~\ref{lem:normalization}), every menu earns at most one, and the optimum has
mean $R$ slightly below $70/29$. Each menu, combined with a pointwise argument about which items
every acceptable report must contain, gives an affine inequality between quantities
of the instance: survival probabilities and stop-loss transforms of the optimum
with values clipped at a few caps, evaluated on a grid, the clipped means, the
expected excess above the cap, and the mean overshoot of a menu above its level. A
covariance inequality for clipped optima (Section~\ref{sec:profile}), several
bounds on the overshoot of a menu above its level (Section~\ref{sec:overshoot}),
and their combination at many levels on a fine grid (Section~\ref{sec:tail})
supply further inequalities of the same kind. The most detailed inequalities follow, in each realization, the
pivotal set of items common to all qualifying reports as the level rises, and
charge every reached level by a dynamic program over the possible paths
(Section~\ref{sec:paths}). The resulting linear system has no solution: for each of
$96$ intervals covering the possible values of the clipped mean $\mu_1$, an explicit
positive combination of rows is positive on a box that contains the instance.

These inequalities are proved by hand, and only the last step is a computation:
the rows and their multipliers are evaluated in exact rational arithmetic,
including the dynamic programs, on every interval. The appendix's upper bound
has the same structure: an explicit family of instances, a reduction of every
rule to a finite dynamic program, and an exact rational computation.

\subsection{Related work}

\paragraph{Delegation.}
The economic literature on delegation studies the optimal set of permitted
actions when the agent's bias is known and transfers are
impossible~\cite{holmstrom1984,melumadshibano1991,alonsomatouschek2008}. Armstrong
and Vickers~\cite{armstrongvickers2010} analyze delegated project choice, in which
the agent finds a set of projects and proposes one. Kleinberg and
Kleinberg~\cite{kleinbergkleinberg2018} related delegated search to threshold
stopping rules and obtained a factor of two for a single choice. This connection
has been extended to delegated stochastic probing~\cite{bechtel2021}, delegated
Pandora's box~\cite{bechteldughmipatel2022}, and online delegated
search~\cite{braun2023}, and Khodabakhsh, Pountourakis and
Taggart~\cite{khodabakhsh2024} study simple delegation mechanisms in the
single-choice setting. Banihashem, Hajiaghayi, Krysta and Shin~\cite{banihashem2025}
introduced delegated choice under combinatorial constraints. They showed that the
reduction to prophet inequalities fails beyond matroids, and gave a mechanism with
guarantee $0.1098$ for downward-closed families. For matroids the optimal
deterministic guarantee of one half was obtained in~\cite{abhm2026matroid}.

\paragraph{Prophet inequalities and contention resolution.}
The single-item prophet inequality with guarantee $1/2$ is due to
Samuel-Cahn~\cite{samuelcahn1984}, and Kleinberg and
Weinberg~\cite{kleinbergweinberg2012} extended it to matroids. Feldman, Svensson
and Zenklusen~\cite{feldman2021} developed online contention resolution schemes,
which give prophet inequalities against an almighty adversary, and Lee and
Singla~\cite{leesingla2018} obtained optimal schemes for matroids under a known
order. For general downward-closed families, Rubinstein~\cite{rubinstein2016}
showed that no constant guarantee is possible, and Rubinstein and
Singla~\cite{rubinsteinsingla2017} studied combinatorial prophet inequalities with
submodular objectives. Our guarantee shows that delegation to an informed agent
avoids this barrier and admits a large explicit constant. The related matroid
secretary problem~\cite{babaioff2007} was recently resolved with an optimal
algorithm~\cite{abdietal2026}.

\paragraph{Probabilistic tools.}
The variance bound of Section~\ref{sec:warmup} is the Efron--Stein
inequality~\cite{efronstein1981,steele1986}. The covariance bounds of the main
proof rely on the Harris--FKG inequality for increasing functions of independent
variables~\cite{harris1960,fkg1971}.

\paragraph{Organization.}
Section~\ref{sec:prelim} defines the model, the score and the normalization.
Section~\ref{sec:hardness-small} proves the upper bounds below one half, and
Appendix~\ref{app:hardness} proves the bound of $0.48105$.
Section~\ref{sec:warmup} proves the $3/10$ guarantee. Section~\ref{sec:menus}
defines the menus and states the main theorem.
Sections~\ref{sec:profile}--\ref{sec:paths} prove the inequalities, and
Section~\ref{sec:certificate} describes the finite computation and completes the
proof.

\section{Model and preliminaries}\label{sec:prelim}

\subsection{Delegation with downward-closed constraints}

Let $E$ be a finite set of items and let $\F\subseteq 2^E$ be nonempty and
downward closed: $T\subseteq S\in\F$ implies $T\in\F$. Item $i$ has a
principal value $X_i\ge0$ and an agent value $Y_i>0$. The pairs $(X_i,Y_i)$
are independent across items; within a pair the dependence is arbitrary. Both
parties value sets additively, and we write $X(S)=\sum_{i\in S}X_i$ and
$Y(S)=\sum_{i\in S}Y_i$.

Before any value is drawn, the principal commits to an acceptance rule $q$,
which assigns to every feasible set $S$ and every vector $x_S$ of principal
values on $S$ a probability $q(S,x_S)\in[0,1]$. The agent observes all values
and submits a set $S\in\F$, or abstains. The principal observes the identities
of the submitted items and their principal values. With probability
$q(S,X_S)$ the whole set $S$ is implemented, and otherwise nothing is. The
agent therefore receives $q(S,X_S)Y(S)$ in expectation and the principal
receives $q(S,X_S)X(S)$. Abstaining gives both parties zero. The agent
best-responds, and ties are broken against the principal.

The rule is \emph{deterministic} if $q$ takes only the values $0$ and $1$. In
that case we call it a \emph{menu} and say that it \emph{accepts} or
\emph{offers} the reports with $q=1$. A rule is \emph{agent-oblivious} if it
is determined by $\F$ and the laws of the principal values; it uses neither the
laws nor the realizations of the agent values. Our algorithms are
deterministic and agent-oblivious. Our impossibility results allow randomized
acceptance and allow the rule to depend on the joint law of all values.

The benchmark is the expected offline optimum
\begin{equation}\label{eq:benchmark}
 W(x)=\max_{S\in\F}\sum_{i\in S}x_i,\qquad W=W(X),\qquad V=\E W,
\end{equation}
and we assume $0<V<\infty$. A rule is an \emph{$\alpha$-guarantee} if, on
every instance, the principal's expected payoff is at least $\alpha V$ under
every best response of the agent. We write $\alpha_D$ and $\alpha_R$ for the
best universal constants achievable by deterministic and by randomized
agent-oblivious rules, so that $\alpha_D\le\alpha_R$.

An item whose singleton is infeasible belongs to no feasible set. Such items
are deleted throughout. Every remaining item then satisfies $X_i\le W$, and
hence $\E X_i\le V<\infty$.

\subsection{The score of a menu}

For a menu $M$ and a realization $x$ of the principal values, let
$\mathcal A_M(x)$ be the family of nonempty feasible reports that $M$ accepts
at $x$. The \emph{score} of $M$ at $x$ is
\[
 b_M(x)=\begin{cases}
 \min\{x(S):S\text{ is inclusion-maximal in }\mathcal A_M(x)\},
   &\mathcal A_M(x)\ne\varnothing,\\
 0,&\mathcal A_M(x)=\varnothing.
 \end{cases}
\]
The score depends only on $\F$, the menu and the principal values. Clearly
$0\le b_M(x)\le W(x)$.

\begin{lemma}\label{lem:score}
For every menu $M$, every realization $(x,y)$ with $y_i>0$ for all $i$, and every
best response of the agent, the principal's payoff is at least $b_M(x)$.
\end{lemma}

\begin{proof}
If $\mathcal A_M(x)$ is empty, the bound is zero. Otherwise some nonempty
report is accepted and gives the agent a positive utility, so the agent does
not abstain and submits an accepted report $S$. If $S$ were strictly contained
in another accepted report $S'$, then $y(S')>y(S)$ because every agent value is
positive, contradicting optimality. Hence $S$ is inclusion-maximal in
$\mathcal A_M(x)$, and the principal receives $x(S)\ge b_M(x)$.
\end{proof}

By Lemma~\ref{lem:score}, a menu whose expected score exceeds $\alpha V$ is an
$\alpha$-guarantee, regardless of the agent-value laws and of how ties are
broken. All our guarantees are proved through the score.

\subsection{Normalization}\label{sec:normalization}

The main proof is by contradiction. It fixes a family $\mathcal M$ of menus,
described in Section~\ref{sec:menus}, and a target ratio $\rho\in(0,1)$, and
assumes that every menu in $\mathcal M$ has expected score at most $\rho V$. Put
$R=1/\rho>1$. The family $\mathcal M$ is closed under multiplying all
dimensional parameters (caps, levels and cutoffs) by a common
positive factor, and the score scales with the values: if $M'$ is obtained from
$M$ by multiplying its dimensional parameters by $\lambda>0$, then
$b_{M'}(\lambda x)=\lambda b_M(x)$.

\begin{lemma}[Safe-singleton normalization]\label{lem:normalization}
Suppose every menu in $\mathcal M$ has expected score at most $\rho V$ on some
instance. Then there is an instance with the following properties, in which
every menu of $\mathcal M$ again has expected score at most one:
\begin{enumerate}[label=(\roman*)]
\item it contains an item $s_0$ with $X_{s_0}=1$ surely, and $\{s_0\}$ is a
 maximal feasible set;
\item the remaining items form the original instance with all principal values
multiplied by $\lambda R/V$, for some $\lambda\in(0,1]$, where $V$ is the
expected optimum of the original instance;
\item $W\ge1$ surely and $\E W=R$.
\end{enumerate}
\end{lemma}

\begin{proof}
Multiply all principal values by $R/V$. The score of every menu scales by the
same factor and $\mathcal M$ is closed under rescaling, so every menu of
$\mathcal M$ now has expected score at most $\rho R=1$, and $\E W=R$. Adjoin
an item $s_0$ with $X_{s_0}=1$ and any positive agent value, feasible only as a
singleton, and multiply the original values by $\lambda\in(0,1]$. The new
optimum is $\max(1,\lambda W)$. The function
$\lambda\mapsto\E\max(1,\lambda W)$ is continuous, equals $1<R$ at
$\lambda=0$, and is at least $\E W=R$ at $\lambda=1$. Choose $\lambda$ with
$\E\max(1,\lambda W)=R$. This gives (i)--(iii).

Let $M\in\mathcal M$. If $M$ accepts $\{s_0\}$, then $\{s_0\}$ is
inclusion-maximal among accepted reports, since no feasible set strictly
contains it, so the score is at most one. Otherwise $M$ accepts no report
containing $s_0$, and its accepted reports and their maximal members are those
of the menu $M'\in\mathcal M$ obtained by dividing the dimensional parameters of
$M$ by $\lambda$, applied to the original values. Its score is therefore
$\lambda$ times the score of $M'$, whose expectation is at most one. In both
cases the expected score is at most one.
\end{proof}

All statements in
Sections~\ref{sec:profile}--\ref{sec:certificate} refer to an instance
normalized as in Lemma~\ref{lem:normalization}, for a fixed $R>1$. The
inequalities derived there hold for every $R>1$; the value of $R$ enters only
through the identity $\E W=R$ and through the hypotheses stated with each
inequality. The final computation uses $R=7000000/2900007$, the reciprocal of
$29/70+10^{-6}$.

\paragraph{Notation.}
For $c\ge1$ we clip the values at $c$ and write
\begin{gather*}
 Z_c=W\bigl((\min(X_i,c))_{i\in E}\bigr),\qquad \mu_c=\E Z_c,\qquad
 S_c(t)=\Prb(Z_c\ge t),\qquad D_c(t)=\E\pos{Z_c-t},\\
 e(c)=\sum_{i}\E\pos{X_i-c},\qquad
 b(y)=\Prb\Bigl(\max_i X_i\le y\Bigr),\qquad
 G(c)=\int_c^\infty\Bigl(\frac1{b(y)}-1\Bigr)dy.
\end{gather*}
The subscript $\infty$ refers to the unclipped optimum, so $Z_\infty=W$.
At cap one we write $w_i=\min(X_i,1)$, $w(S)=\sum_{i\in S}w_i$ and $Z=Z_1$.
In the normalized instance $Z\ge1$ surely, because of the item $s_0$.

\section{Impossibility below one half}\label{sec:hardness-small}

For matroid constraints, a deterministic agent-oblivious menu guarantees $V/2$, and
no deterministic menu does better already when a single item may be
chosen~\cite{abhm2026matroid,bechtel2021}.
This section shows that downward-closed constraints are strictly harder, with one
instance of five items. Appendix~\ref{app:hardness} gives a stronger,
computer-assisted bound.

Consider a knapsack of capacity $3$ with five items. The principal values of $A$,
$B$ and $J$ are independent, the other principal values are constant, and all agent
values are deterministic. Let $0<\eta\le1$.
\begin{center}
\begin{tabular}{cclc}
\toprule
Item & Weight & Principal value & Agent value\\
\midrule
$A$ & 1 & $9$ with probability $1/3$, otherwise $0$ & 3\\
$B$ & 2 & $9$ with probability $1/3$, otherwise $0$ & 2\\
$E$ & 2 & $3$ surely & 4\\
$C$ & 3 & $10$ surely & 8\\
$J$ & 3 & $900$ with probability $1/100$, otherwise $0$ & $\eta$\\
\bottomrule
\end{tabular}
\end{center}
The feasible family is the downward closure of $\{A,B\}$, $\{A,E\}$, $\{C\}$, and $\{J\}$. Write $AB$ and $AE$ for the two pairs. When $X_J=0$, an optimal set is $AB$ if both $A,B$ are high, $AE$ if only $A$ is high, and $\{C\}$ if $A$ is low. Therefore
\begin{equation}\label{eq:hs-offline}
 V=\frac1{100}\,900+\frac{99}{100}
 \left(\frac19\,18+\frac29\,12+\frac23\,10\right)
 =\frac{1011}{50}.
\end{equation}

\subsection{Deterministic menus}
\begin{proposition}\label{prop:hs-det}
For this instance, the largest expected payoff of any deterministic menu is $10$. Its optimal ratio is $500/1011<1/2$.
\end{proposition}
\begin{proof}
The agent strictly ranks the nonempty feasible sets as
\[
 C\succ AE\succ AB\succ E\succ A\succ B\succ J.
\]
Condition on $(X_A,X_B)=(a,b)$. Every non-jackpot report, and hence its acceptance decision, is independent of $X_J$. If any such report is accepted, the agent chooses its favorite regardless of $X_J$. If none is accepted, the conditional expected payoff is at most $\E X_J=9$.

If the menu accepts $C$, the agent always selects it and the payoff is $10$. Suppose it rejects $C$. There are two cases, determined by whether the menu accepts the report $AE$ with values $(9,3)$. Its decision cannot depend on the omitted value $X_B$.

If that report is accepted, then whenever $X_A=9$ it is chosen and pays $12$. When $X_A=0$, every non-jackpot report has value at most $9$, so the preceding conditional bound gives expected payoff at most $9$. The total is at most
\[
 \frac13\,12+\frac23\,9=10.
\]
If that report is rejected, only the state $X_A=X_B=9$ has an available non-jackpot report worth more than $9$. The conditional bound is $18$ there and $9$ elsewhere. The total is at most
\[
 \frac19\,18+\frac89\,9=10.
\]
Accepting only $C$ attains $10$. Equation~\eqref{eq:hs-offline} gives the ratio.
\end{proof}

\subsection{Randomized acceptance}
\begin{theorem}\label{thm:hs-rand}
For $\eta=1/100$, every randomized acceptance rule on the same instance has expected payoff at most $5011/500=10.022$. Thus its ratio is at most
\[
 \frac{5011}{10110}<\frac12,
\]
even with full knowledge of the agent utilities and principal-favorable tie breaking.
\end{theorem}
\begin{proof}
Fix a rule. For a state $(X_A,X_B)=(a,b)$, let $s_{ab}$ be the largest agent utility $q(S,X_S)Y(S)$ offered by a non-jackpot report. Put
\[
 r_{ab}=\max_{\substack{\varnothing\ne S\in\F\\J\notin S}}\frac{X(S)}{Y(S)}.
\]
These quantities do not depend on $X_J$. Their relevant values are
\begin{center}
\begin{tabular}{c cccc}
\toprule
$(a,b)$ & $(0,0)$ & $(0,9)$ & $(9,0)$ & $(9,9)$\\
\midrule
Probability & $4/9$ & $2/9$ & $2/9$ & $1/9$\\
$r_{ab}$ & $5/4$ & $9/2$ & $3$ & $9/2$\\
\bottomrule
\end{tabular}
\end{center}
If $s_{ab}>\eta$, the jackpot cannot be chosen. If $s_{ab}\le\eta$, each non-jackpot report has expected principal payoff
\[
 q(S,X_S)X(S)=q(S,X_S)Y(S)\frac{X(S)}{Y(S)}\le\eta r_{ab}.
\]
In the high jackpot outcome every report has expected payoff at most $900$; in the low outcome the jackpot pays zero. Hence, when $s_{ab}\le\eta$, the conditional expected payoff is at most
\begin{equation}\label{eq:hs-jackpot}
 9+\frac{99}{100}\eta r_{ab}.
\end{equation}
This includes ties with the jackpot.

Let $q_C$ be the acceptance probability of $C$. If $q_C>7/8$, then $8q_C>7$, which exceeds every competing report's agent utility. The agent always chooses $C$, giving payoff at most $10$. Hence assume $q_C\le7/8$, so $C$ pays at most $35/4<9$.

Let $q$ be the acceptance probability of the report $AE$ with principal values $(9,3)$. Again, the same $q$ applies for both values of $X_B$.

\paragraph{Case 1: $q\le5/7$.}
The high-$A$ report $AE$ pays at most $60/7<9$. In each of the first three states in the table, every non-jackpot report therefore pays at most $9$. Combining this with \eqref{eq:hs-jackpot} bounds the conditional payoff by $9+(99/100)\eta r_{ab}$. In state $(9,9)$ the bound is $18$, since $9+(99/100)\eta(9/2)<18$. Averaging gives
\begin{align*}
 \E[\text{payoff}]
 &\le\frac89\,9+\frac19\,18
 +\frac{99\eta}{900}\left(4\cdot\frac54+2\cdot\frac92+2\cdot3\right)\\
 &=10+\frac{11}{5}\eta.
\end{align*}

\paragraph{Case 2: $q>5/7$.}
Whenever $a=9$, report $AE$ offers agent utility $7q>5$. It strictly beats $AB$ and $J$. Every remaining possible winner pays at most $12$, including $C$. When $a=0$, every non-jackpot report pays at most $9$, so \eqref{eq:hs-jackpot} again supplies the bound $9+(99/100)\eta r_{ab}$. Thus
\begin{align*}
 \E[\text{payoff}]
 &\le\frac13\,12+\frac23\,9
 +\frac{99\eta}{900}\left(4\cdot\frac54+2\cdot\frac92\right)\\
 &=10+\frac{77}{50}\eta\le10+\frac{11}{5}\eta.
\end{align*}
Substituting $\eta=1/100$ proves the theorem.
\end{proof}

\begin{proposition}\label{prop:hs-limit}
No universal guarantee greater than $30/61$ is possible for randomized acceptance rules.
\end{proposition}
\begin{proof}
Replace the jackpot by $9/\varepsilon$ with probability $\varepsilon\in(0,1/2)$, and zero otherwise. Its mean remains $9$, and its high value exceeds $18$. The proof of Theorem~\ref{thm:hs-rand}, with $99/100$ replaced by $1-\varepsilon$, bounds every rule by
\[
 10+\frac{20}{9}(1-\varepsilon)\eta.
\]
The offline mean is $9+(1-\varepsilon)34/3$. Letting $\varepsilon,\eta\downarrow0$, the ratio of these two expressions tends to $30/61$. Each instance has five items and strictly positive agent utilities. Hence every constant greater than $30/61$ fails on a finite instance.
\end{proof}

\section{Warm-up: a single total-value threshold}\label{sec:warmup}

This section proves a guarantee of $3/10$ with the simplest possible menu, which
accepts every feasible report whose total principal value reaches a threshold.
The proof is fully analytic. It introduces three ingredients that the main proof
develops much further: forced items, the tail bound implied by an upper bound on
the score, and a variance bound for the clipped optimum.

For $z\ge0$ define
\begin{equation}\label{eq:warmup-B}
 e(z)=\sum_{i\in E}\E\pos{X_i-z},\qquad
 B(z)=z\Prb(W\ge z)+e(z)\Prb(W<z).
\end{equation}
The function $e$ is continuous and nonincreasing, $e(0)>0$, and $e(z)\to0$ as
$z\to\infty$. Hence there is a unique $t>0$ with $e(t)=t$.

\begin{algorithm}[htbp]
\caption{Total-value threshold}\label{alg:threshold}
\begin{algorithmic}[1]
\Require Feasible family $\F$ and the laws of the principal values
\State Compute $e$ and the law of $W$, and find the unique $t>0$ with $e(t)=t$.
\State Let $z^*$ be the smallest maximizer of $B$ on $[t,\infty)$.
\State Accept a verified report $(S,x_S)$ if and only if $S\in\F$ and $\sum_{i\in S}x_i\ge z^*$.
\end{algorithmic}
\end{algorithm}

\begin{theorem}\label{thm:warmup}
Algorithm~\ref{alg:threshold} is well defined, and under every best response the
principal's expected payoff is at least $3V/10$.
\end{theorem}

The verifier needs one feasibility test, one sum and one comparison. For finite
rational supports, the preprocessing can be carried out exactly: $e$ is
piecewise affine, and $\Prb(W\ge z)$ is constant between consecutive support
values of $W$, so $B$ is affine between consecutive breakpoints and it suffices
to evaluate it at $t$ and at the breakpoints of $e$ and of the law of $W$ that
are at least $t$. At a support value of $W$, the weak inequality in
$\Prb(W\ge z)$ includes the atom.

\subsection{Forced items}

For an item $i$, let $W_{-i}=\max\{X(S):S\in\F,\ i\notin S\}$ be the optimum
without $i$.

\begin{lemma}\label{lem:warmup-payoff}
For every $z>0$, the menu accepting exactly the feasible reports of total value at
least $z$ gives expected payoff at least $B(z)$ under every best response.
\end{lemma}

\begin{proof}
Let $P_z$ be the realized payoff. We show that pointwise
\begin{equation}\label{eq:warmup-forced}
 P_z\ge z\ind_{\{W\ge z\}}+\sum_i\pos{X_i-z}\ind_{\{W_{-i}<z\}}.
\end{equation}
If $W<z$, then $W_{-i}\le W<z$ but also $X_i\le W<z$ for every $i$, so the
right side is zero. If $W\ge z$, some nonempty report is accepted, and by
Lemma~\ref{lem:score} the agent submits an accepted report $S$. If $X_i>z$ and
$W_{-i}<z$, then every accepted report contains $i$, since a report without $i$
has value at most $W_{-i}$. Moreover $\sum_{i\in S}\min(X_i,z)\ge z$: either
some summand equals $z$, or clipping changes no summand and the sum is
$X(S)\ge z$. Writing $X(S)=\sum_{i\in S}\min(X_i,z)+\sum_{i\in S}\pos{X_i-z}$
proves \eqref{eq:warmup-forced}.

The variable $W_{-i}$ is a function of the values other than $X_i$, so it is
independent of $X_i$. Since $W_{-i}\le W$,
\[
 \E P_z\ge z\Prb(W\ge z)+\sum_i\E\pos{X_i-z}\Prb(W_{-i}<z)
 \ge z\Prb(W\ge z)+e(z)\Prb(W<z)=B(z).
 \qedhere
\]
\end{proof}

On $[t,\infty)$ we can write
\[
 B(z)=e(z)+(z-e(z))\Prb(W\ge z),\qquad z-e(z)\ge0.
\]
The function $z\mapsto\Prb(W\ge z)$ is upper semicontinuous, so $B$ is upper
semicontinuous on $[t,\infty)$. Moreover
\[
 0\le B(z)\le e(z)+\E[W\ind_{\{W\ge z\}}]\to0\quad(z\to\infty),
 \qquad B(t)=t>0.
\]
Hence the maximizers of $B$ on $[t,\infty)$ form a nonempty compact set and
$z^*$ exists. Put $K=B(z^*)>0$. By Lemma~\ref{lem:warmup-payoff} it remains to
prove $K\ge3V/10$.

Multiply all principal values by $1/K$, and keep the notation for the rescaled
instance; thus $K=1$. Since $B(t)=t$, we have $t\le1$ and therefore
$e(x)\le e(t)=t\le1$ for $x\ge1$. For $x>1$, the inequality $B(x)\le1$ gives
\begin{equation}\label{eq:warmup-tail}
 \Prb(W\ge x)\le\frac{1-e(x)}{x-e(x)}\le\frac1x.
\end{equation}

\subsection{Clipping and variance}

Let $c=3/2$ and define
\[
 Z=\max_{S\in\F}\sum_{i\in S}\min(X_i,c),\qquad
 \mu=\E Z,\qquad a=e(c)\in[0,1].
\]
Clipping an optimal set for $W$ shows
\begin{equation}\label{eq:warmup-clip}
 W\le Z+\sum_i\pos{X_i-c},\qquad V\le\mu+a.
\end{equation}

\begin{lemma}\label{lem:warmup-variance}
$\Var Z\le c\mu$.
\end{lemma}

\begin{proof}
Let $Z_{-i}$ be the clipped optimum without item $i$, and put
$\Delta_i=Z-Z_{-i}$. Fix a set $S$ attaining $Z$. If $i\notin S$, then
$\Delta_i=0$. If $i\in S$, then $S\setminus\{i\}\in\F$ by downward closure, so
$\Delta_i\le\min(X_i,c)$. Therefore
\begin{equation}\label{eq:warmup-deletion}
 0\le\Delta_i\le c,\qquad \sum_i\Delta_i\le Z.
\end{equation}
For a function $f$ of independent variables,
\begin{equation}\label{eq:warmup-efron-stein}
 \Var f(X)\le\sum_i\E\Var\bigl(f(X)\mid X_{-i}\bigr).
\end{equation}
To see this, reveal the coordinates one at a time. The martingale differences
are orthogonal, and by conditional Jensen's inequality the second moment of the
$i$th difference is at most $\E\Var(f(X)\mid X_{-i})$, after averaging over the
coordinates revealed later, which are independent of $X_i$. Summing gives
\eqref{eq:warmup-efron-stein}.

Conditional on $X_{-i}$ the variable $Z_{-i}$ is constant, so
\[
 \Var Z\le\sum_i\E\Var(Z\mid X_{-i})\le\sum_i\E(Z-Z_{-i})^2
 \le c\,\E\sum_i\Delta_i\le c\mu.
\]
All variables are bounded, since $Z\le c|E|$.
\end{proof}

For $1<x<c$ we have $e(x)\ge a$. The function $v\mapsto(1-v)/(x-v)$ is
decreasing on $[0,1]$, so \eqref{eq:warmup-tail} and $Z\le W$ give
$\Prb(Z\ge x)\le(1-a)/(x-a)$. The same bound holds at $x=c$. Together with
\eqref{eq:warmup-tail} and the monotonicity of the survival function, for $a<1$,
\begin{equation}\label{eq:warmup-envelope}
 \Prb(Z\ge x)\le h_a(x):=
 \begin{cases}
 1,&0\le x\le1,\\
 (1-a)/(x-a),&1<x<c,\\
 \min\{(1-a)/(c-a),\,1/x\},&x\ge c.
 \end{cases}
\end{equation}
If $a=1$, then \eqref{eq:warmup-tail} at $x\downarrow1$ forces $Z\le1$ almost
surely, so $\mu+a\le2$.

\subsection{A scalar inequality}

\begin{lemma}\label{lem:warmup-scalar}
Let $a\in[0,1)$ and let $Z$ be a bounded nonnegative random variable satisfying
\eqref{eq:warmup-envelope} and $\Var Z\le\frac32\E Z$. Then $\E Z+a<10/3$.
\end{lemma}

\begin{proof}
Write $\mu=\E Z$. We first show $\mu<5$. For $u>0$, applying Markov's
inequality to $(\mu+\Var(Z)/u-Z)^2$ on the event $\{Z\le\mu-u\}$ gives the
one-sided bound $\Prb(Z<\mu-u)\le\Var(Z)/(\Var(Z)+u^2)$. If $\mu\ge5$, taking
$u=\mu/2$ yields
\[
 \frac\mu2\Prb\Bigl(Z\ge\frac\mu2\Bigr)
 \ge\frac\mu2\cdot\frac{\mu^2/4}{3\mu/2+\mu^2/4}
 =\frac{\mu^2}{2\mu+12}\ge\frac{25}{22}>1,
\]
which contradicts $xh_a(x)\le1$ for $x\ge1$.

Set
\[
 b=\frac{1-a}{3/2-a}\in(0,2/3],\qquad
 d=1-a=\frac{b}{2(1-b)},\qquad H=\frac{13}{2},\qquad L=-\log b.
\]
With $s(x)=\Prb(Z\ge x)$, the identities $\E Z=\int_0^\infty s$ and
$\E Z^2=\int_0^\infty 2xs(x)\,dx$ give
\[
 \E Z^2=2H\mu+2\int_0^\infty(x-H)s(x)\,dx
 \ge2H\mu-2\int_0^H(H-x)h_a(x)\,dx.
\]
Combined with $\E Z^2=\Var Z+\mu^2\le\frac32\mu+\mu^2$, this shows
\begin{equation}\label{eq:warmup-G}
 \Gamma(\mu,b):=13\mu-2\int_0^H(H-x)h_a(x)\,dx-\mu^2-\frac32\mu\le0.
\end{equation}
For $\mu<5$ we have $\partial\Gamma/\partial\mu=23/2-2\mu>0$. If $\mu+a\ge10/3$, then
$\mu\ge\mu_0:=7/3+d$, so it suffices to show $\Gamma(\mu_0,b)>0$ for all
$0<b\le2/3$.

\emph{Case $0<b\le2/13$.} The envelope equals $b$ on $[c,H]$, and integration gives
\begin{equation}\label{eq:warmup-Gsmall}
 \Gamma(\mu,b)=13(\mu-1)+1+d-25b-\mu^2-\frac32\mu-2d(H-a)L.
\end{equation}
For $0<b\le1$ we have $-\log b\le(1-b)/\sqrt b$: with $u=b^{-1/2}\ge1$ the
function $u-u^{-1}-2\log u$ vanishes at $u=1$ and is nondecreasing. Hence
\[
 dL\le\frac{\sqrt b}{2}\le\frac15,\qquad d\le\frac1{11},\qquad
 \mu_0\le\frac{80}{33}.
\]
Using $13(\mu_0-1)\ge52/3$ and $H-a\le H$ in \eqref{eq:warmup-Gsmall},
\[
 \Gamma(\mu_0,b)\ge\frac{52}{3}+1-\frac{50}{13}
 -\Bigl(\frac{80}{33}\Bigr)^2-\frac{40}{11}-\frac{13}{5}
 =\frac{168034}{70785}>0.
\]

\emph{Case $2/13\le b\le2/3$.} The envelope equals $b$ on $[c,1/b]$ and $1/x$
on $[1/b,H]$. Integration gives
\begin{align}\label{eq:warmup-Glarge}
 \Gamma(\mu,b)={}&13\Bigl(\mu-1+\frac32b-\log H\Bigr)
 +1+d-\frac1b-\frac94b-\mu^2-\frac32\mu
 +2a(H+d)(-\log b).
\end{align}
We use
\begin{equation}\label{eq:warmup-logs}
 \log\frac{13}2<\frac{15}{8},\qquad
 -\log b\ge2\Bigl(u+\frac{u^3}{3}\Bigr),\quad u=\frac{1-b}{1+b}.
\end{equation}
The first follows from
$\sum_{k=0}^7(15/8)^k/k!-13/2=3772963/234881024>0$, and the second from
integrating $(1-v^2)^{-1}\ge1+v^2$ over $[0,u]$. Since $a\ge0$, substituting
\eqref{eq:warmup-logs} and $\mu_0=7/3+d$ into \eqref{eq:warmup-Glarge} gives
\begin{equation}\label{eq:warmup-polynomial}
 \Gamma(\mu_0,b)\ge\frac{P(b)}{72b(1-b)^2(1+b)^3},
\end{equation}
where
\[
 P(b)=1242b^7-3593b^6+1795b^5-2948b^4+7546b^3-5459b^2+1345b-72.
\]
To see that $P>0$ on $[2/13,2/3]$, write, on an interval $[\ell,r]$,
\[
 P(\ell+(r-\ell)v)=\sum_{j=0}^7p_jv^j
 =\sum_{k=0}^7\beta_k\binom7k v^k(1-v)^{7-k},\qquad
 \beta_k=\sum_{j=0}^k\frac{\binom kj}{\binom7j}p_j.
\]
A direct computation gives
\begin{center}
\begin{tabular}{cc}
\toprule
$[\ell,r]$ & $\min_{0\le k\le7}\beta_k$\\
\midrule
$[2/13,2/5]$ & $447966/78125$\\
$[2/5,2/3]$ & $6518/118125$\\
\bottomrule
\end{tabular}
\end{center}
The Bernstein basis polynomials are nonnegative on $[0,1]$ and sum to one, so
$P(b)$ is at least the smallest coefficient, which is positive on both intervals.

In both cases $\mu\ge\mu_0$ would contradict \eqref{eq:warmup-G}. Hence
$\mu+a<10/3$.
\end{proof}

\begin{proof}[Proof of Theorem~\ref{thm:warmup}]
The threshold $z^*$ exists and gives payoff at least $K$ by
Lemma~\ref{lem:warmup-payoff}. In the normalization $K=1$,
Lemmas~\ref{lem:warmup-variance} and~\ref{lem:warmup-scalar} together with
\eqref{eq:warmup-clip} give $V<10/3$ when $a<1$, and $V\le2$ when $a=1$.
Undoing the scaling gives $K>3V/10$.
\end{proof}

\section{The menus and the main theorem}\label{sec:menus}

The single threshold of Section~\ref{sec:warmup} loses in two ways. A report can
overshoot the threshold by much more than it needs to, and an item of large value
can be displaced by a set of moderate items that the agent prefers. The menus below
address the first problem with clipped values, minimal qualifying sets and
reserves, and the second with offers for singletons whose value exceeds a
cutoff.

\paragraph{Seeds.}
For nonnegative weights $w$ and a level $t>0$, a \emph{seed at $t$} is an
inclusion-minimal feasible set $S$ with $w(S)\ge t$. Because the weights are
nonnegative and $\F$ is downward closed, a nonempty feasible report $S$ is a seed if
and only if
\begin{equation}\label{eq:seed}
 w(S)\ge t\quad\text{and}\quad w(S)-w_i<t\ \text{for every }i\in S.
\end{equation}
In particular, whether a report is a seed can be decided from the submitted values
alone. If $P$ is a seed at $t$, then removing a lightest member of $P$ drops its
weight below $t$, so
\begin{equation}\label{eq:seed-reserve-scale}
 b_P:=t+\min_{i\in P}w_i-w(P)>0.
\end{equation}

\paragraph{The family $\mathcal M$.}
All menus reject the empty report. The family $\mathcal M$ consists of the
\emph{capped menus} of Algorithm~\ref{alg:capped}, with all real parameters in the
stated ranges. A capped menu clips values at a cap $c$ and adds one of three
\emph{bodies} to a set of singleton offers: all reports reaching a level; the seeds
at a level together with all reports reaching a higher level; or the seeds at a
level together with their extensions that carry a reserve.

\begin{algorithm}[htbp]
\caption{Capped menu: acceptance test}\label{alg:capped}
\small
\begin{algorithmic}[1]
\Require Cap $c>0$; for every item a cutoff $d_i>0$ and a choice of open or closed test
\Require Body: a level $t>c$; or levels $c<t<u$; or a level $t>c$ and a fraction $0<a\le1$
\State Receive a verified nonempty feasible report $(S,x_S)$.
\If{$S=\{i\}$ and $x_i>d_i$ (open test) or $x_i\ge d_i$ (closed test)}
 \State \Return accept
\EndIf
\State Set $w_i=\min(x_i,c)$ for $i\in S$.
\If{the body is a total threshold}
 \State \Return accept if and only if $w(S)\ge t$
\ElsIf{the body is seeds and an upper threshold}
 \State \Return accept if and only if $S$ is a seed at $t$ or $w(S)\ge u$
\Else \Comment{seeds and extensions}
 \For{each $P\subseteq S$ that is a seed at $t$}
  \If{$P=S$ or $w(S\setminus P)\ge a\,b_P$}
   \State \Return accept
  \EndIf
 \EndFor
 \State \Return reject
\EndIf
\end{algorithmic}
\end{algorithm}

Every subset examined in Algorithm~\ref{alg:capped} is feasible by downward closure,
and every test uses only the identities and principal values of the submitted
items. Multiplying the cap and all levels and cutoffs of a menu by
$\lambda>0$ gives another menu of the family, and the reserve fraction $a$ is
dimensionless, so $\mathcal M$ is closed under rescaling as required in
Section~\ref{sec:normalization}.

\begin{theorem}\label{thm:main}
For every instance, some menu in $\mathcal M$ with rational parameters has
expected score strictly greater than $\frac{29}{70}V\approx0.41429\,V$.
Consequently Algorithm~\ref{alg:select} is well defined, and the menu it selects
gives the principal expected payoff strictly greater than $\frac{29}{70}V$ under every
best response. In particular $\alpha_R\ge\alpha_D>29/70$.
\end{theorem}

\begin{algorithm}[htbp]
\caption{Menu selection}\label{alg:select}
\small
\begin{algorithmic}[1]
\Require $\F$ and the laws of the principal values
\State Compute $V=\E\max_{S\in\F}X(S)$.
\For{each menu $M$ in a fixed enumeration of the menus of $\mathcal M$ with rational parameters}
 \If{$\E b_M(X)>\frac{29}{70}V$}
  \State Commit to $M$ and stop.
 \EndIf
\EndFor
\end{algorithmic}
\end{algorithm}

Algorithm~\ref{alg:select} is an existence statement with unrestricted
preprocessing. For finite rational supports each test in the loop is an exact
finite enumeration, and Theorem~\ref{thm:main} guarantees that the loop stops. We
do not claim a polynomial bound on the preprocessing, on the length of the
enumeration, or on the time needed to verify a report for an arbitrary
downward-closed family given by an oracle.

\subsection{Structure of the proof}

The second sentence of Theorem~\ref{thm:main} follows from the first by
Lemma~\ref{lem:score}. We prove the stronger statement that every instance admits
a menu with rational parameters and expected score greater than $\rho_0V$, where
$\rho=29/70$ and $\rho_0=\rho+10^{-6}$. Together with Lemma~\ref{lem:score}, this
gives $\alpha_R\ge\alpha_D\ge\rho_0>29/70$, establishing the third sentence.
Suppose, for a contradiction, that every menu of $\mathcal M$ with real
parameters has expected score at most $\rho_0V$, and normalize as in
Lemma~\ref{lem:normalization} with $R=1/\rho_0$.
Each menu then gives
an upper bound of one on its expected score, while the optimum has mean $R$. The
proof turns many such menus into affine inequalities among finitely many
quantities of the normalized instance: values of the survival functions $S_c$ and
stop-loss functions $D_c$ of clipped optima at grid points, with clipped means
represented by $\mu_c=1+D_c(1)$, the excesses $e(c)$, the large-item term $G(1)$, and the mean overshoots of
certain menus. A finite computation then shows that, for every possible value of
$\mu_1$ in an interval that contains it, a positive combination of these
inequalities is violated by every point of a box that contains the true values.

The inequalities fall into four groups.
\begin{itemize}
\item Section~\ref{sec:profile} proves a covariance bound for clipped optima that
 extends the variance bound of Lemma~\ref{lem:warmup-variance}, and derives from
 it the deletion inequality and the ramp inequalities. With elementary relations
 between the profiles at different levels and caps, these use no menu.
\item Section~\ref{sec:overshoot} bounds the mean overshoot of a body above its
 level, in several ways that the computation can combine. Here the singleton
 offers of the capped menus enter.
\item Section~\ref{sec:tail} combines the deletion inequality with these
 budgets at single levels on a fine grid.
\item Section~\ref{sec:paths} follows, for one realization at a time, the items
 that every qualifying report must contain as the level increases, and converts the
 resulting charges into one inequality per list of levels.
\end{itemize}
Section~\ref{sec:certificate} describes the finite computation and completes the
proof of Theorem~\ref{thm:main}. The computation is carried out in exact rational
arithmetic.

\providecommand{\od}{\operatorname{odds}}
\providecommand{\Cov}{\operatorname{Cov}}

\section{Profile inequalities}\label{sec:profile}

This section proves inequalities between the survival functions, the stop-loss
functions and the means of clipped optima, and the excess $e$. They follow from
the structure of the optimum, from independence and from the normalization, and
they do not involve any menu; the inequalities that use the menus of the
family~$\mathcal M$ are proved in Sections~\ref{sec:overshoot}--\ref{sec:paths}.
The main tool is a covariance bound for clipped optima
(Lemma~\ref{lem:prof-pivcov}), which extends the variance bound of the warm-up.
It bounds the covariance between the optimum and the event that it reaches a
level by the expected weight of the items that every qualifying report must
contain.
Each inequality below becomes an affine inequality in finitely many unknowns, as
described in Section~\ref{sec:prof-rows}, and these affine inequalities are rows
of the finite computation of Section~\ref{sec:certificate}.

\subsection{Standing assumptions and notation}\label{sec:prof-setup}

Throughout this section and the following ones, the instance is normalized as in
Lemma~\ref{lem:normalization}, for a fixed but arbitrary $R>1$. We use only the
following consequences of that lemma.
\begin{enumerate}[label=(N\arabic*)]
\item\label{prof:N1} Every menu of $\mathcal M$ has expected score at most one.
 The same holds for any average of such menus over an auxiliary parameter, for
 instance an independent coin or a random reserve fraction that selects one menu
 of a family.
\item\label{prof:N2} $\E W=R$.
\item\label{prof:N3} There is an item $s_0$ with $X_{s_0}=1$ surely whose singleton
 is a maximal feasible set. In particular $W\ge1$ and $Z_c\ge1$ surely for every
 $c\ge1$.
\end{enumerate}
As in Section~\ref{sec:prelim}, items whose singleton is infeasible have been
deleted, so every singleton is feasible. No inequality below uses the numerical
value of $R$.

Besides the notation of Section~\ref{sec:normalization}, we write
\[
 p_i(s)=\Prb(X_i>s),\qquad q(s)=\sum_ip_i(s),\qquad
 T_c(t)=D_c(t)+tS_c(t)=\E[Z_c\ind_{\{Z_c\ge t\}}],
\]
so that $e(c)=\int_c^\infty q(s)\,ds$. For an item
$i$, $W_{-i}$ denotes the optimum of the instance with $i$ deleted; it depends
only on the other coordinates, and $W_{-i}\le W$. We write $\od(s)=s/(1-s)$ for
$0\le s<1$, and $\E[Y;B]=\E[Y\ind_B]$ for an event $B$ and an integrable
variable $Y$.

\subsection{Association and pivotal covariance}\label{sec:prof-tools}

\begin{lemma}[Association]\label{lem:prof-assoc}
Let $f,g$ be bounded functions of independent real coordinates, both
nondecreasing in every coordinate. Then $\Cov(f,g)\ge0$. If $f$ is
nondecreasing and $g$ nonincreasing, $\Cov(f,g)\le0$. The same holds if $f$
is integrable and $g$ bounded.
\end{lemma}

\begin{proof}
For one coordinate let $X'$ be an independent copy of $X$. Then
$(f(X)-f(X'))(g(X)-g(X'))\ge0$, and its expectation is $2\Cov(f,g)$. For $n$
coordinates condition on the first $n-1$. The conditional covariance is
nonnegative by the one-coordinate case. The conditional means are nondecreasing
functions of the first $n-1$ coordinates, so their covariance is nonnegative by
induction. The law of total covariance adds the two. The second claim follows
by replacing $g$ by $-g$, and the integrable case by truncating $f$ at
$\pm n$ and letting $n\to\infty$.
\end{proof}

For an increasing event $H$ of the values, that is, an event whose indicator is
nondecreasing in every coordinate, we call an item $i$ \emph{pivotal} for $H$ at
a realization if $H$ occurs and does not occur after $i$ is deleted. All events
used below have the form ``some feasible set has weight at least $x$'' for
weights that are nondecreasing functions of the individual values. For such an
event, $i$ is pivotal exactly when $H$ occurs and every qualifying set contains
$i$, by downward closure.

\begin{lemma}[Pivotal covariance]\label{lem:prof-pivcov}
Let $w_i=\phi_i(X_i)$ with each $\phi_i$ nonnegative and nondecreasing and
$\E w_i<\infty$, put $Q=\max_{S\in\F}\sum_{i\in S}w_i$, and let $H$ be an
increasing event of the form just described. Then
\begin{equation}\label{eq:prof-cov}
 \E[Q;H]-\E Q\cdot\Prb(H)\ \le\ \E\sum_{i\text{ pivotal for }H}w_i .
\end{equation}
\end{lemma}

\begin{proof}
Raising $w_i$ by $\epsilon>0$ raises $Q$ by at most $\epsilon$, so
$\sum_iw_i-Q$ is nondecreasing in every coordinate. By
Lemma~\ref{lem:prof-assoc},
$\Cov(Q,\ind_H)\le\sum_i\Cov(w_i,\ind_H)$. Fix $i$ and condition on the other
coordinates $X_{-i}$. If some qualifying set avoids $i$, then $\ind_H=1$ for
every value of $X_i$, and the conditional covariance vanishes. Otherwise $H$
occurs exactly when $i$ is pivotal, and the conditional covariance is at most
$\E[w_i\ind_H\mid X_{-i}]$, since $\E[w_i\mid X_{-i}]\Prb(H\mid X_{-i})\ge0$.
By independence $\E[w_i\mid X_{-i}]=\E w_i$ is constant, so the covariance of
the conditional means vanishes and
$\Cov(w_i,\ind_H)\le\E[w_i;\,i\text{ pivotal}]$. Summing over $i$ proves
\eqref{eq:prof-cov}.
\end{proof}

\subsection{Basic relations}\label{sec:prof-basic}

\begin{lemma}[Basic profile relations]\label{lem:prof-basic}
For $1\le c\le c'\le\infty$ and all levels:
\begin{enumerate}[label=(\alph*)]
\item\label{prof:basic-mon} $1\le Z_c\le Z_{c'}\le W$, so $S_c(t)\le S_{c'}(t)$,
 $D_c(t)\le D_{c'}(t)$, $\mu_c\le\mu_{c'}\le R$, and $S_c(t)=1$, $D_c(t)=\mu_c-t$
 for $t\le1$. In particular $\mu_c=1+D_c(1)$, $D_\infty(1)=R-1$, and
 $S_c(t)\le S_\infty(t)=\Prb(W\ge t)$.
\item\label{prof:basic-loss} $R-\mu_c\le e(c)$.
\item\label{prof:basic-riemann} $S_c$ is nonincreasing and for $x<y$
 \[
  (y-x)S_c(y)\ \le\ D_c(x)-D_c(y)\ \le\ (y-x)S_c(x).
 \]
\item\label{prof:basic-agree} If $1\le y\le c$ then $S_c(y)=S_\infty(y)$ and
 $D_c(1)-D_c(y)=D_\infty(1)-D_\infty(y)$.
\item\label{prof:basic-chord} For fixed $t$ the function $c\mapsto D_c(t)+e(c)$ is
 convex on $[1,\infty)$.
\end{enumerate}
\end{lemma}

\begin{proof}
(a) Clipping is monotone in the cap, and $Z_c\ge1$ by \ref{prof:N3}; the rest
follows by integration and \ref{prof:N2}.

(b) Clipping a maximizer $O$ of $W$ at $c$ shows
$Z_c\ge W-\sum_{i\in O}(X_i-c)_+\ge W-\sum_i(X_i-c)_+$; take expectations.

(c) $D_c(x)-D_c(y)=\int_x^y\Prb(Z_c>u)\,du$, and
$S_c(y)\le\Prb(Z_c>u)\le S_c(x)$ for $x\le u<y$.

(d) A report reaching raw value $y$ either contains an item of value at least
$y$, whose clipped singleton reaches $y$ since $y\le c$, or has all its items
below $y\le c$ and is unchanged by clipping. So $\{Z_c\ge y\}=\{W\ge y\}$, and
likewise $\{Z_c>u\}=\{W>u\}$ for $u<y$. Integrating over $[1,y]$ gives the
second identity.

(e) Pointwise, $(Z_c-t)_++\sum_i(X_i-c)_+$ is the maximum of $\sum_i(X_i-c)_+$
and the finitely many functions
$\sum_{i\in B}X_i-t+\sum_{i\notin B}(X_i-c)_+$, $B\in\F$, because
$\sum_{i\in B}\min(X_i,c)+\sum_i(X_i-c)_+=\sum_{i\in B}X_i+\sum_{i\notin B}(X_i-c)_+$.
Each is convex in $c$, and so are the maximum and its expectation.
\end{proof}

Since $Z_c\ge1$, the convention $D_c(t)=\mu_c-t$ for $t\le1$ in
\ref{prof:basic-mon} extends all stop losses to $[0,1]$; it is used in the ramp
inequalities below. The convexity in \ref{prof:basic-chord} is used through its
chord inequality: for $1\le a<b<c'$,
\begin{equation}\label{eq:prof-chord}
 (c'-a)\bigl(D_b(t)+e(b)\bigr)\le(c'-b)\bigl(D_a(t)+e(a)\bigr)
 +(b-a)\bigl(D_{c'}(t)+e(c')\bigr).
\end{equation}

We next record how the clipped optimum varies with the cap. Fix a realization.
On every bounded interval of caps, $s\mapsto Z_s$ is the maximum of finitely many
piecewise affine functions, hence piecewise affine. At every cap $s$ outside a
finite set, $Z_s$ is differentiable and its derivative is the number of items
$i$ with $X_i>s$ in a maximizing set, for any maximizing set. At such points the
derivative of $(Z_s-t)_+$ is this count if $Z_s>t$ and zero if $Z_s<t$. These
derivatives are bounded by the number of items, which justifies the exchange of
differentiation, integration over caps and expectation below.

\begin{lemma}[Variation along the cap]\label{lem:prof-slice}
For $1\le a<b\le\infty$ and $t\ge1$,
\begin{equation}\label{eq:prof-slice}
 D_b(t)-D_a(t)\ \le\ \int_a^bq(s)S_\infty(t-s)\,ds\ \le\ e(a)-e(b),
\end{equation}
with $S_\infty(v)=1$ for $v\le1$ and $e(\infty)=0$.
\end{lemma}

\begin{proof}
At a cap $s$ where $Z_s$ is differentiable and $Z_s>t$, deleting an item with
$X_i>s$ from a maximizing set leaves a feasible set of clipped value
$Z_s-s>t-s$. So the derivative of $(Z_s-t)_+$ is at most
$\sum_i\ind_{\{X_i>s\}}\ind_{\{W_{-i}\ge t-s\}}$. Independence of $X_i$ and
$W_{-i}$ and $W_{-i}\le W$ bound its expectation by $q(s)S_\infty(t-s)$. Integration
gives the first inequality for finite $b$, and monotone convergence gives
$b=\infty$. The second follows from $S_\infty\le1$ and $e(a)-e(b)=\int_a^bq$.
\end{proof}

\subsection{Deletion}\label{sec:prof-deletion}

\begin{lemma}[Deletion inequality]\label{lem:deletion}
For every cap $c\ge1$ and every $x>0$,
\begin{equation}\label{eq:deletion}
 T_c(x+c)\ \le\ \mu_cS_c(x).
\end{equation}
More precisely, for every collection of disjoint intervals
\[
 I_k\subseteq[(k-1)c,kc)\cap\Bigl[\tfrac{kx}{k-1},\,x+c\Bigr),\qquad k\ge2,
\]
we have
\begin{equation}\label{eq:prof-intdel}
 T_c(x+c)+\sum_{k\ge2}\E\bigl[Z_c-(k-1)c\,;\,Z_c\in I_k\bigr]\ \le\ \mu_cS_c(x).
\end{equation}
\end{lemma}

\begin{proof}
Apply Lemma~\ref{lem:prof-pivcov} with $w_i=\min(X_i,c)$, $Q=Z_c$ and
$H=\{Z_c\ge x\}$; the left side of \eqref{eq:prof-cov} is $T_c(x)-\mu_cS_c(x)$.
Fix a maximizing set $B$. Pivotal items belong to every qualifying set, in
particular to $B$, so their clipped sum is at most $Z_c$. Deleting a member $i$
of $B$ leaves the feasible set $B\setminus\{i\}$ of clipped value at least
$Z_c-c$; so if $Z_c\ge x+c$ there are no pivotal items. At $Z_c=z\ge x$ each
pivotal weight exceeds $z-x$, since deleting it from $B$ would otherwise leave
a qualifying set. If $z\ge kx/(k-1)$ then $z/(z-x)\le k$, so there are at most
$k-1$ pivotal items and their sum is at most $(k-1)c$. On $\{Z_c\in I_k\}$ the
pivotal sum is therefore at most $Z_c-(Z_c-(k-1)c)$, the subtracted quantity
being nonnegative. Substituting these bounds in \eqref{eq:prof-cov} gives
\[
 T_c(x)-\mu_cS_c(x)\le T_c(x)-T_c(x+c)-\sum_{k\ge2}\E\bigl[Z_c-(k-1)c\,;\,Z_c\in I_k\bigr],
\]
which is \eqref{eq:prof-intdel}. With all $I_k$ empty it is \eqref{eq:deletion}.
\end{proof}

The expectations in \eqref{eq:prof-intdel} are affine in the profile: for
$u<v$ and any $\beta$,
\begin{equation}\label{eq:prof-strip}
 \E[Z_c-\beta\,;\,u\le Z_c<v]
 =D_c(u)+(u-\beta)S_c(u)-D_c(v)-(v-\beta)S_c(v).
\end{equation}
In the computation each $I_k$ is either empty or the indicated interval rounded
inward to grid points.

The same covariance argument, with the pivotal set kept, gives the form of the
deletion inequality used in Sections~\ref{sec:tail} and~\ref{sec:paths}. At
cap $c$ and level $t>0$ a report qualifies if its clipped value
$\sum_{i\in S}\min(X_i,c)$ is at least $t$; $I_t$ is the intersection of the
qualifying reports on $\{Z_c\ge t\}$, and $L_t$ is its clipped value, with
$L_t=0$ if $Z_c<t$. The members of $I_t$ are exactly the pivotal items for
$\{Z_c\ge t\}$.

\begin{lemma}[Deletion inequality with the pivotal set]\label{lem:prof-deletion-cap1}
For every $c\ge1$ and $t>0$,
\[
 D_c(t)+\E\bigl[(t-L_t)\ind_{\{Z_c\ge t\}}\bigr]\ \le\ \mu_cS_c(t).
\]
In particular, at cap one, for every $t>1$,
\begin{equation}\label{eq:prof-deletion-cap1}
 D_1(t)+\E\bigl[(t-L_t)\ind_{\{Z\ge t\}}\bigr]\ \le\ \mu_1S_1(t).
\end{equation}
\end{lemma}

\begin{proof}
Lemma~\ref{lem:prof-pivcov} with $w_i=\min(X_i,c)$, $Q=Z_c$ and
$H=\{Z_c\ge t\}$ gives $\E[Z_c;Z_c\ge t]-\mu_cS_c(t)\le\E L_t$. Since $L_t=0$
off $H$ and $Z_c\ind_H=(Z_c-t)_++t\ind_H$, this is the claim.
\end{proof}

Integrating \eqref{eq:deletion} over the level gives a further inequality at cap
one.

\begin{lemma}[Integrated deletion]\label{lem:prof-intdel}
Let $1\le x<y$ and $\mu_1\le r$, and write $D=D_1$. Then
\begin{equation}\label{eq:prof-intdel-int}
 (x+1)D(x+1)-(y+1)D(y+1)+2(y-x)\,D\Bigl(1+\frac{x+y}2\Bigr)\ \le\ r\bigl(D(x)-D(y)\bigr).
\end{equation}
\end{lemma}

\begin{proof}
By \eqref{eq:deletion} at cap one, $D(v+1)+(v+1)S_1(v+1)\le\mu_1S_1(v)\le
rS_1(v)$ for every $v>0$. Integrate over $v\in[x,y]$. The right side gives
$r(D(x)-D(y))$, because $D(x)-D(y)=\int_x^y\Prb(Z>v)\,dv$ and $S_1(v)=\Prb(Z>v)$
for all but countably many $v$. The function $D$ is convex and Lipschitz, with
$D'(u)=-S_1(u)$ for almost every $u$, so integration by parts gives
\[
 \int_{x+1}^{y+1}uS_1(u)\,du=(x+1)D(x+1)-(y+1)D(y+1)+\int_{x+1}^{y+1}D(u)\,du .
\]
Hence the left side of the integrated inequality is
$(x+1)D(x+1)-(y+1)D(y+1)+2\int_{x+1}^{y+1}D$, and convexity gives
$\int_{x+1}^{y+1}D\ge(y-x)D(1+\frac{x+y}2)$.
\end{proof}

\subsection{Ramps}\label{sec:prof-ramps}

The deletion inequality applies the covariance bound to the indicator of a
level. Applying it instead to a ramp, a function that increases linearly from
zero to one over an interval of levels, gives inequalities that relate the
stop-loss function at several levels.

Fix a finite cap $c\ge1$, a foot $a\ge0$ and a width $w>0$, and let
$g(z)=\min(1,\max(0,(z-a)/w))$. For $z>a$ put
\[
 u=z-a,\qquad k=\lfloor z/c\rfloor,\qquad r=z-kc,\qquad s=z-a-w,
\]
let $(\sigma,\tau)=(0,u)$ if $z\le a+w$ and $(\sigma,\tau)=(s,w)$ if $z>a+w$, so
that $\sigma+\tau=u$, and let $\varphi(\beta)=\beta\min\{(\beta-\sigma)_+,\tau\}$
for $\beta\ge0$. Define
\[
 \Lambda(z)=\begin{cases}
 k\varphi(c)+\varphi(r) & \text{if } u\ge c,\\
 \max\{\tau kc+\varphi(z-kc),\ \tau ku+\varphi(z-ku)\}& \text{if } u<c \text{ and } kc<z<(k+1)u,\\
 \tau z & \text{otherwise,}
 \end{cases}
\]
and $\Psi(z)=zg(z)-\Lambda(z)/w$ for $z>a$, $\Psi(z)=0$ for $z\le a$.

\begin{lemma}[Ramps]\label{lem:prof-gramp}
With this notation $\Psi\ge0$ and
\begin{equation}\label{eq:prof-gramp}
 \E \Psi(Z_c)\ \le\ \mu_c\,\E g(Z_c)=\frac{\mu_c}{w}\bigl(D_c(a)-D_c(a+w)\bigr).
\end{equation}
\end{lemma}

\begin{proof}
Write $Z=Z_c$ and $\beta_i=\min(X_i,c)$. Since $g$ is bounded and nondecreasing,
the argument of Lemma~\ref{lem:prof-pivcov} with $\ind_H$ replaced by $g(Z)$
gives $\Cov(Z,g(Z))\le\sum_i\Cov(\beta_i,g(Z))$. Let $Z_{-i}$ be the clipped
optimum with $i$ deleted, which does not depend on $X_i$. Then
$\Cov(\beta_i,g(Z))\le\E[\beta_i(g(Z)-g(Z_{-i}))]$, because the difference of the
two sides is $\E\beta_i\,(\E g(Z)-\E g(Z_{-i}))\ge0$ by independence. Fix a
maximizing set $B$. For $i\notin B$ deletion does not change $Z$; for $i\in B$
downward closure gives $Z_{-i}\ge Z-\beta_i$. Hence
\[
 \mu_c\E g(Z)\ \ge\ \E\Bigl[Zg(Z)-\sum_{i\in B}\beta_i\bigl(g(Z)-g(Z-\beta_i)\bigr)\Bigr].
\]
It suffices to show that for every $z>a$ and all $\beta_i\in[0,c]$ with
$\sum_i\beta_i=z$,
\begin{equation}\label{eq:prof-loss}
 w\sum_i\beta_i\bigl(g(z)-g(z-\beta_i)\bigr)\ \le\ \Lambda(z);
\end{equation}
for $z\le a$ the bracket above is zero. If $z\le a+w$ then $w(g(z)-g(z-\beta))=\min(\beta,u)$,
and if $z>a+w$ then $w(g(z)-g(z-\beta))=\min\{(\beta-s)_+,w\}$. In both cases the
left side of \eqref{eq:prof-loss} is $\sum_i\varphi(\beta_i)$. We use three
properties of $\varphi$: it is convex on $[0,u]$, where it equals
$\beta(\beta-\sigma)_+$; it equals $\tau\beta$ for $\beta\ge u$, in particular
$\varphi(u)=\tau u$; and $\varphi(\beta)\le\tau\beta$ for all $\beta$.

Suppose $u\ge c$. Then $\varphi$ is convex on $[0,c]$. Replacing two coordinates
strictly inside $(0,c)$ by the endpoint of their feasible segment with the same
sum does not decrease $\sum_i\varphi(\beta_i)$. Repeating, the sum is at most its
value at $k$ coordinates $c$ and one coordinate $r$.

Suppose $u<c$. The third property gives $\sum_i\varphi(\beta_i)\le\tau z$. Suppose
also $kc<z<(k+1)u$. Call $\beta_i$ large if $\beta_i\ge u$, let $n$ be the number of
large coordinates and $L\in[nu,nc]$ their sum. The others lie in $[0,u]$ and sum
to $z-L$; by convexity on $[0,u]$ their contribution is at most
$j\varphi(u)+\varphi(\rho)$, where $j=\lfloor(z-L)/u\rfloor$ and
$\rho=z-L-ju\in[0,u)$. Hence
$\sum_i\varphi(\beta_i)\le\tau(L+ju)+\varphi(\rho)=\tau(z-\rho)+\varphi(\rho)=:\psi(\rho)$.
Now $(n+j)u\le z<(k+1)u$ gives $n+j\le k$. If $n+j\le k-1$ then
$\rho\ge z-nc-ju\ge z-(k-1)c\ge c>u$, which is impossible. So $n+j=k$ and
$z-kc\le z-nc-ju\le\rho\le z-nu-ju=z-ku$. Both ends lie in $[0,u)$ and $\psi$ is
convex on $[0,u]$, so $\psi(\rho)\le\max\{\psi(z-kc),\psi(z-ku)\}$, which is the
second line of the definition of $\Lambda$. This proves \eqref{eq:prof-loss}.

Finally, in both regimes $\tau=wg(z)$, and $\varphi(\beta)\le\tau\beta$ shows that
each expression for $\Lambda(z)$ is at most $\tau z=wzg(z)$; so $\Psi\ge0$. The
identity for $\E g(Z_c)$ is the integral of the ramp, using the convention
$D_c(t)=\mu_c-t$ for $t\le1$.
\end{proof}

\begin{corollary}[Rounded ramps]\label{lem:prof-ramp}
For a finite cap $c\ge1$ and an integer $k\ge2$,
\begin{equation}\label{eq:prof-ramp}
 kD_c((k-1)c)-(k-1)D_c(kc)\ \le\
 \frac{\mu_c}{c}\bigl(D_c((k-2)c)-D_c((k-1)c)\bigr).
\end{equation}
\end{corollary}

\begin{proof}
Apply Lemma~\ref{lem:prof-gramp} with $a=(k-2)c$ and $w=c$, and put
$\bar f(z)=k(z-(k-1)c)_+-(k-1)(z-kc)_+$. Then $\Psi\ge\bar f$.
First, $\bar f=0\le\Psi$ for $z\le(k-1)c$. For $(k-1)c<z<kc$ we have $u\ge c$,
$\lfloor z/c\rfloor=k-1$, $\sigma=s=r=z-(k-1)c$ and $\tau=c$, so
$\Lambda(z)=(k-1)c(kc-z)$ and $\Psi(z)=z-(k-1)(kc-z)=\bar f(z)$. For $z\ge kc$
we have $s\ge c$, so $\varphi$ vanishes on $[0,c]$, $\Lambda(z)=0$ and
$\Psi(z)=z=\bar f(z)$. Since $\E\bar f(Z_c)$ is the left side of
\eqref{eq:prof-ramp}, the claim follows from \eqref{eq:prof-gramp}.
\end{proof}

For the computation, the left side of~\eqref{eq:prof-gramp} is bounded below by
an affine form. Let $a,w,c$ be multiples of a grid spacing $h$, and consider a
bin $[\xi,\xi+h)$ between consecutive grid points with $\xi\ge a$. On it put
$k=\lfloor\xi/c\rfloor$; the bin lies entirely in $\{z\le a+w\}$ or entirely in
$\{z\ge a+w\}$, where the two definitions of $\varphi$ agree. Write
$r_1=z-kc$ and $r_2=z-ku$.
\begin{enumerate}[label=(\roman*)]
\item If $\xi-a\ge c$: for $z\le a+w$, $w\Psi=zu-kc^2-r_1^2$, which is affine in
 $z$; for $z\ge a+w$, $w\Psi=wz-kc(c-s)_+-r_1(r_1-s)_+$, where $r_1-s=a+w-kc$ is
 constant and $-(c-s)_+$ is concave in $z$.
\item If $\xi+h-a\le c$, $k\ge1$ and $k\xi\ge(k+1)a$, so that $u\le c$ and
 $kc\le z\le(k+1)u$ on the bin: $w\Psi=\min\{\tau r_1-\varphi(r_1),\tau r_2-\varphi(r_2)\}$.
 For $z\le a+w$ the two terms are $r_1(u-r_1)=(z-kc)(kc-a)$, affine, and
 $r_2(u-r_2)=(ka-(k-1)z)(kz-(k+1)a)$, concave. For $z\ge a+w$ they are
 $wr_1-r_1(r_1-s)_+$ with $r_1-s$ constant, affine, and $wr_2-r_2(r_2-s)_+$,
 where $r_2$ and $r_2-s$ are decreasing affine functions of $z$ with $r_2\ge0$,
 so that $r_2(r_2-s)_+$ is convex. At $z=kc$ and $z=(k+1)u$ the minimum vanishes,
 in agreement with the third line of $\Lambda$, and at $u=c$ it agrees with the
 first line.
\item On every other bin use $\Psi\ge0$.
\end{enumerate}
In each case the displayed expression, with $k$ fixed, is concave on the closed
bin and coincides with $\Psi$ on $[\xi,\xi+h)$. So $\Psi$ lies above the chord
$\alpha+\beta z$ through its two endpoint values there, and
\begin{equation}\label{eq:prof-binned}
 \E \Psi(Z_c)\ \ge\ \sum_{\text{bins}}\Bigl[\alpha\bigl(S_c(\xi)-S_c(\xi+h)\bigr)
 +\beta\bigl(T_c(\xi)-T_c(\xi+h)\bigr)\Bigr]+T_c(a+w+c),
\end{equation}
because $\Psi(z)=z$ for $z\ge a+w+c$ and
$\E[Z_c;\xi\le Z_c<\xi+h]=T_c(\xi)-T_c(\xi+h)$. Atoms of $Z_c$ at grid points
belong to the bin on their right, consistently with $S_c(t)=\Prb(Z_c\ge t)$.

\subsection{The affine rows}\label{sec:prof-rows}

The computation of Section~\ref{sec:certificate} fixes an interval $[l,r]$
assumed to contain $\mu_1$. The rows of this section are the following
inequalities, each written as ``affine expression $\le0$''. All of them are at
cap one except the cap-variation rows, which use the caps
$c\in\{1,\frac65,\frac{27}{20},\frac32,\frac74,2,\frac52\}$ and the unclipped
optimum.
\begin{enumerate}[label=(\roman*)]
\item The relations of Lemma~\ref{lem:prof-basic}\ref{prof:basic-riemann}
 between consecutive levels of the grids $\frac1{40}\mathbb Z$,
 $\frac1{80}\mathbb Z$ and $\frac1{160}\mathbb Z$ in $[1,10]$, and between pairs
 of further rational levels: $S_1(y)-S_1(x)\le0$,
 $D_1(x)-D_1(y)-(y-x)S_1(x)\le0$ and $(y-x)S_1(y)-D_1(x)+D_1(y)\le0$ for $x<y$.
\item The identity of Lemma~\ref{lem:prof-basic}\ref{prof:basic-agree} at cap
 $2$ and $y=\frac{41}{40}$, the chord inequalities \eqref{eq:prof-chord} at
 $t=1$ for triples of consecutive caps in this list (here $D_c(1)=\mu_c-1$), and the
 cap-variation bounds $D_b(t)-D_a(t)\le e(a)-e(b)$ of \eqref{eq:prof-slice} for
 $(a,b,t)=(2,\frac52,1)$ and $(2,\infty,\frac{41}{40})$.
\item The deletion inequality \eqref{eq:deletion} and its refinement
 \eqref{eq:prof-intdel} at cap one, with \eqref{eq:prof-strip}, and the ramps
 \eqref{eq:prof-gramp} with \eqref{eq:prof-binned} and \eqref{eq:prof-ramp} at cap
 one. In each, the product $\mu_1p$ with $p=S_1(x)$ or $p=\E g(Z_1)$ is bounded
 above by $rp$.
\item The integrated deletion \eqref{eq:prof-intdel-int}.
\item The case bounds $l\le\mu_1\le r$, that is, $l\le1+D_1(1)\le r$.
\end{enumerate}

\providecommand{\od}{\operatorname{odds}}
\section{Overshoot budgets}\label{sec:overshoot}

A menu with a body at level $t$ earns at least $t$ whenever the body is nonempty,
and it earns more by its \emph{overshoot}, the amount by which the least
inclusion-maximal member of the body exceeds $t$. Since every menu has expected
score at most one, the mean overshoot of each body in the menu family is bounded, and the bound is
stronger when the singleton offers of the menu earn more. This section proves
several such bounds, which we call \emph{budgets}. They are useful because the
overshoot is large whenever the pivotal mass is large, which is exactly when the
deletion inequality of Lemma~\ref{lem:prof-deletion-cap1} is weak.
Sections~\ref{sec:tail} and~\ref{sec:paths} combine the two.

The standing assumptions and the notation of Section~\ref{sec:prof-setup} remain
in force. In addition, we write
\[
 \omega(y)=\frac1{b(y)}-1,\qquad C(s)=1-s+\frac{s^2}2,\qquad \tau_k(q)=2k-1-k^2(1-q)\quad(k\in\mathbb R).
\]
For $q<1$ we have $\od(q)-\tau_k(q)=(1-q)\bigl(k-1/(1-q)\bigr)^2\ge0$, so
\begin{equation}\label{over:tau}
 \tau_k(q)\le\od(q)\qquad(0\le q<1,\ k\in\mathbb R).
\end{equation}

A menu of Algorithm~\ref{alg:capped} at cap $c$ consists of a \emph{body}, which
is a set of reports described by the clipped weights, together with singleton
offers decided by a separate test for each item. We first isolate the two
properties of bodies on which all overshoot estimates rest. We then prove three
budgets for the mean overshoot of bodies \ref{over:B0}--\ref{over:B2}, and two
further budgets for their capped overshoot.

\subsection{Bodies}\label{over:sec-bodies}
Fix a cap $c\ge1$ and write $w_i=\min(X_i,c)$, $w(S)=\sum_{i\in S}w_i$. At a level $t>c$ a feasible report $S$ \emph{qualifies} if $w(S)\ge t$, and a \emph{seed} is an inclusion-minimal qualifying report. Let $\mathcal Q$ denote the family of qualifying reports. It is nonempty exactly when $Z_c\ge t$, and then every qualifying report contains a seed.

\begin{definition}\label{over:def-body}
A \emph{body} at $(c,t)$ is a rule assigning to every realization a set $\mathcal B$ of feasible reports such that
\begin{enumerate}[label=(\roman*)]
\item every member of $\mathcal B$ qualifies, and every seed is a member of $\mathcal B$;
\item $\mathcal B$ is determined by the family $\mathcal Q$ and by the weights $w_i$ of the items $i\in\bigcup\mathcal Q$.
\end{enumerate}
\end{definition}

The bodies of Algorithm~\ref{alg:capped} used below are the following, at cap one ($c=1$) unless stated otherwise.
\begin{enumerate}[label=(B\arabic*),start=0]
\item\label{over:B0} All qualifying reports.
\item\label{over:B1} The seeds and all reports $S$ with $w(S)\ge t+\eta$, for an increment $\eta>0$.
\item\label{over:B2} For a fixed reserve fraction $0<a\le1$: the seeds, and every report $S$ containing a seed $P$ with $w(S\setminus P)\ge a\,b_P$, where $b_P=t+\min_{i\in P}w_i-w(P)$.
\end{enumerate}
The number $b_P$ is positive, because removing a lightest member of the seed $P$
leaves a report of weight below $t$. The estimates for the capped overshoot
of~\ref{over:B1} assume that $\eta\le1$, and the body~\ref{over:B0} is also used at
caps $c>1$.

\begin{lemma}\label{over:local}
Each of \ref{over:B0}--\ref{over:B2} is a body, at every cap $c\ge1$.
\end{lemma}
\begin{proof}
Members of \ref{over:B0}, \ref{over:B1} and \ref{over:B2} contain a seed, hence qualify, and each of them contains the seeds. For (ii), membership in \ref{over:B0} is membership in $\mathcal Q$. Seeds are the minimal members of $\mathcal Q$. In \ref{over:B1} and \ref{over:B2} the remaining tests concern a qualifying report $S$ and evaluate $w(S)$, $w(P)$, $w(S\setminus P)$ and $\min_{i\in P}w_i$ for $P\subseteq S$; all these weights are weights of items of $\bigcup\mathcal Q$.
\end{proof}

For a body at $(c,t)$, the \emph{overshoot} $O_t$ is the least clipped value of an
inclusion-maximal member of $\mathcal B$, minus $t$, if $\mathcal B\ne\varnothing$,
and $O_t=0$ otherwise. By (i), $\mathcal B\ne\varnothing$ exactly when $Z_c\ge t$, and $O_t\ge0$. We refer
to the events $\{Z_c\ge t\}$ and $\{Z_c<t\}$ as \emph{success} and
\emph{failure}. The \emph{capped overshoot} is $H_t=\min(O_t,1)$. When the body must be named we write $O_t^{\mathcal B}$ and $H_t^{\mathcal B}$.

The structural properties used in the budgets are the following two properties of a body. We
call an item with $w_i=c$ \emph{saturated}.

\begin{lemma}[P1]\label{over:P1}
Let $i$ be a saturated item belonging to a qualifying report. Then some seed contains $i$ and strictly contains $\{i\}$. In particular $\{i\}$ is strictly contained in a member of every body at $(c,t)$.
\end{lemma}
\begin{proof}
Let $Q$ be an inclusion-minimal qualifying report among those containing $i$. Since $w(\{i\})=c<t$, $Q\ne\{i\}$. For $j\in Q\setminus\{i\}$ the report $Q\setminus\{j\}$ does not qualify, by the choice of $Q$, and $w(Q\setminus\{i\})=w(Q)-c\le w(Q)-w_j=w(Q\setminus\{j\})<t$. Hence every one-element deletion from $Q$ fails to qualify, and by monotonicity of $w$ every proper subset fails. Thus $Q$ is a seed, and seeds belong to every body.
\end{proof}

\begin{lemma}[P2]\label{over:P2}
Let $x$ and $x'$ be realizations that differ only on a set $I$ of items, with $x'_i\le x_i$ for $i\in I$, and suppose no item of $I$ belongs to a report qualifying at $x$. Then $x$ and $x'$ have the same qualifying family, the same weights on the items of its members, and the same body members with the same clipped values. In particular $O_t$ and $H_t$ coincide at $x$ and $x'$.
\end{lemma}
\begin{proof}
A report qualifying at $x$ contains no item of $I$, so its weights are unchanged and it qualifies at $x'$. A report qualifying at $x'$ qualifies at $x$, because clipped weights are nondecreasing in the values. Hence the qualifying families coincide, and the items of their members lie outside $I$, so their weights coincide. By (ii) of Definition~\ref{over:def-body} the bodies coincide; their members qualify and therefore have the same clipped values.
\end{proof}

\subsection{Unique seeds and pointwise overshoot}\label{over:sec-pointwise}
In this subsection $c=1$, $Z=Z_1$, and $I_t$, $L_t$ are as in Section~\ref{sec:prof-deletion}: $I_t$ is the intersection of the qualifying reports on $\{Z\ge t\}$ and $L_t=w(I_t)$, with $L_t=0$ if $Z<t$. We say that $t$ has a \emph{unique seed} if $Z\ge t$ and there is exactly one seed.

\begin{lemma}\label{over:unique}
On $\{Z\ge t\}$ the following hold.
\begin{enumerate}[label=(\alph*)]
\item $I_t$ qualifies if and only if $t$ has a unique seed, and then $I_t$ is that seed. Consequently $L_t\ge t$ if and only if $t$ has a unique seed.
\item With a unique seed, $L_t-t<\min_{i\in I_t}w_i\le1$. Hence $\pos{L_t-t}<1$ in all cases.
\item With a unique seed $P$, every qualifying report, in particular every optimizer, contains $P$.
\end{enumerate}
\end{lemma}
\begin{proof}
$I_t$ is feasible by downward closure. If it qualifies, it is contained in every qualifying report, so it is the only minimal one. If $P$ is the unique seed, every qualifying report contains a seed, hence $P$, so $P\subseteq I_t\subseteq P$. This proves (a) and (c), since $L_t\ge t$ means that $I_t$ qualifies. For (b), removing a lightest member of the seed $I_t$ leaves a report that does not qualify, so $L_t-\min_{i\in I_t}w_i<t$. Without a unique seed $L_t<t$ by (a).
\end{proof}

\begin{lemma}\label{over:pointwise}
Suppose $t$ has a unique seed $P=I_t$.
\begin{enumerate}[label=(\alph*)]
\item For every body every member contains $P$, and $O_t\ge L_t-t$.
\item For \ref{over:B1}, if $Z\ge t+\eta$ then $O_t\ge\eta$.
\item For \ref{over:B2} write $s=w(P)$, $b=b_P$ and $y=Z-s$. Then $0\le y<b$ and $O_t\ge s-t+ab\,\ind_{\{y\ge ab\}}$.
\end{enumerate}
\end{lemma}
\begin{proof}
(a) Members qualify, so by Lemma~\ref{over:unique}(c) they contain $P$, and their clipped values are at least $w(P)=L_t$.

(b) Let $Q^*$ be an optimizer. It qualifies with $w(Q^*)=Z\ge t+\eta$, so it is a member. Let $M$ be a maximal member. If $w(M)<t+\eta$, then $M$ is a seed, so $M=P\subseteq Q^*$; maximality forces $M=Q^*$, a contradiction. Hence $w(M)\ge t+\eta$.

(c) Let $Q^*$ be an optimizer and $i\in P$ a lightest member of $P$. Then $Q^*\supseteq P$ and $Q^*\setminus\{i\}$ contains no seed, so it does not qualify: $Z-w_i<t$, that is, $y<b$. Clearly $y\ge0$. Every member contains $P$, and every member other than $P$ is an extension $S\supseteq P$ with $w(S\setminus P)\ge ab$, hence $w(S)\ge s+ab$. If $y\ge ab$, then $Q^*$ is such an extension, and $Q^*\ne P$ since $w(Q^*\setminus P)\ge ab>0$; so $P$ is not maximal and every maximal member has value at least $s+ab$. Otherwise every member has value at least $s$.
\end{proof}

All lower bounds in (a)--(c) are at most one, since $L_t-t<1$ by
Lemma~\ref{over:unique}(b) and $s-t+ab\le s-t+b=\min_{i\in P}w_i\le1$. Without a
unique seed we have $L_t<t$, and $O_t\ge0$ always. We therefore obtain the
following pointwise bounds on the capped overshoot, valid on every realization:
\begin{equation}\label{over:capped-pointwise}
\begin{gathered}
 H_t\ge\pos{L_t-t}\quad\text{for every body},\qquad
 H_t\ge\eta\,\ind_{\{t\text{ has a unique seed},\ Z\ge t+\eta\}}\quad\text{for \ref{over:B1} with }\eta\le1,\\
 H_t\ge s-t+ab\,\ind_{\{Z-s\ge ab\}}\quad\text{for \ref{over:B2}, when $t$ has a unique seed.}
\end{gathered}
\end{equation}
The same bounds hold for $O_t$, which is at least $H_t$.

\paragraph{Averaged reserves.}
For $m>0$ and $1\le K\le e^m$ let $\nu_{m,K}$ be the probability measure on $(0,1]$ with density $1/(ma)$ on $[1/K,1]$ and an atom of mass $1-(\log K)/m\ge0$ at $a=1$. The condition $K\le e^m$ holds whenever $\sum_{i=0}^{12}m^i/i!\ge K$. Averaging the menus with body \ref{over:B2} over $a\sim\nu_{m,K}$ gives a mixture of deterministic menus, each with expected score at most one. Write $\overline O_t$ and $\overline H_t$ for the $\nu_{m,K}$-averages of $O_t$ and $H_t$. For a fixed realization only finitely many reports occur, so $a\mapsto O_t$ is a step function with finitely many steps and the averages are well defined and measurable.

\begin{lemma}\label{over:shape}
If $t$ has a unique seed $P$, with $s$, $b$ as in Lemma~\ref{over:pointwise}(c), then
\[
 \overline O_t\ \ge\ \overline H_t\ \ge\ s-t+\frac{\pos{Z-s-b/K}}m .
\]
\end{lemma}
\begin{proof}
Put $y=Z-s\in[0,b)$. By \eqref{over:capped-pointwise}, $\overline H_t\ge s-t+\int ab\,\ind_{\{a\le y/b\}}\,\nu_{m,K}(da)$. The atom at one does not contribute, since $y/b<1$. The density part gives $\int_{1/K}^{y/b}ab\,\frac{da}{ma}=\frac bm\pos{y/b-1/K}$ when $y/b\ge1/K$ and zero otherwise.
\end{proof}

\subsection{Flag sampling and erasure}\label{over:sec-flags}
We now relate the overshoot to the payoff from singleton offers. The menus used
here offer a body together with the singletons of all items whose values exceed a
cutoff, which we call \emph{flagged}. A flagged item that lies in a qualifying
report is absorbed by the body, by property~(P1); a flag that lies in no
qualifying report can be erased without changing the body, by property~(P2).
Counting flag sets according to their erasures gives the inequalities below.

In this subsection and the subsequent budget arguments, $c\ge1$, $t>c$ and a
body $\mathcal B$ of type \ref{over:B0}--\ref{over:B2} at $(c,t)$ are fixed.
This restriction ensures that adding the singleton offers gives a menu in
$\mathcal M$, to which \ref{prof:N1} applies. Bounds for averages of
\ref{over:B2} over reserve fractions follow by averaging the valid
deterministic-menu bounds. We write $S=S_c(t)$ and $\bar S=1-S$.

\begin{lemma}\label{over:bpos}
$b(c)>0$ and $S\le1/t<1$.
\end{lemma}
\begin{proof}
Offer the body \ref{over:B0} at $(c,t)$ and the singletons with $X_i\ge t$ (closed cutoffs at $t$). On $\{Z_c\ge t\}$ some report is offered, and every offer pays at least $t$: body members have raw value at least their clipped value. Hence the score is at least $t\ind_{\{Z_c\ge t\}}$, and $tS\le1$ by \ref{prof:N1}. If $b(c)=\prod_i\Prb(X_i\le c)$ were zero, some item would satisfy $X_i>c$ almost surely. The menu with body \ref{over:B0} at $(c,t)$ and the singletons with $X_i>c$ would then always have an offer, and every offer pays strictly more than one: singletons pay more than $c\ge1$ and body members pay at least $t>1$. Its expected score would exceed one.
\end{proof}

\paragraph{Movable cutoffs.}
For $\theta\ge c$ and $\pi\in[0,1]$, flag item $i$ if $X_i>\theta$, or if $X_i=\theta$ and an independent coin of bias $\pi$, tossed before the values are observed, shows heads. The menu offers $\mathcal B$ and the singletons of the flagged items. For each outcome of the coins this is a menu of Algorithm~\ref{alg:capped} (each item has an open or a closed test at $\theta$), so the randomized menu has expected score at most one. Let $p_i$ be the probability that $i$ is flagged, $p_i(y)$ the probability that $i$ is flagged and $X_i>y$, and $z=\prod_i(1-p_i)$ the probability that no item is flagged. Put
\[
 \Phi_c(z)=\int_c^\infty\min\Bigl\{\omega(y),\frac1z-1\Bigr\}\,dy .
\]

\begin{lemma}\label{over:cutoff}
For every $z\in[b(c),1)$, and for $z=1$ if $b(c)=1$, there are $\theta\ge c$ and $\pi\in[0,1]$ whose flags have absence probability $z$. For these,
\begin{equation}\label{over:cutoff-id}
 \prod_i\bigl(1-p_i(y)\bigr)^{-1}-1=\min\Bigl\{\omega(y),\frac1z-1\Bigr\}\qquad\text{for almost every }y\ge c .
\end{equation}
Moreover $\Phi_c(z)=G(c)$ for $0<z\le b(c)$.
\end{lemma}
\begin{proof}
If $z=b(c)$ take $\theta=c$ and $\pi=0$. Otherwise $b(c)<z<1$; let $\theta=\inf\{y\ge c:b(y)\ge z\}$. Right continuity of $b$ gives $b(\theta)\ge z$, and $b(y)<z$ for $y<\theta$ gives $b(\theta-)\le z$. The absence probability $\prod_i\bigl(\Prb(X_i<\theta)+(1-\pi)\Prb(X_i=\theta)\bigr)$ is continuous in $\pi$ and runs from $b(\theta)$ to $b(\theta-)$, so some $\pi$ gives exactly $z$. For $y>\theta$ an item has $X_i>y$ only if it is flagged, so $p_i(y)=\Prb(X_i>y)$ and the left side of \eqref{over:cutoff-id} is $\omega(y)\le\omega(\theta)\le1/z-1$. For $c\le y<\theta$ every flagged item has $X_i\ge\theta>y$, so $p_i(y)=p_i$ and the left side is $1/z-1\le1/b(\theta-)-1\le\omega(y)$. The last claim holds because $\omega(y)\le\omega(c)\le1/z-1$ for $y\ge c$ and $z\le b(c)$.
\end{proof}

\paragraph{Provisional draws.}
Fix $\theta,\pi$ as in Lemma~\ref{over:cutoff}. Since $z>0$, every $p_i<1$. Sample independently, for all items, the flag indicators, a \emph{low draw} $\check X_i$ from the law of $X_i$ given that $i$ is unflagged, and a \emph{high draw} $\hat X_i$ from the law of $X_i$ given that $i$ is flagged (omitted if $p_i=0$). Let $J$ be the flag set and set $X_i=\hat X_i$ for $i\in J$, $X_i=\check X_i$ otherwise. This reproduces the joint law of the values and flags. We call $(\check X,\hat X)$ the \emph{draws}. Given the draws the realization is a function of $J$, and $\Prb(J=I)=z\varpi(I)$ with $\varpi(I)=\prod_{i\in I}p_i/(1-p_i)$; thus $\sum_I\varpi(I)=1/z$ and $\varpi(I\cup I')=\varpi(I)\varpi(I')$ for disjoint $I,I'$.

Low draws are at most $\theta$ and high draws at least $\theta\ge c$, so flagged items are saturated and switching on flags only raises values. Let $\check Z=W\bigl((\min(\check X_i,c))_i\bigr)$, $\Lambda=\{\check Z<t\}$ and $f=\Prb(\Lambda)$. Then $Z_c\ge\check Z$, so failure $\{Z_c<t\}$ is contained in $\Lambda$ and $f\ge\bar S>0$ by Lemma~\ref{over:bpos}. The event $\{J=\varnothing\}\cap\Lambda$ implies failure and has probability $zf$, since $\Lambda$ depends only on the low draws. Hence
\begin{equation}\label{over:zf}
 zf\le\bar S,\qquad f\le\min\Bigl(1,\frac{\bar S}z\Bigr).
\end{equation}
Let $q=\Prb(Z_c<t\mid\check X,\hat X)$. Then $q=0$ off $\Lambda$, $q\ge z$ on $\Lambda$, and $\E[q;\Lambda]=\bar S$. Let $A$ be the event that every flag belongs to some qualifying report (in particular $J=\varnothing$ belongs to $A$), and $A^+=A\cap\{J\ne\varnothing\}$. We write $\E_\Lambda$ for conditional expectation given $\Lambda$.

\begin{lemma}\label{over:erasure}
Let $Y\ge0$ be any of $O_t$, $H_t$, $\ind_{\{H_t\ge r\}}$ $(0<r\le1)$, or their averages over reserve fractions. Then, almost surely,
\begin{equation}\label{over:erase-count}
 \E[Y\ind_A\mid\check X,\hat X]\ \ge\ z\,\E[Y\mid\check X,\hat X]\cdot\begin{cases}1/q&\text{on }\Lambda,\\ 1&\text{off }\Lambda.\end{cases}
\end{equation}
Moreover, on $\Lambda$,
\begin{equation}\label{over:witness}
 \frac1q\le1+\frac{\Prb(A^+\mid\check X,\hat X)}z,
\end{equation}
and consequently
\begin{equation}\label{over:cover}
 \frac{\Prb(A^+)}z\ \ge\ \E_\Lambda\Bigl[\frac1q\Bigr]-1\ \ge\ \frac f{\bar S}-1 .
\end{equation}
\end{lemma}
\begin{proof}
Fix the draws. For a flag set $J$ let $\varphi(J)\subseteq J$ consist of the flags that belong to a report qualifying at $J$. Passing from $J$ to $\varphi(J)$ lowers the values of the items of $J\setminus\varphi(J)$ from $\hat X_i$ to $\check X_i$, and these items lie in no qualifying report. By Lemma~\ref{over:P2}, $J$ and $\varphi(J)$ have the same qualifying family and the same body with the same clipped values. Hence $\varphi(J)\in A$ and $Y(\varphi(J))=Y(J)$.

Fix $K\in A$. Every $J$ with $\varphi(J)=K$ has the form $K\cup I$ with $I\cap K=\varnothing$. On $\Lambda$, the set $I$ switched on alone gives failure: a report qualifying with exactly $I$ switched on must contain an item of $I$, because $\check Z<t$, and it still qualifies at $J$ by monotonicity, so that item would belong to $\varphi(J)$. Since $\sum_{I\text{ fails alone}}\varpi(I)=q/z$, the preimages of $K$ have total weight at most $\varpi(K)q/z$ on $\Lambda$, and at most $\varpi(K)/z$ in general. Therefore
\[
\begin{split}
 \E[Y\mid\check X,\hat X]&=\sum_J z\varpi(J)Y(J)=\sum_{K\in A}Y(K)\sum_{\varphi(J)=K}z\varpi(J)\\
 &\le\frac qz\sum_{K\in A}z\varpi(K)Y(K)=\frac qz\,\E[Y\ind_A\mid\check X,\hat X]
\end{split}
\]
on $\Lambda$, with $q$ replaced by one off $\Lambda$. This is \eqref{over:erase-count}.

For \eqref{over:witness}, on $\Lambda$ call a nonempty flag set $I$ a \emph{witness} if some report containing $I$ qualifies when exactly $I$ is switched on. If $J$ succeeds and $Q$ qualifies at $J$, then $Q\cap J\ne\varnothing$ because $\check Z<t$, and $Q\cap J$ is a witness contained in $J$. Split $J$ into the union $V(J)$ of the witnesses contained in $J$ and the remainder $J\setminus V(J)$. The remainder fails when switched on alone, since otherwise it would contain a witness. The union $V(J)$ is empty or belongs to $A^+$: each witness $I\subseteq V(J)$ lies in a report qualifying with $I$ switched on, which still qualifies with $V(J)$ switched on. The two parts determine $J$. Summing $\varpi$ over all $J$,
\[
 \frac1z\le\Bigl(1+\sum_{K\in A^+}\varpi(K)\Bigr)\sum_{I\text{ fails alone}}\varpi(I)=\Bigl(1+\frac{\Prb(A^+\mid\check X,\hat X)}z\Bigr)\frac qz,
\]
which is \eqref{over:witness}.

For \eqref{over:cover}, condition on $J$ and $\hat X$. The event $A^+$ is nondecreasing in the low draws (raising values keeps qualifying reports qualifying), while $\Lambda$ is nonincreasing in them and independent of $(J,\hat X)$. The low draws are independent, so Lemma~\ref{lem:prof-assoc} gives $\Prb(A^+\cap\Lambda\mid J,\hat X)\le f\,\Prb(A^+\mid J,\hat X)$, hence $\Prb(A^+\cap\Lambda)\le f\Prb(A^+)$. By \eqref{over:witness}, $\Prb(A^+\cap\Lambda)\ge z\E[1/q-1;\Lambda]=zf\,\E_\Lambda[1/q-1]$. Finally $\E_\Lambda[1/q]\ge1/\E_\Lambda[q]=f/\bar S$ by Jensen's inequality.
\end{proof}

\begin{lemma}\label{over:score}
For $z$ and the flags of Lemma~\ref{over:cutoff},
\begin{equation}\label{over:score-eq}
 1\ \ge\ 1-z+zf\,\Phi_c(z)+(t-1)\Prb(A^+)+zt(1-f)+\E[O_t;A].
\end{equation}
\end{lemma}
\begin{proof}
We bound the score of the menu pointwise. Suppose first that $J\ne\varnothing$, and let $m=\min_{i\in J}X_i\ge\theta\ge c$. On $\Lambda$ every body member contains a flag, since a qualifying report without flags would have weight at most $\check Z<t$; its raw value therefore exceeds its clipped value by at least $m-c$. On $A^+$, Lemma~\ref{over:P1} shows that every flagged singleton is strictly contained in a body member, so the inclusion-maximal offers are the maximal body members, and the score is at least $t+O_t+(m-c)\ind_\Lambda$. Off $A^+$ the maximal offers are flagged singletons, paying at least $m$, and body members, paying at least $t>1$, and at least $t+m-c\ge m$ on $\Lambda$; so the score is at least $1+(m-c)\ind_\Lambda$, using $c\ge1$. If $J=\varnothing$ the values are the low draws, success means $\check Z\ge t$, and the score is at least $(t+O_t)\ind_{\{\check Z\ge t\}}$. Altogether
\[
 \text{score}\ \ge\ \ind_{\{J\ne\varnothing\}}\bigl(1+(m-c)\ind_\Lambda\bigr)+(t-1)\ind_{A^+}+t\,\ind_{\{J=\varnothing\}\setminus\Lambda}+O_t\ind_A .
\]
The variable $m$ and the event $\{J\ne\varnothing\}$ depend only on the flags and high draws, so they are independent of $\Lambda$. By \eqref{over:cutoff-id},
\[
 \E[(m-c)\ind_{\{J\ne\varnothing\}}]=\int_c^\infty\Prb(J\ne\varnothing,\ m>y)\,dy
 =z\int_c^\infty\Bigl[\prod_i\Bigl(1+\frac{p_i(y)}{1-p_i}\Bigr)-1\Bigr]dy\ \ge\ z\,\Phi_c(z),
\]
because $1+p_i(y)/(1-p_i)\ge1/(1-p_i(y))$ when $p_i(y)\le p_i$. Taking expectations and using that the menu has expected score at most one proves \eqref{over:score-eq}.
\end{proof}

\subsection{Three budgets for the mean overshoot}\label{over:sec-three}
We keep $c\ge1$, $t>c$ and a body of type \ref{over:B0}--\ref{over:B2} at $(c,t)$, and we write $o=\E O_t$, $\kappa=o/\bar S$, $G=G(c)$, $e=e(c)$, and
\[
 B_t=1-(t-1)\od(S).
\]

\begin{proposition}\label{over:area}
For every $0<z<1$,
\begin{equation}\label{over:area-eq}
 \Phi_c(z)+\max(o,\kappa z)\le B_t .
\end{equation}
In particular
\begin{equation}\label{over:pay}
 \E O_t+G(c)+(t-1)\od(S_c(t))\le1 .
\end{equation}
\end{proposition}
\begin{proof}
Let $z\in[b(c),1)$, or $z=1$ if $b(c)=1$. By \eqref{over:erase-count} with $Y=O_t$, $\E[O_t;A]\ge zo$. By \eqref{over:cover}, $(t-1)\Prb(A^+)\ge(t-1)z(f/\bar S-1)$. Insert both into \eqref{over:score-eq} and divide by $zf$:
\[
 0\ge-\frac1f+\Phi_c(z)+\frac{t-1}{\bar S}-\frac{t-1}f+\frac tf-t+\frac of
 =\Phi_c(z)+\frac of-\Bigl(t-\frac{t-1}{\bar S}\Bigr).
\]
Since $t-(t-1)/\bar S=B_t$, this gives $\Phi_c(z)+o/f\le B_t$, and \eqref{over:zf} gives $o/f\ge\max(o,\kappa z)$. If $b(c)<1$ this proves \eqref{over:area-eq} for $z\ge b(c)$; for $z<b(c)$ the function $\Phi_c$ takes the same value $G(c)$ (Lemma~\ref{over:cutoff}) while $\max(o,\kappa z)$ is smaller. If $b(c)=1$ then $\Phi_c\equiv0$, and the case $z=1$ gives $\max(o,\kappa)\le B_t$, which implies \eqref{over:area-eq} for all $z<1$. Finally \eqref{over:pay} is \eqref{over:area-eq} at $z=b(c)$ if $b(c)<1$, and is $o\le B_t$ if $b(c)=1$, when $G(c)=0$.
\end{proof}

\begin{corollary}\label{over:projection}
$e(c)\le G(c)\le1$ and $S_c(t)\le(1-G(c))/(t-G(c))\le1/t$. Moreover $R-\mu_c\le e(c)\le G(c)\le G(1)\le1$ for every $c\ge1$, and in particular $R-1\le\mu_1\le\mu_c\le R$.
\end{corollary}
\begin{proof}
By independence, $-\log b(y)=-\sum_i\log\Prb(X_i\le y)\ge\sum_i\Prb(X_i>y)$, and $-\log b\le1/b-1$; integrating over $y\ge c$ gives $e(c)\le G(c)$. By \eqref{over:pay} with $O_t\ge0$, $G(c)\le1$ and $(1-S)(1-G)\ge(t-1)S$, that is $(t-G)S\le1-G$; the right side $(1-G)/(t-G)$ decreases in $G\in[0,1]$. For the last claim, $R-\mu_c\le e(c)$ and $\mu_c\le R$ by Lemma~\ref{lem:prof-basic}, $G$ is nonincreasing as the integral of a nonnegative function over $[c,\infty)$, and $G(1)\le1$ is the case $c=1$ of what was just shown, applied at any level $t>1$.
\end{proof}

\begin{proposition}\label{over:budgets}
The mean overshoot satisfies the quadratic budget
\begin{equation}\label{over:envelope}
 C(S)\,o+\bar S e+tS\le1,
\end{equation}
and, for every $1/2\le v<1$,
\begin{equation}\label{over:integrated}
 v\,o+\bar S e+tS\le1+\Bigl(\frac1{2(1-v)}-1\Bigr)\bar S\,(G-e).
\end{equation}
\end{proposition}
\begin{proof}
For every $y\ge c$, $\int_0^1\min\{\omega(y),1/z-1\}\,dz=\int_0^{b(y)}\omega(y)\,dz+\int_{b(y)}^1(1/z-1)\,dz=-\log b(y)$. By Tonelli's theorem and the first step in the proof of Corollary~\ref{over:projection},
\begin{equation}\label{over:Phi-int}
 \int_0^1\Phi_c(z)\,dz=\int_c^\infty-\log b(y)\,dy\ \ge\ e .
\end{equation}
Integrate \eqref{over:area-eq} over $z\in(0,1)$. Since $\kappa z\le o$ exactly when $z\le\bar S$,
\[
 \int_0^1\max(o,\kappa z)\,dz=o\bar S+\kappa\,\frac{1-\bar S^2}2=\frac o{\bar S}\Bigl(\bar S^2+\frac{1-\bar S^2}2\Bigr)=\frac{o\,C(S)}{\bar S}.
\]
Multiplying the integrated inequality by $\bar S$ and using $\bar SB_t=1-tS$ gives \eqref{over:envelope}.

For \eqref{over:integrated}, if $\kappa=0$ the claim follows from $\bar Se\le\bar SG\le\bar SB_t=1-tS$ (by \eqref{over:pay}) and $G\ge e$. Let $\kappa>0$ and put $\beta=B_t-G\ge0$. Since $\Phi_c(z)\le G$ and $\Phi_c(z)\le B_t-\kappa z$, we have $G-\Phi_c(z)\ge\pos{\kappa z-\beta}$. Integration and \eqref{over:Phi-int} give
\[
 G-e\ \ge\ \int_0^1\pos{\kappa z-\beta}\,dz\ \ge\ \frac{\pos{\kappa-\beta}^2}{2\kappa}.
\]
If $\kappa-\beta>0$, the inequality between arithmetic and geometric means gives $(1-v)\kappa+(G-e)/(2(1-v))\ge(1-v)\kappa+(\kappa-\beta)^2/(4\kappa(1-v))\ge\kappa-\beta$; if $\kappa\le\beta$ this holds trivially. Thus $v\kappa\le B_t-G+(G-e)/(2(1-v))=B_t-e+(\frac1{2(1-v)}-1)(G-e)$. Multiplying by $\bar S$ gives \eqref{over:integrated}.
\end{proof}

Each of \eqref{over:pay}, \eqref{over:envelope} and \eqref{over:integrated} is affine in $o$ with a positive coefficient, and all other terms are independent of the body. Hence they hold for averages of bodies \ref{over:B2} over reserve fractions, with $o=\E\overline O_t$, and, since $0\le H_t\le O_t$, with $o$ replaced by $h=\E H_t$ or $\E\overline H_t$.

\paragraph{Affine forms.}
From now on $c=1$. Let $U\ge\mu_1$ be an upper bound for the cap-one mean (in the certificate, the right endpoint of the interval assumed for $\mu_1$), and put
\[
 g=R-U,\qquad s_*=\frac{1-g}{t-g},\qquad F(s)=\frac{1-g-(t-g)s}{C(s)} .
\]
By Corollary~\ref{over:projection}, $g\le R-\mu_1\le e(1)\le G(1)\le1$ and $S_1(t)\le s_*$, since $(1-G)/(t-G)$ decreases in $G$.

\begin{corollary}\label{over:affine}
Let $o$ be the mean overshoot of a body of type \ref{over:B0}--\ref{over:B2} at $(1,t)$, or its average over reserve fractions of \ref{over:B2}, or the mean of the corresponding capped overshoot. Then $o+\beta_i\le0$ for $i=0,1,2$, where
\begin{align}
 \beta_0&=G(1)-1+(t-1)\tau_k(S_1(t)),\label{over:beta0}\\
 \beta_1&=-\bigl\{F(x)+F'(x)(S_1(t)-x)\bigr\}+\frac{1-s_*}{C(s_*)}\bigl(e(1)-g\bigr),\label{over:beta1}\\
 \beta_2&=\frac1v\Bigl\{(t-g)S_1(t)+g-1+(1-s_*)\bigl(e(1)-g\bigr)-\Bigl(\frac1{2(1-v)}-1\Bigr)\bigl(G(1)-e(1)\bigr)\Bigr\},\label{over:beta2}
\end{align}
for every $k\in\mathbb R$, $0\le x\le s_*$ and $1/2\le v<1$. Each $\beta_i$ is affine in the unknowns $S_1(t)$, $G(1)$, $e(1)$.
\end{corollary}
\begin{proof}
The bound for $\beta_0$ follows from~\eqref{over:pay} and~\eqref{over:tau}.

For $\beta_2$, we rewrite \eqref{over:integrated} as $vo\le1-g-(t-g)S-\bar S(e-g)+(\frac1{2(1-v)}-1)\bar S(G-e)$, and use $\bar S(e-g)\ge(1-s_*)(e-g)$ and $\bar S(G-e)\le G-e$, valid because $e\ge g$, $G\ge e$ and $\frac1{2(1-v)}\ge1$.

For $\beta_1$, the budget~\eqref{over:envelope} gives $o\le F(S)-\bar S(e-g)/C(S)$. The function $(1-s)/C(s)$ is nonincreasing on $[0,1]$, since the numerator of its derivative is $-C(s)+(1-s)^2=-s(1-s/2)\le0$; with $S\le s_*$ and $e\ge g$ this gives $o\le F(S)-\frac{1-s_*}{C(s_*)}(e-g)$. It remains to show that $F$ is concave on $[0,s_*]$, so that its tangent at $x$ lies above it. Write $N(s)=1-g-(t-g)s\ge0$ on $[0,s_*]$, so $FC=N$. Differentiation gives $F'C^2=-(t-g)C-NC'$ and, with $C'=-(1-s)$,
\[
 (t-g)C+NC'=(t-1)(1-s)+(t-g)\,s\Bigl(1-\frac s2\Bigr)\ge0,
\]
so $F'\le0$. Differentiating $FC=N$ twice gives $F''C=-FC''-2F'C'$ with $C''=1$; since $F\ge0$ and $F'C'\ge0$, $F''\le0$.
\end{proof}

\subsection{Two budgets for the capped overshoot}\label{over:sec-capped}
In this subsection, $c=1$, $t>1$, $S=S_1(t)$ and $\bar S=1-S$, and the body is
of type \ref{over:B0}--\ref{over:B2} at $(1,t)$. We write $h=\E H_t$. Unlike the budgets above, the two budgets
below depend essentially on the bound $H_t\le1$.

\begin{lemma}\label{over:joint}
For $k>0$ and $1\le\gamma\le2k/\bigl(1+\sqrt{1+1/(t-1)}\bigr)$,
\begin{equation}\label{over:b3}
 G(1)-1+(t-1)\tau_k(S)+\gamma h\ \le\ (\sqrt\gamma-1)^2 .
\end{equation}
\end{lemma}
\begin{proof}
Use the flags of Lemma~\ref{over:cutoff} with $z=b=b(1)$, so that $\Phi_1(z)=G(1)$; write $G=G(1)$ and $\alpha=t-1$. Let $\bar h=\E[H_t\mid\check X,\hat X]$ and $h_0=\E[H_t;\Lambda^c]$. By \eqref{over:erase-count} with $Y=H_t$,
\[
 \E[H_t;A]\ \ge\ b\bigl(f\,\E_\Lambda[\bar h/q]+h_0\bigr).
\]
By the first inequality of \eqref{over:cover}, $\Prb(A^+)\ge b\,\E_\Lambda[1/q-1]$. Insert both into \eqref{over:score-eq}, with $O_t\ge H_t$:
\[
 1\ge1-b+bfG+\alpha b\,\E_\Lambda[1/q]-\alpha b+bt(1-f)+bf\,\E_\Lambda[\bar h/q]+bh_0 .
\]
Divide by $bf$; since $-1/f-\alpha/f+t/f=0$, this becomes
\begin{equation}\label{over:joint-1}
 \E_\Lambda\Bigl[\frac{\alpha/f+\bar h}q\Bigr]+\frac{h_0}f\ \le\ 1+\alpha-G .
\end{equation}
For $X>0$ and any $\lambda$, $X/q\ge2\lambda\sqrt X-\lambda^2q$. With $X=\frac\alpha f(1+f\bar h/\alpha)$ and $\lambda=k\sqrt{\alpha f}$ this gives
\[
 \frac{\alpha/f+\bar h}q\ \ge\ 2\alpha k\sqrt{1+f\bar h/\alpha}-\alpha fk^2q .
\]
Because $H_t\le1$ we have $0\le f\bar h\le1$, and therefore
\[
 2\alpha k\Bigl(\sqrt{1+\tfrac{f\bar h}\alpha}-1\Bigr)=\frac{2kf\bar h}{\sqrt{1+f\bar h/\alpha}+1}\ \ge\ \frac{2k}{1+\sqrt{1+1/\alpha}}\,f\bar h\ \ge\ \gamma f\bar h .
\]
Also $f\,\E_\Lambda q=\bar S$ and $f\,\E_\Lambda\bar h=\E[H_t;\Lambda]=h-h_0$. Taking $\E_\Lambda$ in \eqref{over:joint-1},
\[
 2\alpha k-\alpha k^2\bar S+\gamma(h-h_0)+\frac{h_0}f\le1+\alpha-G,
\]
that is, $G-1+\alpha\tau_k(S)+\gamma h\le(\gamma-1/f)h_0$. Since $H_t\le1$, $h_0\le1-f$. If $\gamma\ge1/f$, then $(\gamma-1/f)h_0\le(\gamma-1/f)(1-f)=\gamma+1-(\gamma f+1/f)\le(\sqrt\gamma-1)^2$, because $\gamma f+1/f\ge2\sqrt\gamma$. Otherwise $(\gamma-1/f)h_0\le0$.
\end{proof}

\begin{lemma}\label{over:layer}
For $0<r\le1$ let $d(r)=\Prb(H_t\ge r)$. Then $d(r)\le S<1$ and
\begin{equation}\label{over:layerineq}
 tS+\bar S\,e(1)+\frac12\int_0^1\Bigl[d(r)-\bar S^2\,\frac{\log(1-d(r))}{1-d(r)}\Bigr]dr\ \le\ 1 .
\end{equation}
Consequently
\begin{equation}\label{over:quad}
 tS+\bar S\,e(1)+C(S)\,h+\tfrac34\bar S^2h^2\ \le\ 1 .
\end{equation}
\end{lemma}
\begin{proof}
Since $H_t\ge r>0$ implies success, $d(r)\le S$. Put $\bar d=1-d(r)\ge\bar S>0$. Let $z\in[b(1),1)$, or $z=1$ if $b(1)=1$, and use the flags of Lemma~\ref{over:cutoff}. In the proof of Proposition~\ref{over:area}, keep $\E[O_t;A]\ge\E[H_t;A]$ in place of $zo$; the same algebra gives
\begin{equation}\label{over:layer-1}
 \Phi_1(z)+\frac{\E[H_t;A]}{zf}\le B_t .
\end{equation}
Fix $r$ and write $d=d(r)$, $\delta=\Prb(H_t\ge r,\Lambda)$. Let $\bar p=\Prb(H_t\ge r\mid\check X,\hat X)$. By \eqref{over:erase-count} with $Y=\ind_{\{H_t\ge r\}}$, $\Prb(H_t\ge r,A\mid\check X,\hat X)\ge z\bar p/q$ on $\Lambda$ and $\ge z\bar p$ off $\Lambda$. On $\Lambda$, $q\le1-\bar p$ because $\{H_t\ge r\}$ implies success. Since $p\mapsto p/(1-p)$ is convex, Jensen's inequality gives
\[
 \frac{\Prb(H_t\ge r,A)}z\ \ge\ (d-\delta)+f\,\frac{\delta/f}{1-\delta/f}=d+\frac{\delta^2}{f-\delta}.
\]
Here $f-\delta\ge\Prb(\Lambda,\ J=\varnothing)=zf>0$. Also $\delta\ge\pos{f-\bar d}$, since $\Prb(H_t<r)=\bar d$, and $\delta^2/(f-\delta)$ increases in $\delta\in[0,f)$. Dividing by $f$, the right side is at least $\psi(f)$, where $\psi(f)=d/f$ for $f\le\bar d$ and $\psi(f)=d/f+(f-\bar d)^2/(\bar df)=1/f+f/\bar d-2$ for $f\ge\bar d$. The function $\psi$ is continuous, decreases on $(0,\sqrt{\bar d}]$ and increases afterwards. By \eqref{over:zf}, $f\le\min(1,\bar S/z)$, and $\sqrt{\bar d}\le1$, so $\psi(f)\ge\Psi(z,d)$, where
\[
 \Psi(z,d)=\begin{cases}
 2/\sqrt{\bar d}-2,&z\le\bar S/\sqrt{\bar d},\\
 \bar S/(z\bar d)+z/\bar S-2,&\bar S/\sqrt{\bar d}\le z\le\bar S/\bar d,\\
 dz/\bar S,&\bar S/\bar d\le z.
 \end{cases}
\]
Since $\E[H_t;A]=\int_0^1\Prb(H_t\ge r,A)\,dr$, \eqref{over:layer-1} yields
\begin{equation}\label{over:layer-2}
 \Phi_1(z)+\int_0^1\Psi\bigl(z,d(r)\bigr)\,dr\le B_t .
\end{equation}
The function $\Psi(\cdot,d)$ is continuous and nondecreasing (the middle piece has derivative $1/\bar S-\bar S/(z^2\bar d)\ge0$ there), and $\Phi_1$ is constant on $(0,b(1)]$. Hence \eqref{over:layer-2} holds for every $0<z<1$: for $z<b(1)$ it follows from the case $z=b(1)$, and if $b(1)=1$ from the case $z=1$. A direct integration of the three pieces, split at $\bar S/\sqrt{\bar d}$ and $\bar S/\bar d\le1$, gives
\[
 \int_0^1\Psi(z,d)\,dz=\frac d{2\bar S}-\frac{\bar S\log\bar d}{2\bar d}.
\]
All integrands are nonnegative, so integrating \eqref{over:layer-2} over $z\in(0,1)$, using \eqref{over:Phi-int} and multiplying by $\bar S$ with $\bar SB_t=1-tS$, proves \eqref{over:layerineq}.

For \eqref{over:quad}, the power series $-\log(1-d)/(1-d)=\sum_{n\ge1}(1+\frac12+\dots+\frac1n)d^n\ge d+\frac32d^2$. With $\int_0^1d(r)\,dr=h$ and $\int_0^1d(r)^2\,dr\ge h^2$, the integral in \eqref{over:layerineq} is at least $\frac12\bigl((1+\bar S^2)h+\frac32\bar S^2h^2\bigr)$, and $\frac12(1+\bar S^2)=C(S)$.
\end{proof}

\begin{corollary}\label{over:b4cor}
Let $U\ge\mu_1$, $g=R-U$ and $s_*=(1-g)/(t-g)$ as before. Let $0\le\zeta\le1$, $\gamma=\frac32(1-s_*)^2\zeta$, $\nu=\gamma\zeta/2$, and suppose $t-1+\gamma(t-g)\ge\nu$. Then
\[
 Q(s)=\frac{1-g+\nu-(t-g)s}{C(s)+\gamma}
\]
is nonincreasing and concave on $[0,s_*]$, and
\begin{equation}\label{over:b4}
 h\ \le\ Q(S)-\frac{1-s_*}{C(s_*)+\gamma}\bigl(e(1)-g\bigr).
\end{equation}
In particular $h+\beta_4\le0$ with $\beta_4=-\{Q(x)+Q'(x)(S-x)\}+\frac{1-s_*}{C(s_*)+\gamma}(e(1)-g)$ for every $0\le x\le s_*$, an expression affine in $S_1(t)$ and $e(1)$.
\end{corollary}
\begin{proof}
Since $\bar S\ge1-s_*$ and $h^2\ge2\zeta h-\zeta^2$, we have $\frac34\bar S^2h^2\ge\frac34(1-s_*)^2(2\zeta h-\zeta^2)=\gamma h-\nu$. By \eqref{over:quad}, $(C(S)+\gamma)h\le1-g+\nu-(t-g)S-\bar S(e(1)-g)$. The function $(1-s)/(C(s)+\gamma)$ is nonincreasing on $[0,1]$, the numerator of its derivative being $-s(1-s/2)-\gamma$; with $S\le s_*$ and $e(1)\ge g$ this proves \eqref{over:b4}.

Write $N_\nu(s)=1-g+\nu-(t-g)s$, which is at least $\nu\ge0$ on $[0,s_*]$. Then $(C+\gamma)^2Q'=P(s)$ with $P(s)=-(t-g)(C(s)+\gamma)+(1-s)N_\nu(s)$. Now $P(0)=\nu-(t-1)-\gamma(t-g)\le0$ by hypothesis, and $P'(s)=(t-g)(1-s)-N_\nu(s)-(t-g)(1-s)=-N_\nu(s)\le0$. Hence $Q'\le0$ on $[0,s_*]$. Differentiating $(C+\gamma)Q=N_\nu$ twice gives $(C+\gamma)Q''=-2C'Q'-Q$, which is nonpositive because $C'\le0$, $Q'\le0$ and $Q\ge0$. The tangent bound follows from concavity and $S\le s_*$.
\end{proof}

\paragraph{Averages and rational evaluation.}
Both \eqref{over:b3} and \eqref{over:b4} are affine in $h$, and their other terms do not depend on the body. Hence they hold for every body \ref{over:B0}--\ref{over:B2} at $(1,t)$ (with $\eta\le1$ for \ref{over:B1}), and for averages of \ref{over:B2} over reserve fractions, with $h=\E\overline H_t$. Together with Corollary~\ref{over:affine} and $0\le h\le1$, every such capped mean overshoot therefore satisfies
\begin{equation}\label{over:hbudgets}
 0\le h\le1,\qquad h+\beta_i\le0\quad(i=0,1,2,3,4),
\end{equation}
where \eqref{over:b3} is used with $\gamma=\varrho^2$ for a rational $\varrho\ge1$, in the form $h+\beta_3\le0$ with
\[
 \beta_3=\frac{G(1)-1+(t-1)\tau_k(S_1(t))-(\varrho-1)^2}{\varrho^2}.
\]
Its hypothesis is equivalent to $2k/\gamma-1\ge0$ and $(2k/\gamma-1)^2\ge1+1/(t-1)$, which are rational conditions. The certificate uses \eqref{over:b4} with $\zeta=\min\{1,\max\{0,(1-g-(t-g)x)/C(x)\}\}$. The inequalities \eqref{over:hbudgets} are the budgets for the capped-overshoot unknowns of Section~\ref{sec:paths}. For $i\le2$ they hold also with $h$ replaced by $\E O_t$.

\providecommand{\od}{\operatorname{odds}}
\section{Rows at fractional thresholds}\label{sec:tail}

This section combines the deletion inequality of Section~\ref{sec:profile} with
the overshoot budgets of Section~\ref{sec:overshoot}, one level at a time, at
levels on a fine grid. At a single level, the deletion inequality charges a menu
for the pivotal mass $L_t$ and the budget charges it for the overshoot $O_t$. The
two charges are complementary: a large pivotal mass forces a unique seed and
hence a large overshoot. Bounding their sum pointwise in terms of $\lfloor NZ\rfloor$
gives an inequality between values of $S_1$ on the grid.

The standing assumptions of Section~\ref{sec:prof-setup} remain in force.
Throughout, $[l,r]$ is an interval containing $\mu_1$, as in a case of the
computation of Section~\ref{sec:certificate}, and $g=R-r$.

\subsection{Deletion rows at real levels}\label{sec:tail-deletion}

\begin{lemma}\label{lem:tail-deletion-rows}
Let $r_c\ge\mu_c$ be upper bounds at finitely many caps $c$, and let $x>1$ be
real. Then
\begin{equation}\label{eq:tail-del2}
 D_1(x+1)+(x+1)S_1(x+1)\ \le\ lS_1(x)+\frac{1-v}{x-v}\,(\mu_1-l),
\end{equation}
where $v=\max\bigl(\{0\}\cup\{R-r_c:c\ge x\}\bigr)$.
\end{lemma}

\begin{proof}
The left side is $T_1(x+1)$, which is at most
$\mu_1S_1(x)=lS_1(x)+(\mu_1-l)S_1(x)$ by \eqref{eq:deletion}, and
$\mu_1-l\ge0$. By Corollary~\ref{over:projection}, $S_1(x)\le(1-G)/(x-G)$ with
$G=G(1)$, and $G\ge G(c)\ge R-\mu_c\ge R-r_c$ for every listed cap $c\ge x$;
also $G\ge0$. Hence $0\le v\le G\le1$, and since $(1-G)/(x-G)$ decreases in
$G\in[0,1]$, $S_1(x)\le(1-v)/(x-v)$.
\end{proof}

Since $\mu_1=1+D_1(1)$, \eqref{eq:tail-del2} is affine in the unknowns. It is
used at rational $x\in(1,9]$ with $r_c=R$ for the caps $c>1$, so that $v=0$
and the coefficient is $1/x$.

\subsection{Overshoot rows at fractional thresholds}\label{sec:tail-overshoot}

In this subsection the cap is one: $w_i=\min(X_i,1)$, $Z=Z_1$, and $t>1$. We
use the qualifying reports, seeds, $I_t$ and $L_t$ of
Section~\ref{sec:prof-deletion}, the bodies \ref{over:B0} and \ref{over:B1} of
Section~\ref{over:sec-bodies}, and their overshoot $O_t$.

\paragraph{Pivotal weight on a fine grid.}
Fix a positive integer $N$ and measure values in units of $1/N$. Let $T>N$
be an integer and $t=T/N$. On success, that is, if $Z\ge t$, put $z=NZ$, $j=\lfloor z\rfloor$,
$\lambda=NL_t$ and $\Omega=NO_t$, and let $\kappa=|I_t|$. For \ref{over:B1} let
the increment be $\eta=\iota/N$ with an integer $1\le\iota\le N$.

\begin{lemma}\label{lem:tail-pivots}
Suppose that $Z\ge t$.
\begin{enumerate}
\item Every $i\in I_t$ lies in every maximizer of $Z$ and satisfies $w_i>Z-t$.
 Hence $\lambda\le\min(z,\kappa N)$, and $I_t=\varnothing$ if $Z\ge t+1$.
\item If $j>T$, then $\kappa\le n:=\lfloor(j-1)/(j-T)\rfloor$.
\item If $j>T$ and $L_t<t$, then $\kappa\le m:=\lfloor(T-1)/(j-T)\rfloor$.
\end{enumerate}
\end{lemma}

\begin{proof}
Let $S^*$ be a maximizer of $Z$. It qualifies, so it contains $I_t$. If some
$i\in I_t$ had $w_i\le Z-t$, then the feasible set $S^*\setminus\{i\}$ would
qualify and avoid $i$. Thus each of the $\kappa$ members of $I_t$ has weight
in $(Z-t,1]$, which gives $L_t\le w(S^*)=Z$, $L_t\le\kappa$, and
$I_t=\varnothing$ when $Z-t\ge1$. This is item~1.

For item~2, if $\kappa\ge1$, then $\kappa(z-T)<\lambda\le z$, so
$\kappa<z/(z-T)\le j/(j-T)$, since $z/(z-T)$ decreases for $z>T$ and
$z\ge j>T$. As $\kappa$ is an integer,
$\kappa\le\lceil j/(j-T)\rceil-1=\lfloor(j-1)/(j-T)\rfloor$. Item~3 is the
same argument with $\lambda<T$ in place of $\lambda\le z$:
$\kappa<T/(z-T)\le T/(j-T)$, so
$\kappa\le\lceil T/(j-T)\rceil-1=\lfloor(T-1)/(j-T)\rfloor$.
\end{proof}

Recall from Lemma~\ref{over:unique} that, if $Z\ge t$, then $L_t\ge t$ if and
only if $t$ has a unique seed, and that $I_t$ is then this seed. We refer to this
situation as the \emph{unique case}. By Lemma~\ref{over:pointwise}, the
overshoot is always nonnegative, and in the unique case $O_t\ge L_t-t$ for every
body; moreover $O_t\ge\eta$ for~\ref{over:B1} if $Z\ge t+\eta$.

We now define integer-valued functions $d$ on the integers, which bound the
combined charge from below in terms of $j$. For $T<j<T+N$ we write $n=n(j)$ and
$m=m(j)$ as in Lemma~\ref{lem:tail-pivots}. For the body~\ref{over:B0} at
price one,
\begin{equation}\label{eq:tail-d0}
 d_{\rm B0}(j)=\begin{cases}0,&j\le T,\\ \max(0,T-nN),&T<j<T+N,\\ T,&j\ge T+N.\end{cases}
\end{equation}
For the body~\ref{over:B1} with increment $\iota/N$ at price $p\ge1$, we set
$d(j)=0$ for $j\le T$ and $d(j)=T$ for $j\ge T+N$. For $T<j<T+N$, let
$\delta_{\rm m}=\max(0,T-mN)$ and $\delta_{\rm u}=\max(0,T-nN)$; if
$j\ge T+\iota$, we replace $\delta_{\rm u}$ by
$\max\bigl(\delta_{\rm u},(p-1)\iota,p\iota+T-nN\bigr)$. Then put
\begin{equation}\label{eq:tail-d1}
 d_{\rm B1}(j)=\begin{cases}\min(\delta_{\rm m},\delta_{\rm u}),&nN\ge T,\\
 \delta_{\rm m},&nN<T.\end{cases}
\end{equation}

\begin{lemma}\label{lem:tail-charge}
Let $p=1$ for~\ref{over:B0} and $p\ge1$ for~\ref{over:B1}, and let $d$ be the
corresponding function. Then, pointwise,
\[
 N\bigl[(t-L_t)\ind_{\{Z\ge t\}}+pO_t\bigr]\ \ge\ d(\lfloor NZ\rfloor).
\]
The same holds with $O_t$ replaced by $H_t=\min(O_t,1)$.
\end{lemma}

\begin{proof}
Write $\Phi=(T-\lambda)\ind_{\{Z\ge t\}}+p\Omega$. If $Z<t$, then
$L_t=O_t=0$, so $\Phi=0=d(\lfloor NZ\rfloor)$ because $\lfloor NZ\rfloor<T$.
Assume that $Z\ge t$, so that $j\ge T$. We always have $\Phi\ge0$: if $\lambda\le T$ this is clear, and if
$\lambda>T$ we are in the unique case, so $\Omega\ge\lambda-T$ and $p\ge1$.
This settles the case $j=T$. If $j\ge T+N$, then $Z\ge t+1$, so $\lambda=0$ and
$\Phi\ge T$. Let $T<j<T+N$.

For~\ref{over:B0}, $\lambda\le\kappa N\le nN$ by Lemma~\ref{lem:tail-pivots}.
If $\lambda\le T$, then $\Phi\ge T-\lambda\ge T-nN$, and if $\lambda>T$, then
$T-nN<0$. With $\Phi\ge0$ this gives $\Phi\ge\max(0,T-nN)$.

For~\ref{over:B1}, outside the unique case we have $\lambda<T$ and $\kappa\le m$, so
$\Phi\ge T-\lambda\ge\max(0,T-mN)=\delta_{\rm m}$. In the unique case
$\lambda\ge T$, which requires $nN\ge\kappa N\ge\lambda\ge T$; so when $nN<T$
the unique case does not occur. In the unique case $\Phi\ge0\ge T-nN$, so
$\Phi\ge\max(0,T-nN)$. If moreover $j\ge T+\iota$, then $Z\ge t+\eta$ and
$\Omega\ge\iota$. Then $\Phi\ge T-nN+p\iota$, and
$\Phi=(T-\lambda+\Omega)+(p-1)\Omega\ge(p-1)\iota$ because
$\Omega\ge\lambda-T$. In each case $\Phi$ is at least the minimum of the bounds
of the cases that can occur, which is $d_{\rm B1}(j)$.

The argument uses $O_t$ only through $O_t\ge0$ and, in the unique case,
$O_t\ge L_t-t$ and, for \ref{over:B1}, $O_t\ge\eta$ if $Z\ge t+\eta$. By
\eqref{over:capped-pointwise}, $H_t$ satisfies the same three bounds when
$\eta\le1$, which proves the last claim.
\end{proof}

\begin{proposition}[Fractional-threshold overshoot rows]\label{prop:tail-rows}
Let $N\ge1$ and $T>N$ be integers, $t=T/N$, and let $(d,p)$ be
$(d_{\rm B0},1)$, or $(d_{\rm B1},p)$ with an increment $\iota/N\le1$ and an
integer $p\ge1$. Then
\begin{equation}\label{eq:tail-exact}
 \sum_{j=T+1}^{T+N}\frac{d(j)-d(j-1)}{N}\,S_1\Bigl(\frac jN\Bigr)
 +D_1(t)-\mu_1S_1(t)+p\bigl\{G(1)-1+(t-1)\od(S_1(t))\bigr\}\ \le\ 0,
\end{equation}
and consequently, for every real $k$,
\begin{equation}\label{eq:tail-rows}
 \sum_{j=T+1}^{T+N}\frac{d(j)-d(j-1)}{N}\,S_1\Bigl(\frac jN\Bigr)
 +D_1(t)-rS_1(t)+p\bigl\{G(1)-1+(t-1)\tau_k(S_1(t))\bigr\}\ \le\ 0 .
\end{equation}
\end{proposition}

\begin{proof}
Since $d$ vanishes up to $T$ and is constant from $T+N$ on, and
$\{\lfloor NZ\rfloor\ge j\}=\{Z\ge j/N\}$ even when $Z$ has atoms,
\[
 \E\,d(\lfloor NZ\rfloor)=\sum_{j=T+1}^{T+N}\bigl(d(j)-d(j-1)\bigr)S_1(j/N).
\]
Add \eqref{eq:prof-deletion-cap1} and $p$ times \eqref{over:pay} at $c=1$ with
$1$ moved to the left, and bound $\E[(t-L_t)\ind_{\{Z\ge t\}}+pO_t]$ below by
$\E d(\lfloor NZ\rfloor)/N$ using Lemma~\ref{lem:tail-charge}. This is
\eqref{eq:tail-exact}. For \eqref{eq:tail-rows} replace $\mu_1S_1(t)$ by the
larger $rS_1(t)$, and $\od(S_1(t))$ by the smaller $\tau_k(S_1(t))$, using
\eqref{over:tau} and $t>1$.
\end{proof}

The left side of~\eqref{eq:tail-rows} is affine in the unknowns
$S_1(j/N)$ for $T<j\le T+N$, $S_1(t)$, $D_1(t)$ and $G(1)$. The coefficients of
the values $S_1(j/N)$ are integers divided by $N$. The certificate uses $N=320$, thresholds $t$ that are multiples of $1/320$
in $(1,9]$, increments $\eta\le1$ and prices $p\le2$ for \ref{over:B1}, and
tangent parameters $k\in\frac18\mathbb Z$. It uses the rows only in the
modified forms of the next subsection.

\subsection{Modified rows}\label{sec:tail-modified}

Each of the following four lemmas strengthens~\eqref{eq:tail-rows} by replacing
some of its terms. We write $M=G(1)-1+(t-1)\tau_k(S_1(t))$ for the budget term.

\begin{lemma}[Layer rows]\label{lem:tail-layer}
In \eqref{eq:tail-rows} the term $pM$ may be replaced by $p\beta_4$, where
$\beta_4$ is the expression of Corollary~\ref{over:b4cor} at level $t$ with
$U=r$ and any admissible parameters.
\end{lemma}

\begin{proof}
By the last claim of Lemma~\ref{lem:tail-charge}, the proof of
Proposition~\ref{prop:tail-rows} goes through with $O_t$ replaced by $H_t$, and
it remains to bound $p\,\E H_t$. By Corollary~\ref{over:b4cor},
$p\,\E H_t\le-p\beta_4$.
\end{proof}

\begin{lemma}[Mean-cancelling tangent]\label{lem:tail-cancel}
Let $l\le\mu_1\le r$, $\alpha>0$, $k>0$, and suppose $k_*=k-(r-l)/(\alpha k)>0$.
Then for $0\le s<1$,
\begin{equation}\label{eq:tail-cancel}
 \alpha\,\od(s)-\mu_1s\ \ge\ \alpha\tau_k(s)-rs+(r-\mu_1)\Bigl(1-\frac2{k+k_*}\Bigr).
\end{equation}
Consequently, if $k>0$, $\alpha=p(t-1)$ and $k+k_*>2$, the row
\eqref{eq:tail-rows} remains valid after adding $(r-\mu_1)\bigl(1-2/(k+k_*)\bigr)$
to its left side.
\end{lemma}

\begin{proof}
Put $\delta=r-\mu_1\in[0,r-l]$ and $k_\mu=\sqrt{k^2-\delta/\alpha}$, which is real
since $k^2>(r-l)/\alpha$. Then $\alpha k_\mu^2-\mu_1=\alpha k^2-r$, so
$\alpha\tau_{k_\mu}(s)-\mu_1s$ and $\alpha\tau_k(s)-rs$ have the same coefficient
of $s$, and their constant terms differ by
\[
 \alpha\bigl(2(k_\mu-k)-(k_\mu^2-k^2)\bigr)=\alpha(k_\mu-k)(2-k-k_\mu)=\delta\Bigl(1-\frac2{k+k_\mu}\Bigr),
\]
using $k_\mu-k=-\delta/(\alpha(k+k_\mu))$. Moreover
$k-k_\mu=\delta/(\alpha(k+k_\mu))\le(r-l)/(\alpha k)$, so $k_\mu\ge k_*>0$ and
$1-2/(k+k_\mu)\ge1-2/(k+k_*)$. Now \eqref{over:tau} with parameter $k_\mu$ gives
\eqref{eq:tail-cancel}. The row follows from \eqref{eq:tail-exact} by applying
\eqref{eq:tail-cancel} with $s=S_1(t)$ and $\alpha=p(t-1)$.
\end{proof}

\begin{lemma}[Chord rows]\label{lem:tail-chord}
Let $\alpha>0$, $q\in\mathbb R$ and $P(\xi)=\bigl(\sqrt{\pos{\xi+q}}-\sqrt\alpha\bigr)_+^2$.
Then $\alpha\,\od(s)-\mu_1s\ge qs-P(\mu_1)$ for $0\le s<1$, $P$ is convex, and
for $l<r$
\[
 P(\mu_1)\le\bar P(l)+\frac{\bar P(r)-\bar P(l)}{r-l}(\mu_1-l)
\]
for $l\le\mu_1\le r$ and any upper bounds $\bar P(l)\ge P(l)$,
$\bar P(r)\ge P(r)$. Consequently, with $\alpha=p(t-1)$, the terms
$-rS_1(t)+p(t-1)\tau_k(S_1(t))$ of \eqref{eq:tail-rows} may be replaced by
$qS_1(t)-\bar P(l)-\frac{\bar P(r)-\bar P(l)}{r-l}(\mu_1-l)$.
\end{lemma}

\begin{proof}
For $\xi$ real, $\sup_{0\le s<1}\bigl[(\xi+q)s-\alpha s/(1-s)\bigr]$ is zero if
$\xi+q\le\alpha$ (the bracket is nonincreasing in $s$), and otherwise is
attained at $1-s=\sqrt{\alpha/(\xi+q)}$ with value $(\sqrt{\xi+q}-\sqrt\alpha)^2$.
So this supremum is $P(\xi)$; as a supremum of affine functions of $\xi$ it is
convex, and taking $\xi=\mu_1$ gives the first claim. Convexity bounds
$P(\mu_1)$ by the chord through $(l,P(l))$ and $(r,P(r))$, and replacing the
endpoint values by upper bounds keeps an upper bound at every $\mu_1\in[l,r]$,
because the chord is a convex combination of them. The row follows from
\eqref{eq:tail-exact}.
\end{proof}

In the chord rows, the certificate takes $q=\alpha k^2-(l+r)/2$, and rational
upper bounds $\bar P(l)$ and $\bar P(r)$.

The last lemma bounds a stop-loss increment through the tail bound of
Corollary~\ref{over:projection}.

\begin{lemma}[Trapezoid rows]\label{lem:tail-trap}
Let $1\le x<y$, put $G=G(1)$, and let $V_G(1)=1$ and $V_G(u)=(1-G)/(u-G)$ for
$u>1$. Then $D_1(x)-D_1(y)\le\frac{y-x}2\bigl(V_G(x)+V_G(y)\bigr)$. Each value
$V_G(u)$ with $u>1$ may be bounded above by $(1-g)/(u-g)$, or by the tangent
$\frac{1-g_0}{u-g_0}-\frac{u-1}{(u-g_0)^2}(G-g_0)$ at any $g_0<1$.
\end{lemma}

\begin{proof}
If $G=1$, then $S_1(u)=0$ for $u>1$ by Corollary~\ref{over:projection}, so
$D_1(x)=D_1(y)$ and the claim is trivial. Let $G<1$. By
Lemma~\ref{lem:prof-basic}\ref{prof:basic-riemann},
$D_1(x)-D_1(y)\le\int_x^yS_1(u)\,du$, and $S_1(u)\le V_G(u)$ for $u\ge1$ by
Corollary~\ref{over:projection}. The function $V_G$ is convex on $[1,\infty)$
(it equals $(1-G)/(u-G)$ there, including $u=1$), so its integral over $[x,y]$
is at most the trapezoid value. For fixed $u>1$ the map $G\mapsto(1-G)/(u-G)$ is
decreasing and concave on $(-\infty,1]$, with derivative $-(u-1)/(u-G)^2$. Since
$G\ge R-\mu_1\ge g$ by Corollary~\ref{over:projection}, it is at most its value
at $g$, and by concavity at most its tangent at any $g_0<1$.
\end{proof}

\paragraph{The rows.}
The certificate uses the following rows of this section: the
rows~\eqref{eq:tail-del2}; the rows~\eqref{eq:tail-rows} for \ref{over:B0} and \ref{over:B1} with the modification
of Lemma~\ref{lem:tail-cancel}, of Lemma~\ref{lem:tail-chord}, or of
Lemma~\ref{lem:tail-layer} (the last only for \ref{over:B1}); and the trapezoid
rows of Lemma~\ref{lem:tail-trap} with the tangent bound at
$g_0=R-(l+r)/2$ when this is less than one and at $g_0=g$ otherwise.

\providecommand{\od}{\operatorname{odds}}

\section{Pivotal paths}\label{sec:paths}

The rows of Section~\ref{sec:tail} use one level at a time. The rows of this
section use many levels at once, and they exploit the fact that the pivotal sets
of a single realization are nested. As the level increases, the items common to
all qualifying reports can only accumulate, and they always lie in the same
optimal report. A menu at one level is charged according to the pivotal mass at
that level, so charges at different levels are coupled. We bound the total charge
from below by a minimum over all paths that the pivotal sets could follow, and
compute this minimum by dynamic programming.

The standing assumptions of Section~\ref{sec:prof-setup} remain in force. We
work at cap one, with $w_i=\min(X_i,1)$ and $Z=Z_1$, and we use, at a level
$t>1$, the qualifying reports, the seeds, the intersection $I_t$ and its clipped
weight $L_t$ of Section~\ref{sec:prof-deletion}. The rows combine the deletion
inequality~\eqref{eq:prof-deletion-cap1}, the pointwise overshoot
bounds~\eqref{over:capped-pointwise} and Lemma~\ref{over:shape}, and the
budgets~\eqref{over:hbudgets}. A \emph{reserve shape} $(m,K)$ is the average of
the menus with body~\ref{over:B2} over the reserve fraction $a\sim\nu_{m,K}$, as
in Section~\ref{over:sec-pointwise}. For a reserve shape, $\overline H_t$ takes
the place of $H_t$, and $h$ denotes the average of the means $\E H_t$.

\subsection{The pivotal sets at all reached levels}\label{sec:path-sets}

Fix a positive integer $N$, the number of grid units per unit of value, so that
the cap corresponds to $N$ grid units. Fix a realization and a maximizing report,
and measure clipped weights, levels and overshoots in grid units, that is,
multiply them by $N$. Write $z=NZ$ and $j=\lfloor z\rfloor$. The levels we
consider are the integers $T>N$, corresponding to the levels $t=T/N>1$, and a
level is \emph{reached} if $T\le j$, or equivalently $Z\ge t$.

At reached levels, the family of qualifying reports shrinks as $T$ grows, so the
intersections $I_t$ grow, and all of them lie inside the fixed maximizing report.
We call the members of $I_t$ \emph{pivots}. Each pivot weighs strictly more than
$z-T$ grid units, since removing it from the maximizing report leaves a feasible
report that does not qualify. Put $\Delta=j-T$, let $L=NL_t$ be the pivotal mass
in grid units, and let $k=\min(4,|I_t|)$.

\paragraph{States and restrictions.}
In the \emph{first rounding}, the state at a reached level is the pair $(k,l)$
with $l=\lfloor L\rfloor$. The states satisfy the following restrictions.
\begin{enumerate}[label=(R\arabic*)]
\item\label{it:path-r1} We have $0\le l\le j$. If $k=0$, then the intersection
 is empty and $l=0$. If $k\ge1$, then $\Delta<N$ and $k\Delta\le l$, and if
 moreover $k<4$, then $l\le Nk$. Indeed, each pivot weighs more than $\Delta$
 and at most $N$ grid units, and since $L>k\Delta$ and $k\Delta$ is an integer,
 $l\ge k\Delta$.
\item\label{it:path-r2} Let $T'<T$ be consecutive reached levels, with states
 $(k',l')$ and $(k,l)$. Either the intersection is the same at both levels, or
 $q\ge1$ items enter it. In the second case $l'<T$. Otherwise the old
 intersection would qualify at $T$ by downward closure; the new intersection
 would then be contained in it, and since it also contains the old one, the two
 would be equal. Each new pivot weighs more than $\Delta$ and at most $N$ grid
 units, so $L-L'\in(q\Delta,qN]$. Both endpoints are integers, and therefore
 $l-l'\in[q\Delta,qN]$.
\item\label{it:path-r3} In the second case of~\ref{it:path-r2}, if $k<4$, then
 $k>k'$ and $q=k-k'$. If $k=4$, then $q\ge\max(1,4-k')$, and the upper bound on
 $l-l'$ is replaced by $l-l'\le j$.
\item\label{it:path-r4} Once the intersection is nonempty, it remains nonempty
 at all higher reached levels.
\end{enumerate}
Before the first level the state is $(0,0)$. A sequence of states that starts at
$(0,0)$ and satisfies these restrictions is called \emph{admissible}. Because the
pivotal sets are nested, the true counts and masses of a realization form an
admissible sequence.

In the \emph{second rounding}, we draw one auxiliary variable $\xi$, uniform on
$[0,1)$ and the same for all reached levels, and put $l=\lfloor L+\xi\rfloor$,
so that $\E_\xi l=L$ and $|L-l|<1$. Now $l\le j+1$. In~\ref{it:path-r2}, an
arrival is possible from a state with $l'\le T$, since $L'<T$, and
in~\ref{it:path-r3} the bound for $k=4$ becomes $l-l'\le j+1$. The other
restrictions remain valid. Adding the same $\xi$ to two masses preserves integer
lower and upper bounds on their difference, $L>k\Delta$ gives $L+\xi>k\Delta$,
and $L\le kN$ gives $l\le kN$.

\paragraph{Terms and local charges.}
A \emph{stage} is a level $T>N$ together with a finite list of \emph{terms}. A
term consists of
\begin{itemize}
\item a body at $t=T/N$, which is either~\ref{over:B0},~\ref{over:B1} with an
 increment $\eta=\iota/N$ for an integer $1\le\iota\le N$, or a reserve shape
 $(m,K)$; for~\ref{over:B0} we put $\iota=0$;
\item a positive integer multiplier $w$;
\item a deletion price $a\in\{0,1\}$;
\item an integer overshoot price $p\ge0$, where $p=0$ is allowed only if $a=1$.
\end{itemize}
At a reached level, the \emph{realized charge} of a term is
\[
 w\bigl\{a(T-L)+p\,NH_t\bigr\},
\]
with $\overline H_t$ in place of $H_t$ for a reserve shape. We bound it from
below by a \emph{local charge} $w\chi$, where $\chi$ depends only on the state,
on $T$ and $j$, and on the term.

Recall that $t$ is in the unique case of Lemma~\ref{over:unique} if it has a
unique seed, or equivalently if $L_t\ge t$. In the unique case we write
$d=l-T$. For the bodies~\ref{over:B0} and~\ref{over:B1}, let $\theta=\iota$ if
$j\ge T+\iota$ and $\theta=0$ otherwise. Since $j\ge T+\iota$ implies
$Z\ge t+\eta$, and $\eta\le1$, the bound~\eqref{over:capped-pointwise} gives
$NH_t\ge\max(L-T,\theta)$ in the unique case.

\begin{lemma}\label{lem:path-charges}
In the first rounding, the following functions $\chi$ satisfy
$a(T-L)+p\,NH_t\ge\chi$ at every reached level, with $\overline H_t$ in place
of $H_t$ for a reserve shape.
\begin{enumerate}[label=(\alph*)]
\item If $k=0$, then $\chi=aT$.
\item If $k\ge1$ and $l\le T-1$, then $\chi=a(T-l-1)$.
\item If $l\ge T$ and $p=0$, then $\chi=-d-1$.
\item If $l\ge T$, $p\ge1$ and the body is~\ref{over:B0} or~\ref{over:B1}, then
 \[
  \chi=\max\bigl\{(p-a)d,\ p\theta-a(d+1),\ (p-a)\theta\bigr\}.
 \]
\item If $l\ge T$, $p\ge1$ and the body is a reserve shape $(m,K)$, then, with
 $n=\max(k,\lceil l/N\rceil)$ and $\bar b=\min(N,(l+1)/n)-d$,
 \[
  \chi=(p-a)d+\Big\lfloor\frac pm\pos{j-l-1-\bar b/K}\Big\rfloor .
 \]
\end{enumerate}
\end{lemma}

\begin{proof}
Parts~(a) and~(b) follow from $H_t\ge0$, together with $L=0$ in~(a) and
$L<l+1$ in~(b). In parts~(c)--(e) we have $L\ge l\ge T$, so $t$ is in the
unique case. Write $x=L-T$, so that $x\in[d,d+1)$.

(c) Here $a=1$ and $T-L=-x>-d-1$.

(d) The charge is at least $-ax+p\max(x,\theta)$. This is at least
$(p-a)x\ge(p-a)d$, and at least $p\theta-ax\ge p\theta-a(d+1)$. If $x\ge\theta$,
it is at least $(p-a)x\ge(p-a)\theta$, and if $x<\theta$, it equals
$p\theta-ax\ge(p-a)\theta$.

(e) Let $P=I_t$ be the unique seed. Its cardinality is at least $L/N\ge l/N$ and
at least $k$, hence at least $n$. Its lightest member weighs at most
$\min(N,L/|P|)<\min(N,(l+1)/n)$ grid units, and therefore
$Nb_P=T-L+N\min_{i\in P}w_i\le\bar b$. The mass of the maximizing report outside
$P$ is $z-L>j-l-1$. By Lemma~\ref{over:shape},
$N\overline H_t\ge x+\pos{z-L-Nb_P/K}/m$, so the charge is at least
$(p-a)x+\frac pm\pos{j-l-1-\bar b/K}\ge\chi$.
\end{proof}

In the second rounding, we use the following local charges:
\begin{itemize}
\item $aT$ if $k=0$, and $a(T-l)$ if $k\ge1$ and $l\le T$;
\item if $l>T$, with $d=l-T$: the value $-d$ if $p=0$; the value
 $p\max(d,\theta)-ad$ for the bodies~\ref{over:B0} and~\ref{over:B1}; and, for a
 reserve shape,
 \[
  (p-a)d+\Big\lfloor\frac pm\pos{j-l-1-\bar b'/K}\Big\rfloor,
 \]
 where $n=\max(k,\lceil l/N\rceil)$ and $\bar b'=\min(N,(l+1)/n)-(l-1-T)$.
\end{itemize}
These are not pointwise lower bounds on the realized charge, but their means
over $\xi$ are. The deletion part $a(T-l)$ has mean $a(T-L)$. If $l>T$, then
$L>l-1\ge T$, so the realization is in the unique case; hence on a realization
that is not in the unique case, the overshoot credit vanishes for every $\xi$.
On a realization in the unique case we have $L\ge T$, and the function
$y\mapsto\max(y-T,\theta)$ is affine between consecutive integers, so
$\E_\xi\max(l-T,\theta)=\max(L-T,\theta)\le NH_t$; discarding the credit when
$l\le T$ can only decrease it. Likewise $\E_\xi\pos{l-T}=L-T$. For a reserve
shape, the cardinality bound survives the upward rounding, since $l<L+1\le
N|P|+1$ gives $\lceil l/N\rceil\le|P|$. Since $|L-l|<1$, the quantity $\bar b'$
is an upper bound on $Nb_P$, and $j-l-1$ is a lower bound on $z-L$, so each floor
term is at most $\frac pm\pos{z-L-Nb_P/K}$. Together with
Lemma~\ref{over:shape}, this shows that the mean over $\xi$ of the local charge
is at most the realized charge.

The local charge of a state at a stage with several terms is the sum of the
$w$-weighted local charges of its terms. It is important that this sum is formed
before any minimization over states. All terms of a stage, and all stages, see
the same pivotal sets in a given realization, and minimizing separately would
allow different menus to assign incompatible pivotal masses to the same
realization.

\paragraph{The path minimum.}
Fix a finite list of stages $T_1<\dots<T_r$ and one of the two roundings. For an
integer $j$, let $\pi(j)$ be the minimum, over all admissible sequences of states
at the stages $T_i\le j$, of the sum of the local charges at these stages. In
particular, $\pi(j)=0$ for $j<T_1$. If $j\ge T_r+N$, then $\Delta\ge N$ at every
stage, so every intersection is empty and $\pi(j)=\sum_iT_i\sum_{\text{terms}}wa$
does not depend on $j$. Some admissible sequences may not arise from any
realization, but this only makes $\pi$ smaller. The minimum is computed by
dynamic programming over the stages, with one prefix minimum per stage over the
intervals of predecessor states allowed by~\ref{it:path-r2}
and~\ref{it:path-r3}.

\begin{proposition}\label{prop:path-row}
Let $U\ge\mu_1$. For every list of stages and either rounding,
\begin{equation}\label{eq:path-row}
 \sum_j\frac{\pi(j)-\pi(j-1)}{N}\,S_1(j/N)
 +\sum_{\text{stages}}\ \sum_{\text{terms}}w\bigl[a\{D_1(t)-U\,S_1(t)\}
 -p\,h_{t,\mathrm{body}}\bigr]\le0,
\end{equation}
where $t=T/N$ is the level of the stage and $h_{t,\mathrm{body}}$ is the mean
capped overshoot of the body of the term, averaged over the shape for a reserve
shape.
\end{proposition}

\begin{proof}
Consider first the first rounding. By Lemma~\ref{lem:path-charges}, in every
realization the sum of the realized charges over the reached stages and their
terms is at least the sum of the local charges along the true sequence of
states. This sequence is admissible, so the sum is at least
$\pi(\lfloor NZ\rfloor)$. In the second rounding, for each $\xi$ the rounded
states form an admissible sequence, so the sum of the local charges is at least
$\pi(\lfloor NZ\rfloor)$, and by the discussion above its mean over $\xi$ is at
most the sum of the realized charges. For reserve shapes, the realized charge is
the average over the reserve fraction. Unreached stages contribute nothing,
since $L_t=H_t=0$ when $Z<t$. Taking expectations, we obtain
\[
 \E\,\pi(\lfloor NZ\rfloor)\le N\sum_{\text{stages}}\sum_{\text{terms}}
 w\bigl\{a\,\E[(t-L_t)\ind_{Z\ge t}]+p\,h_{t,\mathrm{body}}\bigr\}.
\]
Since $\pi(j)=0$ for $j<T_1$ and $\pi$ is constant for $j\ge T_r+N$, summation
by parts gives the finite sum
$\E\,\pi(\lfloor NZ\rfloor)=\sum_j(\pi(j)-\pi(j-1))\Prb(\lfloor NZ\rfloor\ge j)$,
and $\{\lfloor NZ\rfloor\ge j\}=\{Z\ge j/N\}$ even when $Z$ has atoms. Dividing
by $N$ and using
$\E[(t-L_t)\ind_{Z\ge t}]\le\mu_1S_1(t)-D_1(t)\le US_1(t)-D_1(t)$, which
is~\eqref{eq:prof-deletion-cap1}, gives~\eqref{eq:path-row}.
\end{proof}

The grid enters only through the lower bound $\pi$; the survival and stop-loss
functions in~\eqref{eq:path-row} are those of the original distribution.
Adding $wp$ copies of a budget $h+\beta_i\le0$ from~\eqref{over:hbudgets}
to~\eqref{eq:path-row} replaces the term $-p\,h$ by $p\beta_i$. We keep $h$ as
an unknown instead, so that a combination can use several budgets for the same
body in fixed proportions, in particular $\beta_3$ and $\beta_4$.

\paragraph{Single-stage rows.}
For a single stage $T$ in the first rounding, we may relax the minimum over
states to the minimum of $aT$, the charge of the empty state, and the local
charges over $l$ alone, where $k$ is replaced by $\max(1,\lceil l/N\rceil)$
in~\ref{it:path-r1} and in Lemma~\ref{lem:path-charges}. The result is at most
$\pi(j)$, because the true state satisfies the restrictions with this smaller
count, and a smaller count can only lower the charge of a reserve shape. We
combine this relaxation with a single fixed budget $\E O_t+\beta_1\le0$ or
$\E O_t+\beta_2\le0$ of Corollary~\ref{over:affine} for the one term, using
$O_t$ in place of $H_t$ in the local charges, which is allowed since $O_t\ge
H_t$. The resulting inequality is a \emph{single-stage row}.

\subsection{The grid unknowns}\label{sec:path-grid}

The rows~\eqref{eq:path-row} are used with $N=80$ and $N=160$, together with the
budgets~\eqref{over:hbudgets} at the corresponding levels, and the single-stage
row uses $N=40$. All levels lie in $[1,10]$. The rows involve $S_1(J/160)$ and
$D_1(J/160)$ for the integers $160\le J\le1600$. Consecutive values are related
by the Riemann relations of Lemma~\ref{lem:prof-basic}\ref{prof:basic-riemann}
(Section~\ref{sec:prof-rows}). A level that is a multiple of $1/40$ gives the
same unknown on every grid.

The mean capped overshoots are further unknowns, one for each combination of
grid, level and body that occurs, and they lie in $[0,1]$. The same body on two
different grids gives two different unknowns, which only weakens the system.

\section{The finite computation}\label{sec:certificate}

\subsection{The linear system}

Let $\rho_0=29/70+10^{-6}=2900007/7000000$ and $R=1/\rho_0=7000000/2900007$,
and consider an instance normalized as
in Section~\ref{sec:normalization}. By Corollary~\ref{over:projection},
$\mu_1\in[R-1,R]$. A \emph{case} of the certificate is a closed interval $[l,r]$
with rational endpoints, together with a list of pairs, each consisting of a row
and a positive integer multiplier. If $\mu_1\in[l,r]$, then the quantities of
the instance satisfy the following inequalities, each of the form ``affine
expression $\le0$''.
\begin{enumerate}[label=(\alph*)]
\item The rows of Section~\ref{sec:prof-rows}: the Riemann relations; the
 agreement, chord and cap-variation rows; the deletion and ramp rows at cap one;
 the integrated deletion rows; and $l\le\mu_1\le r$.
\item The rows of Section~\ref{sec:tail}: the deletion
 rows~\eqref{eq:tail-del2}, the fractional-threshold rows~\eqref{eq:tail-rows}
 in the forms of Lemmas~\ref{lem:tail-layer}--\ref{lem:tail-chord}, and the
 trapezoid rows of Lemma~\ref{lem:tail-trap}.
\item The path rows~\eqref{eq:path-row} for $N\in\{80,160\}$ with $U=r$, the
 single-stage row of Section~\ref{sec:path-sets} with $N=40$, and the
 budgets $h+\beta_i\le0$ for $i=0,\dots,4$ of~\eqref{over:hbudgets}, with $U=r$,
 for the capped overshoot unknowns.
\end{enumerate}
The independent coordinates are the values $S_1(x)$ and $D_1(x)$ at the levels $x\in[1,10]$
that occur; the stop-loss values $D_c(1)$ at the caps
$c\in\{\frac65,\frac{27}{20},\frac32,\frac74,2,\frac52\}$; the values
$D_2(\frac{41}{40})$ and $D_\infty(\frac{41}{40})$; the excesses $e(c)$ at the
caps $c\in\{1,\frac65,\frac{27}{20},\frac32,\frac74,2,\frac52\}$; $G(1)$; and
the mean capped overshoots $h$. The clipped means are represented by
$\mu_c=1+D_c(1)$, not by separate coordinates. Some coefficients depend on the case through the
bounds $\mu_1\le r$ and $\mu_c\le R$ for $c>1$, and through $g=R-r$.

The \emph{coordinate box} places every independent stop-loss coordinate $D_c(x)$
in $[0,R]$ and every remaining independent coordinate in $[0,1]$, with the fixed
identities $S_1(1)=1$ and $D_\infty(1)=R-1$. The instance lies in this box: $D_c(x)\le\mu_c\le R$ by
Lemma~\ref{lem:prof-basic}, $0\le e(c)\le G(1)\le1$ by
Corollary~\ref{over:projection}, and $0\le h\le1$.

For the case whose interval contains $\mu_1$, every listed row is nonpositive at
the instance, and hence so is their weighted sum. The weighted sum is an affine
function of the unknowns. Its minimum over the coordinate box is obtained by
setting each unknown to the lower or upper end of its range according to the
sign of its coefficient. If this minimum is positive, we have a contradiction.

\subsection{The certificate}

We computed a certificate consisting of $96$ cases whose intervals are
consecutive and cover $[R-1,R]$ exactly. Their widths range from $1/1024$ to
$1/32$. For every case, the minimum of the weighted sum over the coordinate box
is positive; in fact it exceeds $1.3\cdot10^{-8}$ times the total multiplier of
the case. All quantities were computed in exact rational arithmetic. In
particular, every path minimum $\pi(j)$ was computed exactly by the dynamic
program of Section~\ref{sec:path-sets}, and every row is used only with
parameters that satisfy the hypotheses of the lemma that gives it, such as the
hypothesis on $\gamma$ in~\eqref{over:b3} and the condition
$t-1+\gamma(t-g)\ge\nu$ of Corollary~\ref{over:b4cor}. The certificate, with
the intervals, the parameters of every row and all multipliers, is available as
an ancillary file of the arXiv version of this paper.

The cases use $3\,837$ distinct rows. Of these, $1\,791$ are rows of
Section~\ref{sec:profile}, including $1\,750$ Riemann relations; $167$ are rows
of Section~\ref{sec:tail}; and $1\,879$ are rows of Section~\ref{sec:paths},
namely six path rows (one with $N=80$ and five with $N=160$, all in the second
rounding), one single-stage row and $1\,872$ budget rows. The cases contain
$205\,239$ multipliers in total. The only reserve shape that occurs is
$(m,K)=(3/2,22/5)$, which is admissible because $e^{3/2}>4.48>22/5$.

\subsection{Proof of the main theorem}

\begin{proof}[Proof of Theorem~\ref{thm:main}]
Fix an instance with $0<V<\infty$, and put $\rho=29/70$, $\rho_0=\rho+10^{-6}$ and
$R=1/\rho_0$. Suppose that every menu of Algorithm~\ref{alg:capped} with real
parameters has expected score at most $\rho_0V$.

\emph{Normalization.} By Lemma~\ref{lem:normalization}, applied with $\rho_0$ in
place of $\rho$, there is a normalized instance on which every menu of the family
has expected score at most one, $\E W=R$ and $W\ge1$. All results of
Sections~\ref{sec:profile}--\ref{sec:paths} apply to this instance.

\emph{Contradiction.} By Corollary~\ref{over:projection}, $\mu_1\in[R-1,R]$, so
$\mu_1$ lies in the interval $[l,r]$ of some case of the certificate. Every row
listed in this case is valid at the instance, and the quantities of the instance
lie in the coordinate box, so the weighted sum of the listed rows is nonpositive
there. On the other hand, the minimum of this sum over
the coordinate box is positive. This contradiction shows that some menu $M$ of
the family with real parameters has $\E b_M(X)>\rho_0V$ on the original instance.

\emph{Rational parameters.} We approximate the parameters of $M$ by sequences of
rationals: the cap from above; the body levels $t$ and $u$ and the reserve
fraction from below; and each singleton cutoff separately, from above for an
open test and from below for a closed test. Strict orderings between parameters
persist for all large indices. Fix a realization; we show that each elementary
test converges.
\begin{itemize}
\item Consider a closed test $y_n\ge\theta_n$, where the clipped values satisfy
 $y_n\downarrow y$ under the decreasing caps and the levels satisfy
 $\theta_n\uparrow\theta$. It passes for all $n$ if $y\ge\theta$, and it
 eventually fails if $y<\theta$.
\item The strict inequalities $w(S)-w_i<t$ of a seed test hold eventually if
 they hold in the limit. If $w(S)-w_i\ge t$, then also $w_n(S)-w_{n,i}\ge t_n$,
 since the clipped weights of the other members do not decrease.
\item Open cutoff tests $x_i>d_n$ with $d_n\downarrow d$, and closed cutoff tests
 $x_i\ge d_n$ with $d_n\uparrow d$, converge in the same way.
\item Consider a reserve test $w(S\setminus P)\ge a\,b_P$ for a fixed seed $P$.
 Under a larger cap, the increase of $w(P)$ is at least the increase of its
 lightest weight, as one sees by taking a member that is lightest under the
 original cap. Hence $b_P$ can only decrease under the approximation, while the
 sum over $S\setminus P$ can only increase. A test that passes in the limit
 therefore passes for all $n$, and a test that fails strictly eventually fails.
\end{itemize}
As there are finitely many reports, every acceptance decision, the family of
inclusion-maximal accepted reports, and hence the score converge pointwise. The
scores are dominated by $W$, which has finite expectation, so by dominated
convergence some menu $M'$ of the family with rational parameters has
$\E b_{M'}(X)>\rho_0V$. Thus every instance admits such a rational-parameter
menu. By Lemma~\ref{lem:score}, this stronger existence statement gives
$\alpha_R\ge\alpha_D\ge\rho_0>29/70$.

\emph{Selection.} The menus of the family with rational parameters form a
countable set. Algorithm~\ref{alg:select} therefore stops, at the first menu in
the fixed enumeration whose expected score exceeds $\rho V$, and this choice
depends only on $\F$ and the laws of the principal values.

\emph{Best responses.} Agent values are strictly positive, so by
Lemma~\ref{lem:score}, every best response of the agent to a deterministic menu
gives the principal at least the score $b_M(X)$, under any tie breaking. Hence
the principal's expected payoff is at least $\E b_M(X)>\rho V$.
\end{proof}

\paragraph{AI Use Statement.}
This work resulted from extensive interaction between the authors and AI systems, in particular OpenAI GPT-6.2 Astra, Anthropic Claude
Fable~5.2, Anthropic Claude Opus~5.5 and Google Gemini Neon, through iterative
exchanges in which the authors set the research directions, supplied suggestions,
and critically checked the proposed arguments. Anthropic Claude Opus~5.5 was also
used to assemble and revise the manuscript based on thorough discussion with the authors.

\begingroup
\small
\interlinepenalty=10000
\bibliographystyle{abbrvnat}
\bibliography{draft/refs}

\begin{thebibliography}{23}
\providecommand{\natexlab}[1]{#1}
\providecommand{\url}[1]{\texttt{#1}}
\expandafter\ifx\csname urlstyle\endcsname\relax
  \providecommand{\doi}[1]{doi: #1}\else
  \providecommand{\doi}{doi: \begingroup \urlstyle{rm}\Url}\fi

\bibitem[Abdi et~al.(2026{\natexlab{a}})Abdi, Banihashem, Hajiaghayi, Huang,
  Mittal, and Singla]{abdietal2026}
H.~Abdi, K.~Banihashem, M.~Hajiaghayi, Z.~Huang, D.~Mittal, and S.~Singla.
\newblock As easy as rank one: Optimal algorithms for matroid secretary and
  single-sample matroid prophet.
\newblock arXiv:2611.08869, 2026{\natexlab{a}}.
\newblock URL \url{https://arxiv.org/abs/2611.08869}.

\bibitem[Abdi et~al.(2026{\natexlab{b}})Abdi, Banihashem, Hajiaghayi, and
  Mittal]{abhm2026matroid}
H.~Abdi, K.~Banihashem, M.~Hajiaghayi, and D.~Mittal.
\newblock Optimal matroid prophets against almighty adversaries and delegation.
\newblock arXiv:2611.30933, 2026{\natexlab{b}}.
\newblock URL \url{https://arxiv.org/abs/2611.30933}.

\bibitem[Alonso and Matouschek(2008)]{alonsomatouschek2008}
R.~Alonso and N.~Matouschek.
\newblock Optimal delegation.
\newblock \emph{Review of Economic Studies}, 75\penalty0 (1):\penalty0
  259--293, 2008.

\bibitem[Armstrong and Vickers(2010)]{armstrongvickers2010}
M.~Armstrong and J.~Vickers.
\newblock A model of delegated project choice.
\newblock \emph{Econometrica}, 78\penalty0 (1):\penalty0 213--244, 2010.
\newblock \doi{10.3982/ECTA7965}.

\bibitem[Babaioff et~al.(2007)Babaioff, Immorlica, and Kleinberg]{babaioff2007}
M.~Babaioff, N.~Immorlica, and R.~Kleinberg.
\newblock Matroids, secretary problems, and online mechanisms.
\newblock In \emph{Proceedings of the 18th Annual ACM-SIAM Symposium on
  Discrete Algorithms (SODA)}, pages 434--443, 2007.

\bibitem[Banihashem et~al.(2025)Banihashem, Hajiaghayi, Krysta, and
  Shin]{banihashem2025}
K.~Banihashem, M.~Hajiaghayi, P.~Krysta, and S.~Shin.
\newblock Delegated choice with combinatorial constraints.
\newblock In \emph{Proceedings of the 26th ACM Conference on Economics and
  Computation (EC)}, pages 275--303, 2025.
\newblock \doi{10.1145/3736252.3742530}.
\newblock URL
  \url{https://livrepository.liverpool.ac.uk/3193865/1/Combinatorial_Delegated_Choice-1.pdf}.

\bibitem[Bechtel and Dughmi(2021)]{bechtel2021}
C.~Bechtel and S.~Dughmi.
\newblock Delegated stochastic probing.
\newblock In \emph{12th Innovations in Theoretical Computer Science Conference
  (ITCS)}, volume 185 of \emph{Leibniz International Proceedings in
  Informatics}, pages 37:1--37:19, 2021.
\newblock \doi{10.4230/LIPIcs.ITCS.2021.37}.

\bibitem[Bechtel et~al.(2022)Bechtel, Dughmi, and
  Patel]{bechteldughmipatel2022}
C.~Bechtel, S.~Dughmi, and N.~Patel.
\newblock Delegated {P}andora's box.
\newblock In \emph{Proceedings of the 23rd ACM Conference on Economics and
  Computation (EC)}, 2022.

\bibitem[Braun et~al.(2023)Braun, Hahn, Hoefer, and Schecker]{braun2023}
P.~Braun, N.~Hahn, M.~Hoefer, and C.~Schecker.
\newblock Delegated online search.
\newblock In \emph{Proceedings of the 32nd International Joint Conference on
  Artificial Intelligence (IJCAI)}, 2023.

\bibitem[Efron and Stein(1981)]{efronstein1981}
B.~Efron and C.~Stein.
\newblock The jackknife estimate of variance.
\newblock \emph{Annals of Statistics}, 9\penalty0 (3):\penalty0 586--596, 1981.

\bibitem[Feldman et~al.(2021)Feldman, Svensson, and Zenklusen]{feldman2021}
M.~Feldman, O.~Svensson, and R.~Zenklusen.
\newblock Online contention resolution schemes with applications to {Bayesian}
  selection problems.
\newblock \emph{SIAM Journal on Computing}, 50\penalty0 (2):\penalty0 255--300,
  2021.
\newblock \doi{10.1137/18M1226130}.

\bibitem[Fortuin et~al.(1971)Fortuin, Kasteleyn, and Ginibre]{fkg1971}
C.~M. Fortuin, P.~W. Kasteleyn, and J.~Ginibre.
\newblock Correlation inequalities on some partially ordered sets.
\newblock \emph{Communications in Mathematical Physics}, 22:\penalty0 89--103,
  1971.

\bibitem[Harris(1960)]{harris1960}
T.~E. Harris.
\newblock A lower bound for the critical probability in a certain percolation
  process.
\newblock \emph{Proceedings of the Cambridge Philosophical Society},
  56:\penalty0 13--20, 1960.

\bibitem[Holmstr{\"o}m(1984)]{holmstrom1984}
B.~Holmstr{\"o}m.
\newblock On the theory of delegation.
\newblock In M.~Boyer and R.~Kihlstrom, editors, \emph{Bayesian Models in
  Economic Theory}, pages 115--141. North-Holland, 1984.

\bibitem[Khodabakhsh et~al.(2024)Khodabakhsh, Pountourakis, and
  Taggart]{khodabakhsh2024}
A.~Khodabakhsh, E.~Pountourakis, and S.~Taggart.
\newblock Simple delegated choice.
\newblock In \emph{Proceedings of the 2024 Annual ACM-SIAM Symposium on
  Discrete Algorithms (SODA)}, 2024.

\bibitem[Kleinberg and Kleinberg(2018)]{kleinbergkleinberg2018}
J.~Kleinberg and R.~Kleinberg.
\newblock Delegated search approximates efficient search.
\newblock In \emph{Proceedings of the 19th ACM Conference on Economics and
  Computation (EC)}, pages 287--302, 2018.
\newblock \doi{10.1145/3219166.3219205}.
\newblock URL \url{https://arxiv.org/abs/1806.06933}.

\bibitem[Kleinberg and Weinberg(2012)]{kleinbergweinberg2012}
R.~Kleinberg and S.~M. Weinberg.
\newblock Matroid prophet inequalities.
\newblock In \emph{Proceedings of the 44th Annual ACM Symposium on Theory of
  Computing (STOC)}, 2012.
\newblock \doi{10.1145/2213977.2213991}.
\newblock URL \url{https://arxiv.org/abs/1201.4764}.

\bibitem[Lee and Singla(2018)]{leesingla2018}
E.~Lee and S.~Singla.
\newblock Optimal online contention resolution schemes via ex-ante prophet
  inequalities.
\newblock In \emph{26th Annual European Symposium on Algorithms (ESA)}, volume
  112 of \emph{Leibniz International Proceedings in Informatics}, pages
  57:1--57:14, 2018.
\newblock \doi{10.4230/LIPIcs.ESA.2018.57}.

\bibitem[Melumad and Shibano(1991)]{melumadshibano1991}
N.~D. Melumad and T.~Shibano.
\newblock Communication in settings with no transfers.
\newblock \emph{RAND Journal of Economics}, 22\penalty0 (2):\penalty0 173--198,
  1991.

\bibitem[Rubinstein(2016)]{rubinstein2016}
A.~Rubinstein.
\newblock Beyond matroids: Secretary problem and prophet inequality with
  general constraints.
\newblock In \emph{Proceedings of the 48th Annual ACM Symposium on Theory of
  Computing (STOC)}, pages 324--332, 2016.

\bibitem[Rubinstein and Singla(2017)]{rubinsteinsingla2017}
A.~Rubinstein and S.~Singla.
\newblock Combinatorial prophet inequalities.
\newblock In \emph{Proceedings of the 28th Annual ACM-SIAM Symposium on
  Discrete Algorithms (SODA)}, pages 1671--1687, 2017.

\bibitem[Samuel-Cahn(1984)]{samuelcahn1984}
E.~Samuel-Cahn.
\newblock Comparison of threshold stop rules and maximum for independent
  nonnegative random variables.
\newblock \emph{The Annals of Probability}, 12\penalty0 (4):\penalty0
  1213--1216, 1984.
\newblock \doi{10.1214/aop/1176993150}.

\bibitem[Steele(1986)]{steele1986}
J.~M. Steele.
\newblock An {E}fron--{S}tein inequality for nonsymmetric statistics.
\newblock \emph{Annals of Statistics}, 14\penalty0 (2):\penalty0 753--758,
  1986.

\end{thebibliography}
\endgroup

\clearpage
\appendix
\section{Impossibility below $0.48105$}\label{app:hardness}
\begingroup

\begin{theorem}\label{hard:main}
There is a sequence of finite instances, whose joint principal- and agent-value
laws are fixed before the acceptance rule, with $0<V_n<\infty$, such that
\[
 \limsup_{n\to\infty}\ \sup_\pi
 \frac{\E[\text{principal payoff under }\pi]}{V_n}
 \le \beta < \frac{9621}{20000} = 0.48105.
\]
Here the supremum ranges over all randomized acceptance rules, and in
particular over all deterministic rules, that know the joint laws of all items,
even under tie breaking in favor of the principal. Consequently, the universal
approximation constants in the stated model satisfy
$\alpha_D\le\alpha_R\le\beta<9621/20000=0.48105$.
\end{theorem}

The constant $\beta$ is defined in \eqref{hard:beta}. All item laws, including
the random agent values used to rank coupons, are fixed before the menu. The
theorem allows the menu to know these laws, but the acceptance test observes
only the principal values in the submitted report.

We first outline the construction and the proof. Each instance contains $n$
independent rare \emph{centers}, each active with probability $a/n$, and every
center owns a family of private alternatives, built from gates and disposable
offers, whose principal value exceeds $1$ only when the center is active. Every
pair of centers shares a pair coupon of deterministic value $b+d$. The instance
also contains rare tails, a deterministic safe item worth slightly more than
$\E[\max(1,N_A+N_B)]$, the expected value of a natural baseline choice, and an independent jackpot that
raises the benchmark by $1$ but that the agent ranks lowest. The agent
utilities are fixed in advance and separate the items by priority, so the agent
always proposes the highest-priority report that is accepted with
non-negligible probability. In Section~\ref{hard:sec:trigger}, we use this
priority structure to bound the principal's payoff by a baseline plus the gains
from two trigger events, one private and one coming from a coupon.
Section~\ref{hard:sec:priv} bounds the private gain: a dynamic program over the
gate configurations bounds the reward of each center by a concave function of
its trigger probability, and a Lagrangian duality argument turns these
per-center bounds into a bound on the total private gain.
Section~\ref{hard:sec:coupon} bounds the coupon gain through a race between
uniformly random coupon ranks. Section~\ref{hard:bench} computes a lower bound
on the benchmark $V_n$, which determines $\beta$, and Section~\ref{hard:finite}
lists the finite data and the remaining numerical inequalities.

\subsection{The instances}\label{hard:sec:instances}
This subsection defines the instances: the item values, the feasible family,
and the agent utilities.

We fix the rational parameters
\[
 a=\frac{67}{100},\qquad b=\frac{9}{200},\qquad d=\frac{9}{50},\qquad
 \Delta=\frac{1}{400},\qquad q=\frac{7}{20}.
\]
Each center $i\in\{1,\dots,n\}$ is equipped with five independent biased gates
$\gamma_{i,0},\dots,\gamma_{i,4}$, each active with probability $q=7/20$, and
with two ordered sequences of nonnested private items, each coupled to a subset
of the gates:
\begin{itemize}
\item Phase~$0$ consists of $L_0^{\rm ph}=45$ items
 $e^{(0)}_{i,1},\dots,e^{(0)}_{i,45}$. Item $r$ is coupled to a subset
 $J^{(0)}_r\subseteq\{0,1,2,3\}$, encoded by a binary mask in
 $\{0,\dots,15\}$, and never to $\gamma_{i,4}$; it has activation
 probability $p^{(0)}_r\in\{1/8,1/4,3/8,1/2,3/4,1\}$ and active principal
 value $v^{(0)}_r>0$.
\item Phase~$1$ consists of $L_1^{\rm ph}=39$ items
 $e^{(1)}_{i,1},\dots,e^{(1)}_{i,39}$. Item $r$ is coupled to a subset
 $J^{(1)}_r\subseteq\{0,1,2,3\}$, with mask $J^{(1)}_r\le 14$, together with
 $\gamma_{i,4}$; it has activation probability
 $p^{(1)}_r\in\{1/16,1/8,1/4,1/2,3/4,1\}$ and active principal value
 $v^{(1)}_r>0$.
\end{itemize}
An item with $p=1$ is a \emph{deterministic subset marker}, and an item with
$p\in(0,1)$ is an independent \emph{disposable offer}, worth $v$ with
probability $p$ and $0$ with probability $1-p$. In both phases, the masks $J_r$
are nondecreasing: each distinct subset starts with one deterministic subset
marker, which is optionally followed by one to four independent disposable
offers on the same subset with strictly ascending active values. In total,
Phase~$0$ has $16$ deterministic markers and $29$ disposable offers, and
Phase~$1$ has $12$ deterministic markers and $27$ disposable offers, so each
center has $28$ deterministic markers and $56$ independent dyadic disposable
offers.

The rational gate values $g_0,\dots,g_4$ and all $84$ phase entries
$(J^{(0)}_r,p^{(0)}_r,v^{(0)}_r)$ and $(J^{(1)}_r,p^{(1)}_r,v^{(1)}_r)$ are
listed in Section~\ref{hard:finite}. They satisfy
\begin{gather}\label{hard:sanity}
 0<a<\frac{7}{10},\qquad b>0,\qquad a+b<1,\qquad 0<d\le\frac{1}{3},\qquad b+d<1,\nonumber\\
 0\le\Delta<1,\qquad 0<q<1,\qquad g_j>0\ (0\le j<5),\qquad \sum_{j=0}^4 g_j<1,\\
 v^{(0)}_r>0,\quad v^{(0)}_r+\sum_{j\in J^{(0)}_r}g_j<1\ (1\le r\le 45),\qquad
 v^{(1)}_r>0,\quad v^{(1)}_r+g_4+\sum_{j\in J^{(1)}_r}g_j<1\ (1\le r\le 39).\nonumber
\end{gather}

For every integer $n\ge2$, we put $p=a/n$, $\eta=b/n$, $H=n^3$, and
$\delta=n^{-4}$. The instance consists of the following items, whose principal
values are mutually independent:
\begin{itemize}
\item Centers $c_1,\dots,c_n$, each worth $1$ with probability $p$ and $0$
 otherwise. We write $N_A=\sum_{i=1}^n X_{c_i}$ for the number of active
 centers.
\item Tails $t_1,\dots,t_n$, each worth $1$ with probability $\eta$ and $0$
 otherwise, and deterministic tail markers $\tau_k$ worth $(n-k)\eta$ for
 $0\le k<n$. We write $N_B=\sum_{j=1}^n X_{t_j}$ and $S_k=\sum_{j=1}^k X_{t_j}$.
\item A deterministic \emph{raised safe item} $s_0$ of value
\[
 C_n := v_n + \Delta,\qquad \text{where } v_n := a+b+(1-p)^n(1-\eta)^n = \E[\max(1,N_A+N_B)],
\]
 and an independent jackpot item $J_{\rm jp}$ worth $H$ with probability $1/H$
 and $0$ otherwise.
\item For each center $i\in\{1,\dots,n\}$, the five gates $\gamma_{i,j}$
 ($0\le j<5$), each worth $g_j$ with probability $q=7/20$ and $0$ otherwise;
 the Phase-$0$ items $e^{(0)}_{i,r}$, worth $v^{(0)}_r$ with probability
 $p^{(0)}_r$ and $0$ otherwise ($1\le r\le 45$); and the Phase-$1$ items
 $e^{(1)}_{i,r}$, worth $v^{(1)}_r$ with probability $p^{(1)}_r$ and $0$
 otherwise ($1\le r\le 39$).
\item For every unordered pair $\{i,j\}$ of distinct centers, a pair coupon
 $\kappa_{ij}$ with deterministic principal value $b+d$.
\end{itemize}

The feasible family $\mathcal F_n$ is the downward closure of the following
generators:
\begin{enumerate}
\item the singleton safe set $\{s_0\}$ and the singleton jackpot set $\{J_{\rm jp}\}$;
\item for every center $i$, the Phase-$0$ generators
 $\{e^{(0)}_{i,r},c_i\}\cup\{\gamma_{i,j}:j\in J^{(0)}_r\}$ ($1\le r\le 45$),
 the Phase-$1$ generators
 $\{e^{(1)}_{i,r},c_i,\gamma_{i,4}\}\cup\{\gamma_{i,j}:j\in J^{(1)}_r\}$
 ($1\le r\le 39$), and the terminal five-gate generator
 $\{c_i,\gamma_{i,0},\dots,\gamma_{i,4}\}$;
\item the pair-coupon generators $\{\kappa_{ij},c_i,c_j\}$ for every $1\le i<j\le n$;
\item the tail-marker generators $\{\tau_k,t_1,\dots,t_k,c_1,\dots,c_n\}$ for
 $0\le k<n$, and the combined set $\{c_1,\dots,c_n,t_1,\dots,t_n\}$.
\end{enumerate}

\paragraph{Agent utilities, fixed before the rule.}
Let $T_n$ be the total number of items, set $M=n^{10}$, and choose the integer
$B=2(T_n+1)n^4$. We order the noncoupon items from lowest to highest priority
as follows: the jackpot; the tails $t_n,\dots,t_1$; the centers
$c_n,\dots,c_1$; the tail markers $\tau_{n-1},\dots,\tau_0$; all private items,
in reverse of the priority order given below; and finally the raised safe item
$s_0$, which has the highest priority. The items up to and including the tail
markers receive distinct consecutive integer exponents $0,1,\dots,L-1$. Each
pair coupon $\kappa_{ij}$ independently draws a uniform random rank
$R_{ij}\in\{1,\dots,M\}$ and receives exponent $L+R_{ij}$. All private items
and $s_0$ receive distinct consecutive exponents above $L+M$, respecting their
stated order. Each item has agent utility $Y_e=B^{\mathrm{exp}(e)}>0$. In the
private priority order, from highest to lowest, center $1$ comes first, then
center $2$, and so forth through center $n$; within each center $i$, the order
from highest to lowest priority is
\[
 e^{(0)}_{i,1},\dots,e^{(0)}_{i,45},\qquad
 e^{(1)}_{i,1},\dots,e^{(1)}_{i,39},\qquad
 \gamma_{i,0},\gamma_{i,1},\gamma_{i,2},\gamma_{i,3},\gamma_{i,4}.
\]
The coupon ranks are independent of one another and of all principal draws.
Hence the item pairs $(X_e,Y_e)$ are mutually independent with $X_e\ge0$ and
$Y_e>0$, and their joint law is fixed before the acceptance rule.

Let $\mathcal E$ be the event that all $\binom{n}{2}$ coupon ranks $R_{ij}$ are
pairwise distinct. By the union bound,
$\PP(\mathcal E^c)\le \binom{\binom{n}{2}}{2}/M = O(n^{-6})$. On $\mathcal E$,
all item exponents are distinct. Any report that is accepted with probability
at least $\delta=n^{-4}$ and contains an item of exponent $e$ has expected
agent utility at least $\delta B^e$, whereas any report whose items all have
exponents strictly below $e$ has agent utility at most
$T_n B^{e-1}<\delta B^e$. Moreover, $s_0$ outranks every other item
deterministically, even off $\mathcal E$. We refer to these two facts as
\emph{priority separation}; they hold uniformly over all acceptance rules.

\subsection{Trigger reduction and failure salvage}\label{hard:sec:trigger}
In this subsection, we use priority separation to reduce the principal's
expected payoff to the baseline $v_n$ plus two gain terms, one from private
triggers and one from coupon triggers, which are bounded in the following two
subsections.

We call a nonempty report that does not contain the jackpot \emph{regular}.
Since $C_n=v_n+\Delta<2+1=3$, every regular report has principal value at most
$K_n=2n+3$. A regular report is \emph{substantial} if its acceptance
probability is at least $\delta$, and it is \emph{qualifying} if it is
substantial and has principal value strictly greater than $1$. Because the
acceptance rule observes only item identities and principal values,
qualification depends only on the principal realizations and never on the
random coupon ranks.

If $\{s_0\}$ is substantial, then by priority separation every best response of
the agent is $\{s_0\}$, since $s_0$ deterministically has the unique highest
exponent, and the expected payoff is at most $C_n=v_n+\Delta$. We assume
henceforth that $\{s_0\}$ is not substantial.

We partition the regular reports into priority classes according to their
highest-exponent item. In decreasing order of priority, the classes are
$\{s_0\}$, which is not substantial by assumption; the Phase-$0$ and Phase-$1$
private classes $e^{(0)}_{i,1},\dots,e^{(0)}_{i,45}$ and
$e^{(1)}_{i,1},\dots,e^{(1)}_{i,39}$ of the centers $1,\dots,n$, in the
displayed order, where each center also has one terminal private gate class
without markers following its Phase-$1$ classes; one class per pair coupon
$\kappa_{ij}$; the tail-marker classes in the order $\tau_0,\dots,\tau_{n-1}$;
and the bottom class of centers and tails. On $\mathcal E$, the coupon classes
are ordered by descending random rank.

If some qualifying report exists, let $c$ be the highest-priority class
containing one. By priority separation, no report from a strictly lower class,
including the jackpot, can be a best response. Hence any best response is a
nonsubstantial regular report, with expected payoff at most $\delta K_n$, a
substantial nonqualifying report, with value at most $1$, or a qualifying
report in class $c$, with value at most the largest principal value in
class $c$.

For center $i$, we write $(\omega,\omega_4)\in\{0,1\}^4\times\{0,1\}$ for its
five gate outcomes, where $\omega=(\omega_0,\omega_1,\omega_2,\omega_3)$, and
we write $\xi^{(0)}_{i,r}=\ind\{X_{e^{(0)}_{i,r}}>0\}$ and
$\xi^{(1)}_{i,r}=\ind\{X_{e^{(1)}_{i,r}}>0\}$ for its Phase-$0$ and Phase-$1$
activation indicators, with $\xi\equiv 1$ for deterministic markers. No
generator of $\mathcal F_n$ contains more than one item from
$\{e^{(0)}_{i,r},e^{(1)}_{i,r'}\}$. By \eqref{hard:sanity}, when center $i$ is
inactive ($X_{c_i}=0$), every private generator of center $i$ has principal
value strictly below $1$, so no private class of center $i$ can qualify. When
center $i$ is active ($X_{c_i}=1$), the maximum principal values of its
Phase-$0$, Phase-$1$, and terminal private classes are
$1+b+R^{(0)}_r(\omega,\xi^{(0)}_{i,r})$,
$1+b+R^{(1)}_r(\omega,\omega_4,\xi^{(1)}_{i,r})$, and
$1+b+R_*(\omega,\omega_4)$, respectively, where
\begin{align}
 R^{(0)}_r(\omega,\xi) &= v^{(0)}_r\xi + \sum_{j\in J^{(0)}_r} g_j \omega_j - b \qquad (1\le r\le 45),\nonumber\\
 R^{(1)}_r(\omega,\omega_4,\xi) &= v^{(1)}_r\xi + g_4\omega_4 + \sum_{j\in J^{(1)}_r} g_j \omega_j - b \qquad (1\le r\le 39),\label{hard:privrewards}\\
 R_*(\omega,\omega_4) &= \sum_{j=0}^4 g_j \omega_j - b.\nonumber
\end{align}

When a disposable offer fails ($\xi=0$), a report in its class
$\{e_{i,r},c_i\}\cup\{\text{observed gates}\}$ still has principal value
$1+\sum_{j\in J}g_j\omega_j$ in Phase~$0$ and
$1+g_4\omega_4+\sum_{j\in J}g_j\omega_j$ in Phase~$1$. Thus a failed
disposable offer retains what we call \emph{failure salvage}: its class can
qualify if and only if the observed gate sum is strictly positive. If the
observed gate sum is $0$, the maximum principal value in the class is $1$,
which does not strictly exceed $1$, so the class cannot qualify. Whenever a
private class qualifies, its maximum principal value $1+b+R$ strictly exceeds
$1$. A pair-coupon class $\kappa_{ij}$ has maximum value $b+d+r$, where
$r\in\{0,1,2\}$ counts its active endpoints; since $b+d<1$, qualification
requires $r\ge1$.

We call the highest-priority qualifying private class, if there is one, the
\emph{private trigger}. If there is no private trigger, we call the qualifying
coupon of largest random rank, if there is one, the \emph{coupon trigger}.
Neither trigger event involves the tails or the jackpot.

Now condition on all centers, all private gate and offer outcomes, and the
coupon ranks, and suppose that neither trigger occurs. Let
$\mathcal G_k=\sigma(t_1,\dots,t_k)$ for $0\le k\le n$, and define the Doob
martingale $M_k=\E[\max(1,N_A+N_B)\mid\mathcal G_k]$. Let
$\tau\in\{0,\dots,n\}$ be the index of the earliest qualifying tail-marker
class, with $\tau=n$ if none qualifies. Since reports in class $\tau_k$
observe only $t_1,\dots,t_k$ among the tails, $\tau$ is a bounded
$\mathcal G_k$-stopping time. For $\tau<n$, the maximum value of class
$\tau_k$ is $N_A+S_k+(n-k)\eta=\E[N_A+N_B\mid\mathcal G_k]$, so by Jensen's
inequality, $\max(1,N_A+S_k+(n-k)\eta)\le M_k$. For $\tau=n$, we fix the
regular realization. If a substantial regular report exists, it strictly
dominates the jackpot in agent utility for either jackpot realization, and
every substantial regular choice has principal value at most
$\max(1,N_A+N_B)$, since only a bottom-class report can qualify. If no
substantial regular report exists, then by independence the payoff is bounded
in conditional expectation by $\delta K_n+\E[X_{J_{\rm jp}}]=\delta K_n+1$. In
both cases, the conditional expected payoff is at most $\delta K_n+M_n$. By
optional stopping, the conditional expected payoff on the no-trigger event is
therefore at most $\delta K_n + M_0$. Whenever a private or coupon trigger
occurs, we have $N_A\ge1$, and so $M_0=\E[N_A+N_B\mid N_A]=N_A+b$.

Finally, on $\mathcal E^c$, the realized payoff is at most
$K_n + X_{J_{\rm jp}}$. Since $\mathcal E$ is independent of all principal
values, its expectation over $\mathcal E^c$ is
$(K_n+1)\PP(\mathcal E^c)=O(n^{-5})=o(1)$, uniformly over all rules.
Conditioning on $\mathcal E$ leaves all principal distributions unchanged and
makes the coupon ordering uniformly random and independent of the principal
draws. Taking expectations gives
\begin{equation}\label{hard:red}
 \E[\text{payoff}] \le v_n + \Gamma_{\rm priv} + \Gamma_{\rm coup} + o(1),
\end{equation}
uniformly over all acceptance rules for which $\{s_0\}$ is not substantial,
where
\begin{align*}
 \Gamma_{\rm priv} &= \E\big[(1+R-N_A)\,\ind\{\text{private trigger}\}\big],\\
 \Gamma_{\rm coup} &= \E\big[(d+r-N_A)\,\ind\{\text{no private trigger, coupon trigger}\}\big],
\end{align*}
and the coupon order in $\Gamma_{\rm coup}$ has the uniform law conditional on
$\mathcal E$. The baseline in \eqref{hard:red} is $v_n$ rather than $C_n$.
Hence, certifying
$\limsup(\Gamma_{\rm priv}+\Gamma_{\rm coup})\le\Delta+\varepsilon$ bounds the
payoff by $v+\Delta+\varepsilon$ both when $\{s_0\}$ is substantial and when
it is not.

\subsection{The private bound}\label{hard:sec:priv}
In this subsection, we bound $\Gamma_{\rm priv}$. We first show that the
expected reward of a single center is at most a concave function $F$ of its
trigger probability, by solving a dynamic program over a relaxed class of
menus, and we then combine the centers through a duality argument.

Fix an acceptance rule and condition on center $i$ being active. Let
$P_i\in[0,1]$ be the probability that some private class of center $i$
qualifies, and let $r_i$ be the expected reward $R$, as in
\eqref{hard:privrewards}, of its earliest qualifying class, where the reward is
$0$ if no class qualifies.

Every report in Phase-$0$ class $e^{(0)}_{i,r}$ is a subset of
$\{e^{(0)}_{i,r},c_i\}\cup\{\gamma_{i,j}:j\in J^{(0)}_r\}$. Whether this class
qualifies therefore depends only on $(\omega|_{J^{(0)}_r},\xi^{(0)}_{i,r})$;
it is independent of $\omega_4$, of $\{\omega_j:j\notin J^{(0)}_r\}$, and of
all other offer indicators; and it is false whenever $\xi^{(0)}_{i,r}=0$ and
$\sum_{j\in J^{(0)}_r}g_j\omega_j=0$. Similarly, every report in Phase-$1$
class $e^{(1)}_{i,r}$ is a subset of
$\{e^{(1)}_{i,r},c_i,\gamma_{i,4}\}\cup\{\gamma_{i,j}:j\in J^{(1)}_r\}$, so
whether it qualifies depends only on
$(\omega|_{J^{(1)}_r},\omega_4,\xi^{(1)}_{i,r})$, is independent of all other
offer indicators, and is false whenever $\xi^{(1)}_{i,r}=0$ and
$g_4\omega_4+\sum_{j\in J^{(1)}_r}g_j\omega_j=0$.

We relax the rule to the larger class of \emph{adaptive local menus}, in which
each class observes only its local gates but may adapt to the realized
higher-priority offers of the same center. Precisely, an adaptive local menu
specifies the following:
\begin{itemize}
\item in each Phase-$0$ class $r$, given the realizations of the
 higher-priority Phase-$0$ offer indicators
 $\xi^{(0)}_{i,1},\dots,\xi^{(0)}_{i,r-1}$, an arbitrary Boolean qualification
 predicate of $(\omega|_{J^{(0)}_r},\xi^{(0)}_{i,r})$ that is false whenever
 $\xi^{(0)}_{i,r}=0$ and $\sum_{j\in J^{(0)}_r}g_j\omega_j=0$;
\item in each Phase-$1$ class $r$, given $\omega_4\in\{0,1\}$ and the
 realizations of all higher-priority offer indicators (all of Phase~$0$ and
 $\xi^{(1)}_{i,1},\dots,\xi^{(1)}_{i,r-1}$), an arbitrary Boolean
 qualification predicate of $(\omega|_{J^{(1)}_r},\omega_4,\xi^{(1)}_{i,r})$
 that is false whenever $\xi^{(1)}_{i,r}=0$ and
 $g_4\omega_4+\sum_{j\in J^{(1)}_r}g_j\omega_j=0$;
\item in the terminal class, given all offer realizations, an arbitrary
 Boolean predicate of $(\omega,\omega_4)$ with reward $R_*(\omega,\omega_4)$.
\end{itemize}
Every globally fixed acceptance rule, deterministic or randomized, induces an
adaptive local menu that ignores the earlier offer realizations. Hence, for
every acceptance rule and every real fee $t$, the quantity $r_i-tP_i$ is at
most its maximum over adaptive local menus.

\subsubsection{A two-phase dynamic program}\label{hard:sec:twophase}
We now compute the maximum of $r_i-tP_i$ over adaptive local menus by a
backward recursion over the private classes, whose state is the set of core
gate worlds that have not yet triggered.

Let $0=P_0<P_1<\dots<P_N=1$ and $0=F_0,F_1,\dots,F_N$ have nonincreasing chord
slopes, and let $F:[0,1]\to\mathbb R$ be their concave piecewise-linear
interpolant. For each chord $j\in\{1,\dots,N\}$, we define its slope and
intercept
\[
 t_j = \frac{F_j-F_{j-1}}{P_j-P_{j-1}},\qquad e_j = F_{j-1}-t_j P_{j-1}.
\]

Write $q=q_{\rm num}/q_{\rm den}=7/20$ and $\bar q = 1-q = 13/20$. Every core
world $\omega\in\{0,1\}^4$ with Hamming weight $|\omega|=\sum_{j=0}^3\omega_j$
has probability
\[
 \mu_4(\omega) = q^{|\omega|}(1-q)^{4-|\omega|} = \frac{W_4(\omega)}{20^4},\qquad W_4(\omega):=7^{|\omega|}13^{4-|\omega|},
\]
and for a subset $A\subseteq\{0,1\}^4$, we write
$\mu_4(A)=\sum_{\omega\in A}\mu_4(\omega)$. The fifth gate has probability
$p_4(0)=\bar q=13/20$ for $b_4=0$ and $p_4(1)=q=7/20$ for $b_4=1$, so each
five-gate world $(\omega,b_4)$ has probability $p_4(b_4)\mu_4(\omega)$.

For a subset $J\subseteq\{0,1,2,3\}$ and a local assignment $u\in\{0,1\}^J$,
we write $C_{J,u}=\{\omega\in\{0,1\}^4:\omega|_J=u\}$ for its core support and
$g_J(u)=\sum_{j\in J}g_j u_j$. Consider a real fee $t$, a branch factor
$\alpha\in\{\bar q,q,1\}$, an extra gate value $g_{\rm add}\in\{0,g_4\}$, an
active bonus $v\ge 0$, a salvage flag $\sigma\in\{0,1\}$, and a value function
$U:\mathcal P(\{0,1\}^4)\to\mathbb R$. We define the pattern-update operator
$\mathcal T_{J,v,\sigma,\alpha,g_{\rm add},t}(U)$ by applying, sequentially
over all patterns $u\in\{0,1\}^J$ except those with $\sigma=1$ and
$g_{\rm add}+g_J(u)=0$, the update
\begin{equation}\label{hard:patupdate}
 U_{\rm next}(M) = \max\!\Bigl\{U(M),\ \alpha\,\mu_4(M\cap C_{J,u})\,\big(v+g_{\rm add}+g_J(u)-b-t\big) + U(M\setminus C_{J,u})\Bigr\}.
\end{equation}
Since the core supports $C_{J,u}$ are pairwise disjoint, the order of the
patterns does not matter.

Fix a real fee $t$. For each branch $b_4\in\{0,1\}$, we initialize at the
terminal class over subsets $M\subseteq\{0,1\}^4$ by
\begin{equation}\label{hard:dp_term}
 V^{(b_4)}_t(L_1^{\rm ph}+1,M) = p_4(b_4)\sum_{\omega\in M}\mu_4(\omega)\big(R_*(\omega,b_4)-t\big)_+.
\end{equation}
For $r=L_1^{\rm ph},\dots,1$, let $g_{\rm add}=b_4 g_4$, $\alpha=p_4(b_4)$,
$J=J^{(1)}_r$, $p_r=p^{(1)}_r$, and $v_r=v^{(1)}_r$, and define
\begin{equation}\label{hard:dp_ph1}
 V^{(b_4)}_t(r,M) = (1-p_r)\,\mathcal T_{J,0,1,\alpha,g_{\rm add},t}\!\bigl(V^{(b_4)}_t(r+1,\cdot)\bigr)(M)
 + p_r\,\mathcal T_{J,v_r,0,\alpha,g_{\rm add},t}\!\bigl(V^{(b_4)}_t(r+1,\cdot)\bigr)(M),
\end{equation}
where the $(1-p_r)$ term is omitted when $p_r=1$. We join the two Phase-$1$
value functions into Phase~$0$ by setting
\begin{equation}\label{hard:dp_bridge}
 V^{\rm ph0}_t(L_0^{\rm ph}+1,M) = V^{(0)}_t(1,M) + V^{(1)}_t(1,M),
\end{equation}
and for $r=L_0^{\rm ph},\dots,1$, with $J=J^{(0)}_r$, $p_r=p^{(0)}_r$,
$v_r=v^{(0)}_r$, and $\alpha=p_4(0)+p_4(1)=1$, we define
\begin{equation}\label{hard:dp_ph0}
 V^{\rm ph0}_t(r,M) = (1-p_r)\,\mathcal T_{J,0,1,1,0,t}\!\bigl(V^{\rm ph0}_t(r+1,\cdot)\bigr)(M)
 + p_r\,\mathcal T_{J,v_r,0,1,0,t}\!\bigl(V^{\rm ph0}_t(r+1,\cdot)\bigr)(M),
\end{equation}
again omitting the $(1-p_r)$ term when $p_r=1$.

\begin{lemma}[Factorization]\label{hard:lem:twophase}
Fix a real fee $t$. The quantity $V^{\rm ph0}_t(1,\{0,1\}^4)$ equals the
maximum of $r_i-tP_i$ over all adaptive local menus on
$\{0,1\}^5\times\prod_r\{0,1\}$ under the biased gate distribution $q$.
Therefore, $r_i-tP_i\le V^{\rm ph0}_t(1,\{0,1\}^4)$ for every globally fixed
acceptance rule. Consequently, if $V^{\rm ph0}_{t_j}(1,\{0,1\}^4)\le e_j$ for
every chord $j=1,\dots,N$, then $r_i\le F(P_i)$ for every center $i$ and
every acceptance rule.
\end{lemma}
\begin{proof}
Fix a realization of all higher-priority Phase-$0$ offer
indicators. Because Phase-$0$ predicates never observe $\omega_4$, they remove
the same core worlds on both branches $\omega_4=0$ and $\omega_4=1$. Hence the
worlds that have not triggered upon entering Phase~$1$ have the product form
$M\times\{0,1\}$ for a single subset $M\subseteq\{0,1\}^4$. In Phase~$1$ and
the terminal class, every predicate observes $\omega_4$, so an adaptive local
menu chooses its Phase-$1$ and terminal bits for $\omega_4=0$ and $\omega_4=1$
independently.

For fixed $b_4\in\{0,1\}$, we show by backward induction on
$r=L_1^{\rm ph}+1,\dots,1$ that $V^{(b_4)}_t(r,M)$ is the maximum contribution,
with each five-gate world weighted by its probability $p_4(b_4)\mu_4(\omega)$,
over all adaptive Phase-$1$ and terminal choices from class $r$ onward, given
the surviving core set $M$ on the branch $\omega_4=b_4$. The base case
$r=L_1^{\rm ph}+1$ is immediate from \eqref{hard:dp_term}. At step $r$, we
condition on the draw $\xi^{(1)}_{i,r}\in\{0,1\}$, which is independent of
$(\omega,\omega_4)$ and of all other offers. Given $\xi^{(1)}_{i,r}=1$, which
has probability $p_r$, the menu chooses one Boolean acceptance bit for each
pattern $u\in\{0,1\}^J$. Since the core supports $C_{J,u}$ partition
$\{0,1\}^4$, accepting pattern $u$ collects
$p_4(b_4)\mu_4(M\cap C_{J,u})(v_r+b_4 g_4+g_J(u)-b-t)$ on
$(M\cap C_{J,u})\times\{b_4\}$ and passes $M\setminus C_{J,u}$ to class $r+1$,
where the optimal continuation value is
$V^{(b_4)}_t(r+1,M\setminus C_{J,u})$. Applying \eqref{hard:patupdate}
sequentially over the disjoint supports $C_{J,u}$ therefore gives the
conditional optimum
$\mathcal T_{J,v_r,0,p_4(b_4),b_4 g_4,t}(V^{(b_4)}_t(r+1,\cdot))(M)$. Given
$\xi^{(1)}_{i,r}=0$, which has probability $1-p_r$, the same reasoning applies
with active bonus $0$ and with qualification forbidden whenever
$b_4 g_4+g_J(u)=0$, which gives
$\mathcal T_{J,0,1,p_4(b_4),b_4 g_4,t}(V^{(b_4)}_t(r+1,\cdot))(M)$. Taking the
expectation over $\xi^{(1)}_{i,r}$ yields \eqref{hard:dp_ph1}.

Summing the two independent branch optima at $r=1$ shows that
\eqref{hard:dp_bridge} is the optimal continuation value from Phase~$1$ and
the terminal class, given the surviving core set $M$ after Phase~$0$. The same
backward induction through Phase~$0$ proves \eqref{hard:dp_ph0}; here each core
world $\omega\in M$ represents both $(\omega,0)$ and $(\omega,1)$, with
combined probability $(p_4(0)+p_4(1))\mu_4(\omega)=\mu_4(\omega)$, that is,
$\alpha=1$. Setting $r=1$ and $M=\{0,1\}^4$ shows that
$V^{\rm ph0}_t(1,\{0,1\}^4)$ is the maximum over adaptive local menus, and
hence it bounds $r_i-tP_i$ for every globally fixed acceptance rule. If
$V^{\rm ph0}_{t_j}(1,\{0,1\}^4)\le e_j$ for all $j$, then
$r_i\le\min_j(e_j+t_j P_i)=F(P_i)$.
\end{proof}

\subsubsection{Offline private value}
We next record the offline value of the private alternatives, which enters the
benchmark in Section~\ref{hard:bench}.

For an active center, fix $(\omega,\omega_4)\in\{0,1\}^5$ and let
$v_{\det}(\omega,\omega_4)$ be the maximum of $R_*(\omega,\omega_4)$ and the
rewards of all deterministic Phase-$0$ and Phase-$1$ markers ($p_r=1$).
Whenever a disposable offer fails ($\xi=0$), its class achieves reward at most
$g_4\omega_4+\sum_{j\in J}g_j\omega_j-b\le R_*(\omega,\omega_4)\le
v_{\det}(\omega,\omega_4)$, so failed offers never improve the offline maximum.
Let $(r_1,p_1),\dots,(r_s,p_s)$ be the active rewards and activation
probabilities of all disposable offers, across both phases, that satisfy
$r_k>v_{\det}(\omega,\omega_4)$, ordered so that $r_1\ge r_2\ge\dots\ge r_s$.
Since the $56$ offers are mutually independent, the conditional offline premium
is
\begin{equation}\label{hard:L0cond}
 L(\omega,\omega_4) = \sum_{k=1}^s \Bigl(\prod_{\ell=1}^{k-1}(1-p_\ell)\Bigr)p_k r_k + \Bigl(\prod_{\ell=1}^s (1-p_\ell)\Bigr)v_{\det}(\omega,\omega_4),
\end{equation}
and the offline expectation above $1+b$ under the biased five-gate law is
\begin{equation}\label{hard:L0def}
 L_0 = \sum_{(\omega,\omega_4)\in\{0,1\}^5} p_4(\omega_4)\,\mu_4(\omega)\,L(\omega,\omega_4).
\end{equation}

\subsubsection{The private potential and duality}\label{hard:sec:control}
We now sum the per-center bounds $r_i\le F(P_i)$ over the centers. A time
multiplier $\nu$ decouples the centers, and the resulting potential $\Phi_\nu$
has a closed form.

Put $x_i=p\sum_{j<i}P_j$ and $x=p\sum_{j=1}^n P_j\in[0,a]$. Because the
branches of distinct centers are independent, the probability that no center
before $i$ triggers privately is $\Pi_i=\prod_{j<i}(1-pP_j)$. Conditionally on
this event and a private trigger at center $i$, the expected number of other
active centers is
\[
 (n-i)p + \sum_{j<i}\frac{p(1-P_j)}{1-pP_j} \ge a - p - x_i.
\]
Using $r_i\le F(P_i)$ from Lemma~\ref{hard:lem:twophase}, together with
$\log\Pi_i=-x_i+O(1/n)$ uniformly over all rules, we obtain
\begin{equation}\label{hard:privsum}
 \Gamma_{\rm priv}
 = \sum_{i=1}^n p\,\Pi_i\big[r_i - P_i\,\E(\text{other active centers})\big]
 \le \sum_{i=1}^n p\,e^{-x_i}\big[F(P_i) - (a-x_i)P_i\big] + O(1/n),
\end{equation}
with an implied constant depending only on $a$ and $\max F$.

For $\nu\ge0$ and $X\in[0,a]$, we define
\[
 f_{\nu,j}(X)=\frac{\nu}{P_j}+e^{-X}\Bigl(a-X-\frac{F_j}{P_j}\Bigr),\qquad
 \varphi_\nu(X)=\min_{1\le j\le N} f_{\nu,j}(X),\qquad
 \Phi_\nu(x)=\int_0^x \varphi_\nu(X)\,dX.
\]

\begin{lemma}[Time-multiplier inequality]\label{hard:lem:mult}
For all $\nu\ge0$, $X\in[0,a]$, and $P\in[0,1]$,
\[
 e^{-X}\big[F(P)-(a-X)P\big] \le \nu - P\,\varphi_\nu(X).
\]
\end{lemma}
\begin{proof}
Let $\psi(P)=\nu-e^{-X}[F(P)-(a-X)P]$, which is convex and piecewise-linear on
$[0,1]$ with $\psi(0)=\nu\ge0$ and $\psi(P_j)=P_j f_{\nu,j}(X)$. On $[0,P_1]$,
linearity gives
$\psi(P)=(1-P/P_1)\psi(0)+(P/P_1)\psi(P_1)\ge P f_{\nu,1}(X)\ge P\varphi_\nu(X)$.
On each segment $[P_{j-1},P_j]$ with $j\ge2$, we have $\psi(P)=\alpha+\beta P$,
so $\psi(P)/P=\beta+\alpha/P$ is monotone and therefore bounded below by
$\min(f_{\nu,j-1}(X),f_{\nu,j}(X))\ge\varphi_\nu(X)$.
\end{proof}

\begin{lemma}[Private potential bound]\label{hard:lem:privbound}
Assume that Lemma~\ref{hard:lem:twophase} holds for all chords of $F$ and
that $0\le F_j<P_j$ for all $1\le j\le N$. Then for every $\nu\ge0$, the
following hold.
\begin{enumerate}
\item The function $\varphi_\nu$ is strictly decreasing and $(1+a)$-Lipschitz
 on $[0,a]$; in particular, $-\Phi_\nu$ is convex on $[0,a]$.
\item $\Gamma_{\rm priv} \le \nu a - \Phi_\nu(x) + O(1/n)$, with an implied
 constant depending only on $a$ and $\max F$.
\end{enumerate}
\end{lemma}
\begin{proof}
For every $1\le j\le N$ and $X\in[0,a]$, we have $0\le a-X\le a$ and
$0\le F_j/P_j<1$, so $1-F_j/P_j \le 1+a-X-F_j/P_j \le 1+a$. Differentiating
$f_{\nu,j}$ gives
\[
 f'_{\nu,j}(X) = -e^{-X}\Bigl(1+a-X-\frac{F_j}{P_j}\Bigr)
 \in \Bigl[-(1+a),\ -e^{-a}\Bigl(1-\frac{F_j}{P_j}\Bigr)\Bigr].
\]
Hence $f'_{\nu,j}(X)<0$ and $|f'_{\nu,j}(X)|\le 1+a$ throughout $[0,a]$ for
every $j$. Taking the pointwise minimum over $1\le j\le N$ preserves strict
monotonicity and the Lipschitz constant $1+a$, so $\varphi_\nu$ is strictly
decreasing and $(1+a)$-Lipschitz. Consequently, $(-\Phi_\nu)'=-\varphi_\nu$ is
strictly increasing, and $-\Phi_\nu$ is convex on $[0,a]$. Applying
Lemma~\ref{hard:lem:mult} at $(X,P)=(x_i,P_i)$ to each summand of
\eqref{hard:privsum} and using $np=a$ yields
$\Gamma_{\rm priv}\le \nu a-\sum_{i=1}^n pP_i\varphi_\nu(x_i)+O(1/n)$. Since
the points $x_i$ partition $[0,x]$ into intervals of lengths $pP_i\le p$, the
left Riemann sum $\sum_i pP_i\varphi_\nu(x_i)$ differs from $\Phi_\nu(x)$ by at
most $(1+a)px/2\le (1+a)pa/2=O(1/n)$.
\end{proof}

Setting $A_j=\nu/P_j$ and $r_j=F_j/P_j$ and substituting $t=e^{-X}$, so that
$dX=-dt/t$, yields the representation
\begin{equation}\label{hard:exactpotential}
 \Phi_\nu(x) = a - 1 + (x+1-a)e^{-x} + \int_{e^{-x}}^1 \min_{1\le j\le N}\Bigl(\frac{A_j}{t}-r_j\Bigr)\,dt.
\end{equation}
For $t>0$, minimizing $A_j/t-r_j$ over $j$ is equivalent to minimizing the
affine function $A_j-r_jt$. All pairwise intersections of these affine
functions are rational when $\nu$, $P_j$, and $F_j$ are rational. On any
subinterval $[u,v]\subseteq[e^{-x},1]$ without interior intersections, a single
index $j$ is minimal throughout $[u,v]$, and it contributes the integral
\begin{equation}\label{hard:logpiece}
 \int_u^v \Bigl(\frac{A_j}{t}-r_j\Bigr)\,dt = A_j\log\frac{v}{u} - r_j(v-u).
\end{equation}

\subsection{Independent coupon ranks}\label{hard:sec:coupon}
In this subsection, we bound $\Gamma_{\rm coup}$. The key point is that the
coupon ranks are uniformly random and independent of the principal values, so
a coupon with two active endpoints rarely outranks the coupons with a single
active endpoint.

Fix an acceptance rule, and recall that, conditionally on center $i$ being
active, $P_i$ is its private trigger probability. Define
\[
 \lambda_i = \frac{p(1-P_i)}{1-p},\qquad s_n = \sum_{i=1}^n \lambda_i = \frac{a-x}{1-p}.
\]
Integrating out all private outcomes, namely the gates and the disposable
offers, the probability that no private trigger occurs and the active center
set equals $A\subseteq\{1,\dots,n\}$ is
\begin{equation}\label{hard:weights}
 (1-p)^n \prod_{i\in A}\lambda_i.
\end{equation}
Coupon qualification depends only on the endpoint principal values and the
fixed rule, so it is independent of the private outcomes and of the random
coupon ranks.

Fix the cutoff $J_0=12$. Every active set $A$ with $|A|>J_0$ has coupon excess
$d+r-|A|\le d+2-13<0$, so discarding the sets with $|A|>J_0$ gives an upper
bound on $\Gamma_{\rm coup}$. Along any sequence of acceptance rules, we pass
to a subsequence on which $x\to x_*\in[0,a]$, and we set $s=a-x_*$. If $s=0$,
then $\limsup\Gamma_{\rm coup}\le0$ directly from \eqref{hard:weights}.
Otherwise, $s_n\to s>0$, and the normalized probability weights
$\varpi_i=\lambda_i/s_n$ satisfy $\max_i\varpi_i=O(1/n)$. For every
$1\le m\le J_0$, the sum of \eqref{hard:weights} times the expected coupon
excess $\gamma(A)$ over supports $A$ with $m$ elements equals
\[
 (1-p)^n \frac{s_n^m}{m!}\,\E\big[\gamma(\{V_1,\dots,V_m\})\,\ind\{V_1,\dots,V_m\text{ distinct}\}\big],
\]
where $V_1,\dots,V_m$ are i.i.d.\ with law $\varpi$. Because
$|\gamma(A)|\le J_0+3$ is bounded and the probability of a collision among
$V_1,\dots,V_{J_0}$ is $O(1/n)$, the distinctness restriction changes the sum
by $o(1)$.

Let $U\subseteq\{1,\dots,n\}$ be the set of centers $i$ for which some coupon
report qualifies when $\{i\}$ is the only active center. For each $i\in U$, we
fix one such qualifying report, independently of the random ranks, and call it
the \emph{witness} of $i$; its coupon is incident to $i$ and has at most one
other endpoint $j(i)$. Among $V_1,\dots,V_{J_0}$, the probability that some
sampled pair equals $(i,j(i))$ for a sampled $i\in U$ is at most
$J_0(J_0-1)\max_\ell\varpi_\ell=O(1/n)$. Outside this collision event, the
witness of every sampled member of $U$ still qualifies on the full active
support $A$, because any inactive endpoint it includes remains inactive, so it
still has $r=1$ active endpoint. Moreover, the witness coupons of distinct
sampled members of $U$ are distinct, since a shared coupon would have these
two members as its endpoints and would create the excluded collision.

\begin{lemma}[Rank race]\label{hard:lem:race}
Suppose that an active support $A$ of size $m\ge1$ contains $k\ge1$ elements of
$U$ and avoids the witness-collision event. With $C_m=\binom{m}{2}$, its
expected coupon excess satisfies
\[
 \gamma(A) \le h_m(k) := 1 + d - m + \frac{C_m}{C_m+k},
\]
where the last fraction is interpreted as $0$ for $m=k=1$.
\end{lemma}
\begin{proof}
The $k$ distinct witness coupons force a coupon trigger, and each has $r=1$
active endpoint. At most $C_m$ coupons have both endpoints in $A$, and none of
them coincides with a witness coupon. Under the uniform random ordering of all
coupons on $\mathcal E$, each of these at most $C_m+k$ distinct coupons is
equally likely to rank highest, so the probability that some coupon with two
active endpoints outranks all $k$ witnesses is at most $C_m/(C_m+k)$.
Nonqualifying pair coupons and additional qualifying coupons with one active
endpoint can only lower this probability. Thus the triggering coupon has
expected active-endpoint count at most $1+C_m/(C_m+k)$, and its excess $d+r-m$
has expectation at most $h_m(k)$.
\end{proof}

Now consider supports $A\subseteq\{1,\dots,n\}\setminus U$. Any qualifying
coupon on such a support must have both endpoints active, because a qualifying
report with one active endpoint would place that endpoint in $U$, and a report
with no active endpoint has value $b+d<1$. Let $G$ be the graph on $U^c$ whose
edges are the pairs $ij$ for which $\{\kappa_{ij},c_i,c_j\}$ qualifies when
both endpoints are active. Then a coupon trigger occurs on $A\subseteq U^c$
precisely when $A$ contains an edge of $G$. Let $r_m$ be the probability that
$m$ i.i.d.\ samples from $\varpi|_{U^c}$ span an edge of $G$. Coupling all $m$
on one infinite i.i.d.\ sequence gives $0\le r_2\le r_3\le\dots\le1$. Writing
$w$ for the limiting weight of the complement, the total contribution of the
complement through $J_0$ is at most
\begin{equation}\label{hard:Kbound}
 \sum_{m=2}^{J_0}\frac{w^m}{m!}(d+2-m)r_m \le r_2 K(w) \le \max\{0,K(w)\},\qquad
 K(w):=\sum_{m=2}^{J_0}\frac{w^m}{m!}(d+2-m),
\end{equation}
because $d+2-m>0$ for $m=2$ and $d+2-m<0$ for all $m\ge3$, so replacing $r_m$
by $r_2\le r_m$ for $m\ge3$ can only increase the sum.

We pass to a further subsequence along which $\sum_{i\in U}\lambda_i\to
u\in[0,s]$. Splitting $m$ i.i.d.\ labels binomially between $U$, with mass $u$,
and $U^c$, with mass $s-u$, and applying Lemma~\ref{hard:lem:race} and
\eqref{hard:Kbound} gives
\begin{equation}\label{hard:SD}
 \limsup_{n\to\infty}\Gamma_{\rm coup} \le e^{-a}\big(S(u,s)+\max\{0,K(s-u)\}\big),
\end{equation}
where
\[
 S(u,s) = \sum_{m=1}^{J_0}\sum_{k=1}^m \frac{u^k(s-u)^{m-k}}{k!(m-k)!} h_m(k).
\]

\begin{lemma}[Endpoint convexity]\label{hard:lem:convex}
For every $0\le s\le a$ and $0\le u\le s$,
\[
 S(u,s)+\max\{0,K(s-u)\} \le D(s):=\max\{0,K(s),A(s)\},
\]
where
\[
 A(s) := \sum_{m=1}^{J_0}\frac{s^m}{m!}\Bigl(2+d-m-\frac{2}{m+1}\Bigr).
\]
\end{lemma}
\begin{proof}
The case $s=0$ is immediate. For $s>0$, write $t=u/s\in[0,1]$. For each
$m\ge2$, the degree-$m$ part of $S(u,s)$ is
$(s^m/m!)\sum_{k=0}^m\binom{m}{k}a_k t^k(1-t)^{m-k}$, with $a_0=0$ and
$a_k=h_m(k)$ for $k\ge1$. The second derivative in $t$ of this Bernstein
polynomial, without the factor $s^m/m!$, is
$m(m-1)\sum_{k=0}^{m-2}\binom{m-2}{k}(a_{k+2}-2a_{k+1}+a_k)t^k(1-t)^{m-2-k}$.
For $k\ge1$, the inequality $a_{k+2}-2a_{k+1}+a_k\ge0$ follows from convexity
of $k\mapsto C_m/(C_m+k)$. For $k=0$, we have
\[
 a_2 - 2a_1 + a_0 = m - 2 - d + \frac{2}{(C_m+1)(C_m+2)} \ge 0,
\]
since for $m=2$ this equals $1/3-d\ge0$ by \eqref{hard:sanity}, and for
$m\ge3$ it is strictly positive. The $m=1$ term is linear. Hence $S(\cdot,s)$
is convex on $[0,s]$.

For $S(u,s)+K(s-u)$, the $k=0$ coefficient of degree $m\ge2$ becomes
$a_0=d+2-m=1+d-m+C_m/C_m=h_m(0)$, so the sequence $(a_0,\dots,a_m)$ follows the
same convex formula, and $S(\cdot,s)+K(s-\cdot)$ is also convex. The maximum
of these two convex functions on $[0,s]$ is bounded by its values at the
endpoints. At $u=0$, it equals $\max\{0,K(s)\}$, and at $u=s$, it equals
$A(s)$, because $C_m/(C_m+m)=(m-1)/(m+1)$ and the $m=1$ coefficient is $d$.
\end{proof}

Combining \eqref{hard:red}, Lemma~\ref{hard:lem:privbound}, \eqref{hard:SD},
and Lemma~\ref{hard:lem:convex}, it suffices to certify that for every
$x\in[0,a]$ there is a multiplier $\nu\ge0$ in a finite set such that
\begin{equation}\label{hard:certificatecondition}
 \rho_\nu(x) := e^{-a}D(a-x) + \nu a - \Phi_\nu(x) < \Delta + \varepsilon,
 \qquad \Delta = \frac{1}{400},\quad \varepsilon = 10^{-6}.
\end{equation}
Indeed, for each $n$, choose a rule whose expected payoff is within $1/n$ of
the supremum. We take a subsequence attaining the limsup, and then a further
infinite subsequence on which the substantiality status of $\{s_0\}$ is
constant. In the substantial case, the expected payoff is at most
$C_n\to v+\Delta$, where $v=a+b+e^{-(a+b)}$. In the nonsubstantial case, we
extract further so that $x\to x_*$ and $\sum_{i\in U}\lambda_i\to u$. Choosing
the certified multiplier at $x_*$ and using the continuity of $D$ and
$\Phi_\nu$ gives the bound $v+\Delta+\varepsilon$ on the same limsup.

\subsection{The benchmark and the constant $\beta$}\label{hard:bench}
In this subsection, we bound the benchmark $V_n$ from below by exhibiting a
feasible choice on each realization, and we define the constant $\beta$.

Put $z=a+b=143/200$, $Q=e^{-z}$, $v=z+Q$, and $C=v+\Delta$. Let
$\ell\in\{0,\dots,n\}$ be the largest index of a successful tail, with $\ell=0$
if $N_B=0$. The feasible set $\{c_1,\dots,c_n,t_1,\dots,t_\ell,\tau_\ell\}$,
where $\tau_\ell$ is omitted if $\ell=n$, has principal value
$N_A+N_B+(n-\ell)\eta$. Because
$\E[(n-\ell)\eta]=\eta\sum_{j=1}^n(1-\eta)^j\to 1-e^{-b}$, this base choice
achieves limiting expectation $z+1-e^{-b}$. We improve it on four disjoint
events:
\begin{itemize}
\item If $N_A=N_B=0$, we choose the raised safe singleton $\{s_0\}$, whose
 value is $C_n=v_n+\Delta$ instead of the base value $b$; this adds
 $Q(v+\Delta-b)$ in the limit.
\item If $N_A=0$ and $N_B=1$, we choose $\{s_0\}$. The index of the unique
 successful tail is uniform on $\{1,\dots,n\}$, so the expected base value is
 $1+(n-1)\eta/2\to 1+b/2$, and the limiting improvement is
 $Qb(v+\Delta-1-b/2)$. By Section~\ref{hard:finite}, $v+\Delta-1>b$, which
 ensures that $\{s_0\}$ strictly beats the base choice at every tail index
 $\ell$ for all large $n$.
\item If $N_A=1$ and $N_B=0$, we choose the best private generator of the
 unique active center, whose expected value is $1+b+L_0$ instead of the base
 value $1+b$; the limiting improvement is $QaL_0$.
\item If $N_A=2$ and $N_B=0$, we choose the pair-coupon generator
 $\{\kappa_{ij},c_i,c_j\}$ for the active pair, whose value is $2+b+d$ instead
 of the base value $2+b$; the limiting improvement is $Q a^2 d/2$.
\end{itemize}
Consequently, the regular offline optimum satisfies
$\liminf_{n\to\infty}\E[\mathrm{OPT}_{\rm reg}]\ge W$, where
\begin{equation}\label{hard:Wdef}
 W = z + 1 - e^{-b} + Q\left[v + \Delta - b + b\Bigl(v + \Delta - 1 - \frac{b}{2}\Bigr) + a L_0 + \frac{a^2 d}{2}\right].
\end{equation}
Because the jackpot $J_{\rm jp}$ is independent and worth $H$ with probability
$1/H$, choosing $\{J_{\rm jp}\}$ whenever it is active and
$\mathrm{OPT}_{\rm reg}$ otherwise yields
$V_n\ge 1 + (1-1/H)\E[\mathrm{OPT}_{\rm reg}]$, so
$\liminf_{n\to\infty} V_n \ge 1+W$. We define
\begin{equation}\label{hard:beta}
 \beta = \frac{v + \Delta + 10^{-6}}{1 + W}.
\end{equation}
Section~\ref{hard:finite} certifies that $\beta < 9621/20000 = 0.48105$, which,
together with \eqref{hard:certificatecondition}, proves
Theorem~\ref{hard:main}.

\subsection{The finite certificate}\label{hard:finite}
This subsection lists the numerical data of the construction and the finite
inequalities used above, namely the chord bounds of
Lemma~\ref{hard:lem:twophase}, the condition \eqref{hard:certificatecondition},
and the bound on $\beta$. All of them were verified in exact rational
arithmetic.

\subsubsection{Parameters}
We use $a=67/100$, $b=9/200$, $d=9/50$, $\Delta=1/400$, and five independent
biased gates, each active with probability $q=7/20$. In units of $10^{-10}$,
the gate values are
\[
 g = (1063860742,\ 1322186165,\ 1613310011,\ 2140778234,\ 3059764500),
\]
with sum $9199899652/10^{10}<1$. Tables~\ref{hard:tab:ph0}
and~\ref{hard:tab:ph1} list all $45$ Phase-$0$ items and all $39$ Phase-$1$
items in priority order, from highest to lowest, with active principal values
in units of $10^{-10}$.

\begin{table}[htbp]\centering\small
\begin{tabular}{@{}rrl@{\qquad}rrl@{\qquad}rrl@{}}
\toprule
$J$ & $p$ & $v\cdot 10^{10}$ & $J$ & $p$ & $v\cdot 10^{10}$ & $J$ & $p$ & $v\cdot 10^{10}$\\
\midrule
$0$ & $1$ & $4202791876$ & $7$ & $3/8$ & $1395208298$ & $11$ & $1/2$ & $1395613379$\\
$1$ & $1$ & $3213031524$ & $7$ & $1/4$ & $1651051908$ & $12$ & $1$ & $1401807157$\\
$2$ & $1$ & $3011116304$ & $7$ & $1/8$ & $1667466912$ & $12$ & $1/2$ & $1612386952$\\
$2$ & $3/4$ & $3049627003$ & $8$ & $1$ & $2307491578$ & $12$ & $1/4$ & $1621205876$\\
$3$ & $1$ & $2084974000$ & $8$ & $3/4$ & $2361836720$ & $13$ & $1$ & $1017614262$\\
$4$ & $1$ & $2723663716$ & $8$ & $1/2$ & $2425463429$ & $13$ & $3/4$ & $1043090462$\\
$4$ & $3/4$ & $2759691658$ & $8$ & $1/4$ & $2471039066$ & $13$ & $1/2$ & $1077163296$\\
$4$ & $1/2$ & $2848146280$ & $9$ & $1$ & $1633976535$ & $13$ & $1/4$ & $1125511004$\\
$5$ & $1$ & $1954139616$ & $9$ & $3/4$ & $1676157594$ & $14$ & $1$ & $975640596$\\
$5$ & $3/4$ & $1993045243$ & $10$ & $1$ & $1527086959$ & $14$ & $1/2$ & $1005560602$\\
$5$ & $1/2$ & $2014825864$ & $10$ & $3/4$ & $1543846787$ & $14$ & $1/4$ & $1043178169$\\
$6$ & $1$ & $1784516708$ & $10$ & $1/2$ & $1627406940$ & $14$ & $1/8$ & $1157094029$\\
$6$ & $3/4$ & $1790945508$ & $10$ & $1/4$ & $1656728646$ & $15$ & $1$ & $759131158$\\
$6$ & $1/2$ & $1931791608$ & $11$ & $1$ & $1189630358$ & $15$ & $3/4$ & $858001920$\\
$7$ & $1$ & $1354734356$ & $11$ & $3/4$ & $1227035267$ & $15$ & $1/2$ & $894276990$\\
\bottomrule
\end{tabular}
\caption{Ordered Phase-$0$ items $(J^{(0)}_r,p^{(0)}_r,v^{(0)}_r)$, read
column by column: $16$ deterministic markers and $29$ dyadic disposable
offers.}\label{hard:tab:ph0}
\end{table}

\begin{table}[htbp]\centering\small
\begin{tabular}{@{}rrl@{\qquad}rrl@{\qquad}rrl@{}}
\toprule
$J$ & $p$ & $v\cdot 10^{10}$ & $J$ & $p$ & $v\cdot 10^{10}$ & $J$ & $p$ & $v\cdot 10^{10}$\\
\midrule
$0$ & $1$ & $1970668731$ & $4$ & $1/2$ & $1324256220$ & $8$ & $1$ & $569333828$\\
$0$ & $3/4$ & $2027048314$ & $4$ & $1/4$ & $1333509509$ & $8$ & $1/2$ & $591275305$\\
$0$ & $1/2$ & $2044082689$ & $5$ & $1$ & $730944143$ & $8$ & $1/4$ & $772702104$\\
$0$ & $1/4$ & $2080453071$ & $5$ & $3/4$ & $752793850$ & $8$ & $1/8$ & $955152635$\\
$1$ & $1$ & $1383777685$ & $5$ & $1/2$ & $836361223$ & $8$ & $1/16$ & $1076454790$\\
$2$ & $1$ & $1346782929$ & $5$ & $1/4$ & $903257677$ & $9$ & $1$ & $352908862$\\
$2$ & $3/4$ & $1352350844$ & $5$ & $1/8$ & $981704606$ & $11$ & $1$ & $265750800$\\
$2$ & $1/2$ & $1433651994$ & $6$ & $1$ & $689428306$ & $11$ & $1/2$ & $377632371$\\
$2$ & $1/4$ & $1514685701$ & $6$ & $3/4$ & $708834566$ & $11$ & $1/4$ & $428281060$\\
$2$ & $1/8$ & $1697281450$ & $6$ & $1/2$ & $784005826$ & $11$ & $1/8$ & $499188937$\\
$3$ & $1$ & $1013285156$ & $6$ & $1/4$ & $945286049$ & $13$ & $1$ & $267809261$\\
$3$ & $3/4$ & $1081281321$ & $7$ & $1$ & $308171511$ & $13$ & $1/2$ & $420075553$\\
$4$ & $1$ & $1130431825$ & $7$ & $3/4$ & $412771243$ & $13$ & $1/4$ & $462857287$\\
\bottomrule
\end{tabular}
\caption{Ordered Phase-$1$ items $(J^{(1)}_r,p^{(1)}_r,v^{(1)}_r)$, read
column by column: $12$ deterministic markers and $27$ dyadic disposable
offers.}\label{hard:tab:ph1}
\end{table}

With these rational values, \eqref{hard:L0cond}--\eqref{hard:L0def} give
\[
 L_0 = \frac{15733850124938532977}{32768000000000000000} \approx 0.480159000395.
\]

The envelope $F$ has the $N=21$ dyadic knots
\[
 P = \Bigl(\frac{1}{128},\frac{1}{64},\frac{3}{128},\frac{1}{32},\frac{3}{64},\frac{1}{16},\frac{5}{64},\frac{3}{32},\frac{7}{64},\frac{1}{8},\frac{5}{32},\frac{3}{16},\frac{1}{4},\frac{5}{16},\frac{3}{8},\frac{7}{16},\frac{1}{2},\frac{5}{8},\frac{3}{4},\frac{7}{8},1\Bigr);
\]
and rational values $F_j$, with $F_1/P_1=875061/1000000=0.875061$ and
$F_{21}=24019923/64000000$; exactly $8$ knots satisfy $F_j/P_j>a$. The
multipliers in \eqref{hard:certificatecondition} are $1024$ rational numbers
$\nu_i$, multiples of $10^{-7}$, one for each of the cells of $[0,a]$ of width
$a/1024$. The values $F_j$ and $\nu_i$, together with all other data of this
subsection, are available as an ancillary file of the arXiv version of this
paper. With these data, the following hold.
\begin{enumerate}
\item all parameter inequalities \eqref{hard:sanity}, the ordering and strict
 monotonicity of every cascade, the concavity of the envelope $F$ with $N=21$
 knots, and the strict bound $0\le F_j<P_j$ for all $1\le j\le 21$;
\item for every chord $j\in\{1,\dots,21\}$, the inequality
 $V^{\rm ph0}_{t_j}(1,\{0,1\}^4)\le e_j$ of Lemma~\ref{hard:lem:twophase},
 where $V^{\rm ph0}_{t_j}$ is given by the recursion of
 Section~\ref{hard:sec:twophase};
\item the inequality \eqref{hard:certificatecondition}, uniformly on all
 $1024$ mesh cells covering $[0,a]$, with worst certified upper bound
 $\le 2.499128087\times 10^{-3}<\Delta+\varepsilon=2.501\times 10^{-3}$;
\item $v+\Delta-1>b$, the positivity of the bracket in \eqref{hard:Wdef}, and
 the strict rational upper bound
 $\beta\le 0.481043999746<9621/20000=0.48105$.
\end{enumerate}

\subsubsection{Exponentials, logarithms, and convexity on cells}
The last two claims involve exponentials and logarithms, and the third is a
statement about whole cells. We reduce them to finitely many rational
inequalities as follows.

For every rational $0\le t\le1$, the consecutive Taylor partial sums of
degrees $18$ and $19$ bracket $e^{-t}$ by the alternating series theorem,
since $t^k/k!$ decreases for $k\ge1$. Since $a+b=143/200<1$, every exponential
that occurs, namely $e^{-x}$ for $x\in[0,a]$, $e^{-a}$, $e^{-b}$ and
$e^{-(a+b)}$, has its argument in $[0,1]$. For a rational $x\in[0,a]$, let
$\ell\le e^{-x}\le h$ be this bracket. Then $\ell\ge1/2$, because
$a=67/100<\ln 2$.

To bound the integral in \eqref{hard:exactpotential} from below, partition
$[\ell,1]$ at the pairwise intersections of the affine functions $A_k-r_kt$
that lie in $(\ell,1)$. On each resulting segment $[u,v]$, a single index $j$
is minimal, and \eqref{hard:logpiece} applies. Since $1/2\le\ell\le u<v\le1$,
the ratio $y=v/u$ lies in $[1,2]$, so $z=(y-1)/(y+1)\in[0,1/3]$ and
\[
 \log y = 2\sum_{k=0}^{\infty}\frac{z^{2k+1}}{2k+1}
 \ge 2\sum_{k=0}^{15}\frac{z^{2k+1}}{2k+1}.
\]
Since $A_j\ge0$, substituting this bound into \eqref{hard:logpiece} gives a
rational lower bound on $\int_\ell^1 \min_j(A_j/t-r_j)\,dt$. Moving the lower
endpoint of integration from $\ell$ to $e^{-x}\in[\ell,h]$ subtracts at most
$(h-\ell)\max_j(A_j/\ell+r_j)$, since
$\min_j(A_j/t-r_j)\le\max_j(A_j/\ell+r_j)$ on $[\ell,e^{-x}]$. Moreover, since
$x+1-a>0$, the quantity $a-1+(x+1-a)\ell$ is a lower bound on the elementary
term. Together, these give a rational lower bound on $\Phi_\nu(x)$.

For convexity on cells, each polynomial $q_{\rm poly}\in\{0,K,A\}$ in the
definition of $D(s)$ has coefficient at least $1-m$ on $s^m/m!$ for all
$m\ge2$, so
\[
 q_{\rm poly}''(s) \ge -\sum_{m=2}^\infty (m-1)\frac{s^{m-2}}{(m-2)!} = -(1+s)e^s \qquad (0\le s\le a).
\]
Hence $e^{-a}q_{\rm poly}''(a-x)\ge -e^{-a}(1+a-x)e^{a-x}\ge -(1+a) > -2$,
since $a=67/100<1$. Moreover, by Lemma~\ref{hard:lem:privbound}(1), which
requires only the condition $0\le F_j<P_j$ verified for all $21$ knots,
$\varphi_\nu$ is strictly decreasing and $-\Phi_\nu$ is convex on $[0,a]$.
Therefore $e^{-a}q_{\rm poly}(a-x)+\nu a-\Phi_\nu(x)+x^2$ is convex for each
$q_{\rm poly}$, and taking the maximum over $q_{\rm poly}$ preserves convexity.
It follows that, on any mesh cell $[L,L+H_0]$ with $H_0=a/1024$,
\[
 \max_{x\in[L,L+H_0]} \rho_\nu(x)
 \le \max\big\{\rho_\nu(L),\ \rho_\nu(L+H_0)\big\} + \frac{2H_0^2}{8}.
\]
The third claim therefore follows from upper bounds on $\rho_\nu$ at the
endpoints of each cell, with the multiplier $\nu$ of that cell, obtained from
the upper Taylor bound for $e^{-a}$ (since $D\ge0$) and the lower bound on
$\Phi_\nu$ above.

For the fourth claim, $L_0$ is the rational number computed above, and the
Taylor brackets for $e^{-(a+b)}$ and $e^{-b}$ give bounds
$v_{\rm lo}+\Delta\le v+\Delta\le v_{\rm hi}+\Delta$ and
$W_{\rm lo}\le W\le W_{\rm hi}$, using the opposite exponential endpoints for
the upper bound. Dividing the lower numerator by $1+W_{\rm hi}$ and the upper
numerator by $1+W_{\rm lo}$ gives the stated bound on $\beta$. Together with the analytic reductions of
Sections~\ref{hard:sec:trigger}--\ref{hard:bench}, this completes the proof of
Theorem~\ref{hard:main}.\qed

\endgroup

\end{document}